\documentstyle[twoside,fleqn,espcrc2]{article}
\newcommand{\ft}[2]{{\textstyle\frac{#1}{#2}}}

\def\dop{{\rm d}\hskip -1pt}
\def\bfone{\relax{\rm 1\kern-.35em 1}}
\def\bfzero{\relax{\rm I\kern-.18em 0}}
\def\inbar{\vrule height1.5ex width.4pt depth0pt}
\def\IC{\relax\,\hbox{$\inbar\kern-.3em{\rm C}$}}
\def\ID{\relax{\rm I\kern-.18em D}}
\def\IF{\relax{\rm I\kern-.18em F}}
\def\IH{\relax{\rm I\kern-.18em H}}
\def\II{\relax{\rm I\kern-.17em I}}
\def\IN{\relax{\rm I\kern-.18em N}}
\def\IP{\relax{\rm I\kern-.18em P}}
\def\IQ{\relax\,\hbox{$\inbar\kern-.3em{\rm Q}$}}
\def\IR{\relax{\rm I\kern-.18em R}}
\def\IG{\relax\,\hbox{$\inbar\kern-.3em{\rm G}$}}
\font\cmss=cmss10 \font\cmsss=cmss10 at 7pt
\def\ZZ{\relax\ifmmode\mathchoice
{\hbox{\cmss Z\kern-.4em Z}}{\hbox{\cmss Z\kern-.4em Z}}
{\lower.9pt\hbox{\cmsss Z\kern-.4em Z}}
{\lower1.2pt\hbox{\cmsss Z\kern-.4em Z}}\else{\cmss Z\kern-.4em
Z}\fi}

\def\G{\Gamma} 
\def\L{\Lambda}

\def\cF{{\cal F}} 
 
 \def\cK{{\cal K}}
\def\cL{{\cal L}} \def\cM{{\cal M}}
\def\cN{{\cal N}}

\def\tilde{\widetilde}
\def\bar{\overline}

\def\hat{\widehat}

\def\Coe#1.#2.{{#1\over #2}}

\def\coe#1.#2.{\relax{\textstyle {#1 \over #2}}\displaystyle}

\def\to{\rightarrow}
\def\notin{\hbox{{$\in$}\kern-.51em\hbox{/}}}

\def\del{\partial}

\def\IE{\relax{{\rm I\kern-.18em E}}}

\def\IGam{\relax{{\rm I}\kern-.18em \Gamma}}

\def\IA{\relax{\hbox{{\rm A}\kern-.82em {\rm A}}}}

\def\o#1#2{{{#1}\over{#2}}}
\newcommand{\be}{\begin{equation}}
\newcommand{\ee}{\end{equation}}
\newcommand{\ba}{\begin{eqnarray}}
\newcommand{\ea}{\end{eqnarray}}
\newtheorem{definizione}{Definition}[section]
\newcommand{\bd}{\begin{definizione}}
\newcommand{\ed}{\end{definizione}}
\newtheorem{teorema}{Theorem}[section]
\newcommand{\bth}{\begin{teorema}}
\newcommand{\eth}{\end{teorema}}
\newtheorem{lemma}{Lemma}[section]
\newcommand{\blem}{\begin{lemma}}
\newcommand{\elem}{\end{lemma}}
\newcommand{\brr}{\begin{array}}
\newcommand{\err}{\end{array}}
\newcommand{\nn}{\nonumber}
\newtheorem{corollario}{Corollary}[section]
\newcommand{\bcorol}{\begin{corollario}}
\newcommand{\ecorol}{\end{corollario}}
\def\twomat#1#2#3#4{\left(\begin{array}{cc}
 {#1}&{#2}\\ {#3}&{#4}\\
\end{array}
\right)}
\def\twovec#1#2{\left(\begin{array}{c}
{#1}\\ {#2}\\
\end{array}
\right)}

\newcommand{\AmS}{{\protect\the\textfont2
  A\kern-.1667em\lower.5ex\hbox{M}\kern-.125emS}}
\title{LECTURES ON SPECIAL KAHLER GEOMETRY \\
AND \\
ELECTRIC--MAGNETIC DUALITY ROTATIONS}
\author{Pietro Fr\'e\address{Dipartimento di Fisica Teorica,
Universit\'a di Torino, \\
Via P. Giuria 1, I-10125 TORINO, Italy}%
        \thanks{Email: Fre@toux40.to.infn.it } }
\begin{document}
\begin{abstract}
In these lecture I review the general structure of
electric--magnetic duality rotations in every even
space--time dimension. In four dimensions, which is
my main concern, I discuss the general issue
of symplectic covariance and how it relates to the typical
geometric structures involved by   N=2 supersymmetry, namely
Special K\"ahler geometry for the vector multiplets
and either HyperK\"ahler or Quaternionic geometry for the
hypermultiplets.
I discuss classical continuous dualities versus non--perturbative discrete
dualities. How the moduli space geometry of an auxiliary dynamical
Riemann surface  (or Calabi--Yau threefold) relates
to exact space--time dualities is exemplified in detail
for the Seiberg Witten model of an $SU(2)$ gauge theory.
\end{abstract}
\maketitle
\section{Introduction and Historical Remarks }
\label{intro}
Electric--Magnetic Duality is an old idea. It was introduced by
Dirac \cite{DIRMON} in the context of an {\it abelian gauge theory}
and lead to the derivation of the Dirac quantization rule for the magnetic
charge $q_m$ relative to the electric charge $q_e$:
\begin{equation}
\o{q_e \, q_m}{4 \pi \, \hbar} \, = \, \o{n}{2} \qquad n \, \in \, \ZZ
\label{dirquant}
\end{equation}
In the seventies, after the discovery of the 't Hooft--Polyakov
non--abelian monopole solution \cite{HOFPOLMON}, the studies on
electro--magnetic dualities were resurrected in a new context and
lead to generalizations of eq.~\ref{dirquant} where the
interpretation of the magnetic charge as a topological quantum number
plays a crucial role. The theory of monopoles made a substantial
progress in the context of {\it non--abelian gauge theories}, spontaneously
broken to an abelian phase by {\it Higgs fields} that transform in the
{\it adjoint representation} of the gauge group. The Montonen--Olive
conjecture \cite{MONTOLIVE} developed in this context establishes a
well--defined correspondence between an electric spontaneously
broken {\it non abelian} gauge theory and a dual magnetic gauge
theory which is abelian. In the dual theory the unbroken electric
subgroup ${\cal H} \, \subset \, {\cal G}$
is replaced by a dual magnetic group ${\cal H}^v$, such that the
weight--lattice of
the former is dual to the weight--lattice of the latter, (this is the
suitable generalization of eq.~\ref{dirquant} ), the magnetic
monopoles appear as elementary excitations rather than solitary
waves while it is the massive non--abelian gauge bosons that are
generated as soliton solutions of the non linear field equations. Finally
the original coupling constant is replaced by its inverse:
\begin{equation}
g^2 \, \longrightarrow \, \o{1}{g^2}
\label{strongweak}
\end{equation}
relating, in this way, the {\it weak coupling regime} of the magnetic theory
to the {\it strong coupling regime} of the electric one and viceversa.
\par
The very fact that scalars fields in the adjoint representation play
a fundamental role  shows the relation of
electro--magnetic duality with supersymmetry, in particular with
extended supersymmetry. Indeed it is a feature of $N \ge 2$
SUSY that the vector multiplets including the gauge bosons
contain also scalar partners which, out of necessity, transform in
the same representation as the vectors, i.e. the adjoint. This
association was discovered early by Witten and Olive who identified the
{\it topological charges} leading to the {\it magnetic charges}
with the {\it central charges} of extended supersymmetry algebras
\cite{WITOL}.
\par
The {\it group} of {\it electric--magnetic}  duality rotations
${\Gamma}_D$ is, by definition, that group of transformations that,
while implementing the exchange of strong--weak regimes
eq.~\ref{strongweak}, takes electric field strengths $F_{\mu\nu}^\Lambda$
into linear combinations of the same with their magnetic duals
\begin{equation}
{\tilde F}^{\mu\nu\,\Lambda}={1 \over 2}\,\varepsilon^{\mu\nu\rho\sigma}\,
F_{\rho\sigma}^\Lambda
\end{equation}
and viceversa. In such a process Bianchi
identities are exchanged with field equations and electric currents
are exchanged with magnetic currents. This has three immediate
consequences:
\par
a) The duality group $\Gamma_D$ is not, in general, a symmetry of the
action since it involves a non--local transformation from electric
gauge fields $A_\mu^\Lambda$ to magnetic ones $B_\mu^\Lambda$. Yet
it is a symmetry of the complete quantum theory including all the
topological sectors.
\par
b) In view of the Dirac quantization condition ~\ref{dirquant}, or of
its non--abelian generalizations,  $\Gamma_D$  should be a discrete
rather than a continuous group. Indeed it should be an automorphism
group for the lattice $\Lambda_{em}$ containing the  electric and the
magnetic charges that are both quantized.
\par
c) The determination of the exact duality group $\Gamma_D$ involves
a complete control over the non--perturbative quantum corrections to
the theory under consideration. Indeed by a duality rotation the
perturbative expansion is mapped into a non--perturbative one.
\par
Recently a great deal of work has been made in developing new ideas
and new strategies apt to determine the exact duality group
$\Gamma_D$, both in field theories and in string theories possessing
an N=2 target supersymmetry \cite{SW1},
\cite{SW2}, \cite{kltold}, \cite{faraggi},
\cite{CDF},\cite{CDFVP}, \cite{AFGNT},\cite{DWKLL},
\cite{HS}, \cite{stromco}, \cite{FHSV}, \cite{kava},
\cite{lopez}, \cite{noi}, \cite{cynoi}.
These efforts have been a follow up of similar
considerations previously developed in N=4 models, (for a review
see in particular \cite{sen4},\cite{senlecture} and references therein)
where the Montonen--Olive conjecture is particularly nice (differently from
the N=2 case, here both the elementary gauge
fields and the magnetic monopoles sit in the same type of supermultiplet
that is in a vector multiplet) and where the
vanishing of the beta--function guarantees the absence of true quantum
corrections.
\par
In these studies the focus of interest is on a non--abelian gauge
group, characterized by a non vanishing coupling constant $g^2 \ne
0$, and on the exact duality group $\Gamma_D$, which
is discrete. Yet, the very reason why so much more of this programme
can be done in extended supersymmetric theories,
with respect to $N \le 1$ theories, resides in the
symbiotic relation existing between $N \ge 2$ supersymmetry
and suitable groups $\Gamma_{cont}$ of {\it continuous duality rotations}
that occurs for abelian (non gauged) $g^2 = 0$ theories.
\par
The line of thought is essentially the following. As we have seen,
crucial for the existence of monopole--states and, hence, for
the realization of
non--perturbative dualities is the presence, in the elementary
spectrum of the theory, of Higgs fields $\phi^I$ transforming in
the adjoint representation. In extended ($N \ge 2$) SUSY, these latter
do indeed exist and
are the supersymmetric scalar partners of the gauge bosons $A_\mu^I$.
Supersymmetry requires that the fields $\phi^I$ should be
interpreted as coordinates of a suitable scalar manifold
${\cal M}_{scalar}^v$   acting as
target space in a 4D $\sigma$--model:
\begin{equation}
\phi : \quad {M}_4  \quad \longrightarrow \quad
{\cal M}_{scalar}^v
\end{equation}
the kinetic term being
 \begin{equation}
{\cal L}_{kin} = g_{IJ}(\phi) \, \partial_\mu \phi^I \partial_\nu \phi^J
g^{\mu\nu}
\end{equation}
where $g_{IJ}$ is a suitable metric on ${\cal M}_{scalar}^v$.
Suitable in the above sentence means appropriate to the fact that
the scalars sit in the same multiplet as the gauge--bosons so that
any transformations on the former should have an image on the
latter. What are the transformations (symmetries or field redefinitions)
one considers on the scalars of a $\sigma$--model ?  Just the
diffeomorphisms of the scalar manifold:
\begin{eqnarray}
\forall t \in Diff[{\cal M}_{scalar}^v]\quad : \quad
\phi^I   &\longrightarrow &  t^{I} (\phi)\nonumber \\
t  \quad : {\cal M}_{scalar}^v &\longrightarrow &
{\cal M}_{scalar}^v
\end{eqnarray}
If the metric is invariant under some diffeomorphism $t$ of ${\cal
M}_{scalar}^v$, namely $t^* g=g$, than that particular diffeomorphism
is an isometry and, as such, it is a global symmetry of the theory.
Otherwise it is just a conceivable field redefinition. In any
case,  in order to commute with supersymmetry the transformation induced
by $t$ on the scalar fields $\phi$ must have a counterpart on
the vector fields. What are the conceivable
transformation of vector fields? Just duality rotations. It follows
that the constraints imposed by supersymmetry on the geometry
of the scalar manifold is that it should be the base--manifold of
a symplectic vector bundle ${\cal SV} \longrightarrow {\cal M}_{scalar}^v$,
whose  structural group $Sp(2n,\IR)$ is the
most general group of duality rotations
(the sections of ${\cal SV}$ transform as  the column vector of electric
plus magnetic field strengths), and such that the tangent
bundle $T{\cal M}_{scalar}^v$ is canonically embedded in ${\cal SV}$.
\par
The isometries of ${\cal M}_{scalar}^v$ constitute therefore a subgroup
${\cal I} \subset Sp(2n,\IR)$, to be identified with the previously
mentioned continuous duality group $\Gamma_{cont}$, which plays a
fundamental role in the construction of the supersymmetric action
and, eventually, also in the understanding of the quantum duality group
$\Gamma_D$.
\par
The scheme is the following:
\begin{itemize}
\item{Having fixed the number $n_V$ of vector multiplets and $m$ of
matter multiplets (if permitted \footnotemark
\footnotetext{N=2 admits scalar multiplets, but
for $N \ge 3$ only vector multiplets are available} ), one sets the
gauge coupling constant to zero ($g^2=0$) so that the gauge group is
abelian and all the matter fields are neutral: there are neither
electric nor magnetic charges}
\item{Next one tries to construct the globally or locally
supersymmetric lagrangian with the above field content. As outlined
above this involves the choice of a target manifold ${\cal M}_{scalar}$
for the 4D $\sigma$-model spanned by the scalar fields. It has a
direct product structure:
\begin{equation}
{\cal M}_{scalar} = {\cal M}_{scalar}^v \, \otimes \,
{\cal M}_{scalar}^s
\end{equation}
the first factor containing the vector multiplet scalars, the second
including those pertaining to the scalar multiplets
(hypermultiplets in the N=2
case that is also the only relevant one, as already remarked).}
\item{
The scalar manifold  ${\cal M}_{scalar}^v$ is restricted by
supersymmetry to carry the symplectic vector bundle structure
${\cal SV} \longrightarrow {\cal M}_{scalar}^v$
previously recalled and will admit a specific continuous group
of isometries ${\cal I} \, \subset Sp(2 {\bar n},\IR)$ where
${\bar n}$ denotes the total number of vector fields in the theory.
One has:
\begin{eqnarray}
 N=2  \quad & {\bar n} = & n_V + 1 ~ \mbox{graviphoton}\nonumber\\
 N=3 \quad & {\bar n} = & n_V + 3 ~ \mbox{graviphotons}\nonumber\\
 N=4 \quad  & {\bar n} = & n_V + 6 ~ \mbox{graviphotons}\nonumber\\
 N=5 \quad  & {\bar n} = & 10 ~ \mbox{graviphotons}\nonumber\\
 N=6 \quad  & {\bar n} = & 16 ~ \mbox{graviphotons}\nonumber\\
 N=7,8 \quad  & {\bar n} = & 56 ~ \mbox{graviphotons}\nonumber\\
 \nonumber\\
\end{eqnarray}
For $N \ge 3$ the solution of the constraints is unique. The
scalar manifold is a uniquely chosen coset manifold ${\cal I}/{\cal
S}$. For $N=2$ the duality constraints requested by supersymmetry
allow more general solutions. They just select a class of complex
manifolds with a peculiar geometry:
{\bf The special K\"ahler manifolds}
that are precisely defined by the existence of a  flat holomorphic
vector bundle with structural group $Sp(2{\bar n},\IR)$, the norm of
a typical section yielding the K\"ahler potential.
There are in fact two kinds of special K\"ahler geometries, the
{\it rigid} and the {\it local} one, depending on whether ${\bar n}$
equals $n_V$ or $n_V +1$. The {\it rigid special manifolds} accommodate
the vector multiplet scalars in globally supersymmetric N=2
Yang--Mills theories, while the {\it local special manifolds}
provide the appropriate description for the same scalar fields
in locally supersymmetric theories, namely in matter coupled $N=2$
supergravity. In the sequel when we say special manifolds without
further specifications we  always refer to the local ones.
Differently from what happens in the $N \ge 3$ theories
the class of special manifolds includes both homogenous spaces as
well as manifolds with no continuous isometries.
It turns out that the structure of special geometry, as defined
by supersymmetry, is also the geometric structure of the {\it moduli
spaces} of {\it Calabi--Yau threefolds} in the local case and of a
special class of {\it Riemann surfaces} in the rigid case. This
mathematical identity is the source of many important physical
considerations developed both in the context of superstring
compactifications and in the present context.}
\item{When the choice of the scalar manifold has been performed
on the basis of the continuous duality invariance and the
corresponding supersymmetric lagrangian has been constructed at $g^2=0$,
{\it gauging} of the model can be considered. Gauging, in this
parlance, corresponds to two independent things: by switching on the gauge
coupling constant $g^2 \ne 0$,
\par
i) the gauge group can be made non--abelian
\par
ii) electric charges can be given to the matter multiplets with
respect to the gauge group, whether abelian or non abelian.
\par
The possible gauge groups ${\cal G}$ are subgroups of the
symplectically embedded isometry group of the scalar manifold
with dimension equal to the number of vector multiplets :
\begin{eqnarray}
{\cal G} & \subset & {\cal I}\left [{\cal M}_{scalar} \right ] \,
\subset \,  Sp(2{\bar n},\IR) \nonumber\\
\mbox{dim}_{\IR} \, {\cal G} &=& n_V
\end{eqnarray}
and satisfying some extra constraints discussed further on.
\par
In relation with this there is a crucial point that
is not all the time fully appreciated by  non
supergravity experts.  Non abelian
gauge groups are introduced only in a second stage to turn
global symmetries of the lagrangian into local
ones. This is a familiar concept also in ordinary field theories.
What is less familiar is that the gauge fields needed to
{\it gauge} such global symmetries are already present
in the lagrangian before gauging, while in ordinary field
theories they are usually introduced at the same time
with the gauging. This happens because vector and scalars
are  superpartners. Hence what one gauges is a subgroup
of the duality symmetry group endowed with the further property
of mixing electric charges only with electric charges. }
\par
\item{
The gauging procedure introduces few more terms, both in the Lagrangian
and in the supersymmetry transformation rules, that are completely
controlled by the geometric structure of the scalar manifold and
by the action of the gauge group on such a manifold.
\par
At $g^2\ne 0$
the continuous duality group $\Gamma_{cont} \, \subset \, Sp(2{\bar
n},\IR)  $ is broken.
The question is: does any discrete subgroup
\begin{equation}
Sp(2{\bar n},\ZZ) \, \supset \, \Gamma_D \,  \subset \, \Gamma_{cont}
\end{equation}
survive in the quantum corrected theory?}
\item{
The recent exciting work on dualities, covered in many
lectures at this Spring School, has been directed toward answering
the above question in the affirmative sense. One considers the
spontaneously broken
phase of the globally or locally supersymmetric gauge theory and
restricts one's attention to the {\it effective quantum action} for
the massless fields. All the perturbative and non perturbative
corrections (space--time instanton contributions) are supposed to
be taken into account. The scalar fields that are included in this action
are the subset of neutral fields in the original lagrangian and are
named {\it moduli}: they correspond to flat directions of the scalar
potential generated by the gauging. Since,
notwithstanding quantum corrections, supersymmetry is unbroken,
the submanifold ${\cal M}_{moduli}$ of the scalar manifold
must satisfy all the constraints of duality requested by
supersymmetry. In the $N\ge 3$ case the fact that there is
a unique choice provides the solution for the local geometry of
${\cal M}_{moduli}$. The covering space of the moduli--space is
just the appropriate coset manifold ${\cal I}/{\cal S}$. There is
only a question of global identifications that corresponds to modding
by the exact duality group $\Gamma_D$. This latter is nothing else
but the discrete part of ${\cal I}$. In a rather symbolic way we can
write:
\begin{equation}
 \Gamma_D \, = \, {\cal I} \, \cap \, \ZZ
\end{equation}
The restriction to integers is due to the quantization of electric
and magnetic charges, the latter being introduced by
non--perturbative effects.
In the $N=2$ case things are more complicated by the fact that
the duality constraints  requested by supersymmetry allow wider
choices. What we know a priori is that ${\cal M}_{moduli}$ is
a special K\"ahler manifold. Its global and local geometry, however,
is quantum--corrected with respect to the geometry of the classical
${\cal M}_{scalar}$ restricted to neutral fields. Since the
exact duality group corresponds, essentially, to the group of global
identifications for ${\cal M}_{moduli}$, it follows that it cannot
be just the restriction to integers of ${\cal I}$, as it happens
for the $N\ge 3$. In this case the solution for  the geometry
of ${\cal M}_{moduli}$ and for $\Gamma_D$ has been found relying
on the previous observation that special geometry is the
moduli--space geometry for a class of Riemann surfaces (in the rigid
case) and for Calabi--Yau threefolds (in the local case). Solving
the problem consists of guessing the right Riemann surface in the
rigid case \cite{SW1,SW2,kltold,faraggi,lopez,cynoi,kltnew}
and the right Calabi--Yau manifold in the local case
\cite{CDF,CDFVP,cynoi,kava}.  What remains is a straightforward
application of algebraic geometry techniques, in particular a
derivation of the relevant Picard--Fuchs differential equations
for the periods. The duality group $\Gamma_D$ corresponds to the
substitution group of integer homology bases and it has a semidirect
product structure:
\begin{equation}
\Gamma_D = \Gamma_{\cal W} \, \times   \, {\cal M}on
\end{equation}
where ${\cal M}on$ is the monodromy group of the Picard--Fuchs
differential system, while  $\Gamma_{\cal W}$ is the symmetry
group of the polynomial constraint $ {\cal W}(X)=0$ defining
the Riemann surface or Calabi--Yau threefold.}
\end{itemize}

Having summarized the logical development that relates the issue
of electric--magnetic duality to the geometry of special
K\"ahler manifolds from the perspective of the most recent
developments let me briefly summarize the history of the subject from
the viewpoint of supergravity.

\begin{itemize}
\item{ {\bf 1978: Hidden symmetries} The prototype of duality symmetries
of supergravity theories is given by the non--compact
group $E_{7(-7)}$ governing the structure of the $N=8$
Lagrangian. It is discovered in 1978 by Cremmer and
Julia \cite{cremmerjulia}. At the time the duality symmetries are
named hidden symmetries.}
\item{ {\bf 1981: General structure of the duality group} In this
year
Gaillard and Zumino \cite{gaizum} solve the general problem of
constructing a lagrangian for a set of ${\bar n}$ abelian vector fields
interacting with $m$ scalar fields having $\sigma$--model
interactions, in such a way that a group of electric--magnetic
duality rotations is realized.}
\item{{\bf 1981--1986: Duality and the construction of supergravities}
In these
years  the role of duality rotations is fully
exploited in the construction of all extended supergravity models.
In particular just to quote few items of a very extensive
literature: $N=4$ \cite{bergkoh}, $N=5$ \cite{dewitnic5},
$N=3$ \cite{petrusetal}, $N=8$ \cite{dewitnic8}}
\item{{\bf 1985-1986: Special Geometry in special coordinates}
is discovered in the construction of the vector
multiplet coupling to  N=2
supergravity by using the so called superconformal tensor calculus
\cite{specspec}.
In this approach the definition of special manifolds emerges in
a preferred coordinate frame and it involves the existence of a
holomorphic prepotential $F(X)$. It will be only later on that
the notion of special geometry will be freed from these limitations.}
\item{{\bf 1986: Classification of Homogeneous Special Manifolds}
The complete classification of those Special manifolds
that are homogeneous spaces ${\cal G}/{\cal H}$ is achieved
in \cite{cremvanp}.}
\item{{\bf 1989-1990: Relation of Special Geometry with Calabi--Yau
manifolds}
The special geometry structure of Calabi--Yau moduli spaces is
first pointed out by Ferrara and Strominger in {1989}
\cite{ferstrom} with a supersymmetry argument. Subsequently it
is rederived from (2,2) conformal theories by Kaplunovsky, Dixon
and Luis \cite{kapdiluis} in { 1990}.}
\item{{\bf 1990: Intrinsic definition of Special Geometry.} Relying
on the relation of special geometry with Calabi Yau moduli spaces,
Strominger arrives at its intrinsic definition.\cite{intrinstrom}.}
\item{{\bf 1990-1991: Geometric Reformulation of N=2 Supergravity.} Using the
intrinsic definitions of Special Geometry and of quaternionic
geometry \cite{bagwit,galicki} $N=2$ supergravity is reformulated
geometrically by Castellani, D'Auria, Ferrara and Fr\'e \cite{specnonspec1},
\cite{specnonspec2}.
This allows the extension of the tensor calculus results to the most
general case of couplings and gaugings.}
\item{{\bf 1991-1994: Picard--Fuchs equations, Mirror Symmetry and
Special Geometry} In these years  there is an intense
activity in studying the Special Geometry of Calabi--Yau
moduli spaces and their physical and mathematical implications.
Following the suggestion of mirror symmetry Candelas and
collaborators \cite{candelinas} show that by means of algebraic
geometry techniques one can derive the exact quantum moduli space
of K\"ahler class deformations including all the world--sheet instanton
corrections. At the same time one obtains the exact duality group.
This result deals with what, in modern parlance, is named $T$--duality,
yet the same mathematics is at work when a Calabi--Yau threefold
is used to describe the $S$--duality invariance.
The profound relation between Special Geometry and Picard Fuchs
equations is investigated in {\bf 1993-1994} in\cite{cerferlui} }
\item{{\bf 1994-95: Trading $T$--duality for $S$--duality and
second quantized mirror symmetry}. The
paper by Seiberg and Witten \cite{SW1} on the solution of
rigid N=2 Yang--Mills theory by means of an auxiliary Riemann
surface opens new perspectives on the use of a well established
lore on Special Geometry and Duality Groups in a new context.
$T$--duality is traded for $S$--duality and world--sheet instantons
are traded for space--time instantons. The concept of second quantized
mirror symmetry is introduced in \cite{FHSV}.}
\end{itemize}
\section{Plan of these Lectures}
\label{pianino}
In the present lectures my goal is to present the geometrical
structures underlying matter coupled $N=2$ supergravity and
$N=2$ Yang--Mills theories, emphasizing their role in the discussion
of electric--magnetic duality rotations and strong--weak duality.
\par
The very purpose is to show how all the recent exciting results
that go under the collective name of S--duality or string--string
duality are founded in these geometric structures and could not
be properly understood without them. What is actually true is that
the very formulation of N=2 supersymmetry, both in the local and
in the global case, incorporates duality rotations as a necessary
ingredient. This is most likely related with the other intriguing
aspect of N=2 supersymmetry, namely its symbiotic relation with the
topological twist to topological field--theories~
\cite{topfgen_7},\cite{topfgen_6},\cite{topftwist_1},\cite{topftwist_2},
\cite{topf4d_4},\cite{topf4d_8}. As these latter are the appropriate
framework to single out instanton and monopole dominance
in the path integral, duality symmetries that exchange elementary
excitations with solitonic states could not fail to be deeply related
with $N=2$ theories and that from the very beginning.
\par
The geometric structures underlying $D=4$, $N=2$ theories
are:
\begin{enumerate}
\item{{\bf Special K\"ahler geometry} in its two versions, {\it rigid
and local}. This is the geometry that pertains to the scalar fields
belonging to the vector multiplets.}
\item{{\bf Hypergeometry} in its two versions: {\it HyperK\"ahler
and Quaternionic geometry}. This is the structure of the manifolds
spanned by the scalar fields that sit in the hypermultiplets.}
 \item{{\bf The momentum map.} This is the geometrical construction
 that lifts the action of a Lie group on a symplectic manifold to
 a dual hamiltonian realization. Well known in classical mechanics,
 the concept of momentum map is the key--ingredient in the gauging
 of $N=2$ supersymmetric theories}
\end{enumerate}
The momentum map explains the nature of the auxiliary fields and it
is used to write down the scalar potential of N=2 theories. It also
plays a major role in the topological twist. From the viewpoint
taken in these lectures, gauging is that procedure that through
the introduction of electric charges spoils a primeval continuous
duality symmetry. Yet, at the same time, it introduces those
non--linear interactions (typically the non--abelian couplings)
that are responsible for the existence of classical non perturbative
solutions of the field equations with magnetic--like charges.
Hence, after gauging, the primeval continuous duality symmetry is
lost, but a new discrete one can emerge, due to the introduction
of magnetic charges.
\par
A major issue is to understand these quantum duality symmetries
that emerge after gauging and their relations with the primeval
continuous dualities. In some cases, like in the $N=4$ theory
the quantum dualities are just a discrete subgroup of the classical
primeval duality group. In the $N=2$ theories they are also a
subgroup of the primeval classical ones, yet with a different
symplectic embedding. In this difference resides non--perturbative
physics.
\par
As it might be expected the geometrical structures 1) 2) and 3), that
so nicely marry with electric--magnetic dualities, are an yield
of the supersymmetry algebra. So to see their origin one should deal
with the fermions and work hard with Dirac algebra and gamma
matrices. This has been done elsewhere \cite{specnonspec2,newpap} and
it is completely accomplished. I do not see any reason to repeat it
here. Let us take it for given and let us rather illustrate the
various geometrical items involved in the construction of the $N=2$
theory. Fermions, although at the heart of the whole matter, obscure
with their technicalities the profound meaning of the involved
geometry and its relation with the physics of electric--magnetic
duality rotations.
\par
Hence in these lectures I shall never mention the fermions:
of all supersymmetric lagrangians I shall write only the
bosonic part. The structures of these bosonic lagrangians is
determined by
intercourse with the fermions, yet, once their form is established,
it is their properties we want to investigate.
\par
The final form of the bosonic action  for $N=2$ supergravity
is as follows:
\begin{eqnarray}
&{\cal L}^{SUGRA} \, =& \nonumber\\
& \sqrt{-g}\Bigl  [ \,  R[g]
\, + \, g_{i {j^\star}}(z,\bar z )\, \nabla^{\mu} z^i \,
\nabla_{\mu} \bar z^{j^\star} & \nonumber \\
& - \, 2 \,\lambda\, h_{uv}(q) \, \nabla^{\mu} q^u \,
 \nabla_{\mu} q^v
\nonumber\\
&  + \,{\rm i} \,\left(
\bar {\cal N}_{\Lambda \Sigma} {\cal F}^{- \Lambda}_{\mu \nu}
{\cal F}^{- \Sigma \vert {\mu \nu}}
\, - \,
{\cal N}_{\Lambda \Sigma}
{\cal F}^{+ \Lambda}_{\mu \nu} {\cal F}^{+ \Sigma \vert {\mu \nu}} \right )
\, \Bigr ] & \nonumber \\
 & \, - \, g^2 \, {\cal V}
\label{gausugra}
\end{eqnarray}
where:
\begin{eqnarray}
g_{i {j^\star}}(z,\bar z )&=& \mbox{ special K\"ahler
metric}\nonumber\\
h_{uv}(q)&=& \mbox{quaternionic metric}\nonumber\\
\bar {\cal N}_{\Lambda \Sigma}(z,\bar z)&=& \mbox{period matrix in
spec.geom.}\nonumber\\
\nabla_\mu z^i &=& \partial_\mu z^i + g A_\mu^\Lambda \,
k_\Lambda^i(z)\nonumber \\
k_\Lambda^i(z)&=& \mbox{holomorphic Killing vec.} \nonumber\\
\nabla_\mu q^u &=&\partial_\mu z^i + g A_\mu^\Lambda \,
k_\Lambda^u(q)\nonumber \\
k_\Lambda^u(q)&=& \mbox{triholomorphic Killing vec.}\nonumber\\
{\cal V}&=& {\bar L}^\Lambda \left (4 k^u_\Lambda   k^v_\Sigma
h_{uv} + k^i_\Lambda   k^{j^\star}_\Sigma g_{ij^\star} \right)
L^\Sigma\nonumber\\
& &+ \left ( U^{\Lambda\Sigma} - 3 {\bar L}^\Lambda L^\Sigma \right )
\,{\cal P}^x_\Lambda {\cal P}^x_\Sigma \nonumber \\
 {\cal P}^x_\Lambda&=&\mbox{triholomorphic mom. map }\nonumber\\
{\bar L}^\Lambda&=& \mbox{cov. hol. sect.}
\label{filastrocca}
\end{eqnarray}
Giving a precise geometrical meaning to the list of words appearing
in eq.~\ref{filastrocca} is the purpose of the following lectures.
They are organized as follows.
\par
Lecture~\ref{LL1} is devoted to a study of the general structure, in any
even space--time dimensions of those lagrangians,
containing as elementary fields differential forms
and scalars,  that display duality covariance.
The role of symplectic or pseudo--orthogonal transformations is
explained.
\par
Lecture ~\ref{LL2} concentrates on the D=4 case where symplectic
covariance is the relevant one and discusses the symplectic
embedding  of the scalar manifold diffeomorphism group. A very
general formula for the {\it period matrix} is given in the case of
homogeneous scalar manifolds: the Gaillard Zumino master formula.
\par
Lecture ~\ref{LL4} introduces the notion of Special K\"ahler geometry
both in its local and rigid version.
\par
Lecture ~\ref{LL5} is devoted to a case study of some particular
Special K\"ahler manifold of relevance in the discussion of non
perturbative dualities both of the target type and of the S--type.
An exemplification of the concepts previously introduced is given.
\par
Lecture~\ref{LL6} explains hypergeometry, that is the geometry
of the scalar manifold in the hypermultiplet sector. A complete
parallelism is drawn between the two instances of rigid and local
hypergeometries and the corresponding rigid and local special geometries.
\par
Lecture ~\ref{LL7} discusses the gauging after having drawn
the distinction between classical, perturbative and non perturbative
dualities.
\par
Lecture~\ref{LL8} presents the last necessary geometrical ingredient
of the N=2 construction, that is the momentum map.
\par
Finally, as an exemplification of the whole
set of ideas discussed individually in previous lectures,
lecture~\ref{LL9} presents
a detailed derivation and discussion of the rigid non perturbative
special geometry associated with the effective lagrangian of an
$SU(2)$ theory. This is the famous Seiberg Witten model
already mentioned several times.

\section{The Group of Continuous Duality Rotations in $D=2p$
Space--Time Dimensions }
\label{LL1}
In this lecture I am going to review the general structure of an
abelian theory of vectors and scalars displaying duality symmetries.
The basic reference is the 1981 paper by Gaillard and Zumino
\cite{gaizum} my presentation, however, will be a little more
general. Rather than restricting my attention to ordinary one--index
gauge fields $A^\Lambda_\mu$ in $D=4$ I will consider a theory of
antisymmetric gauge tensor with $(p-1)$--indices $A^\Lambda_{\mu_1
\dots\ \mu_{p-1}}$, in a $D=2 p$ dimensional space--time
$M^{space-time}$:
\begin{equation}
D \, \equiv \, \mbox{dim} \,
M^{space-time} \, = \, 2 p
\end{equation}
The gauge tensors
$A^\Lambda_{\mu_1 \dots\ \mu_{p-1}}$ correspond to a set of
differential $(p-1)$--forms:
\begin{eqnarray}
A^\Lambda & \equiv &
A^\Lambda_{\mu_1 \dots\ \mu_{p-1}} \, dx^{\mu_1} \, \wedge \, \dots
\, \wedge dx^{\mu_{p-1}} \nonumber\\
&\null & \left ( \Lambda = 1,
\dots , {\bar n} \right )
\end{eqnarray}
The corresponding field strengths and their Hodge duals are defined
as it follows:
\begin{eqnarray}
{ F}^\Lambda & \equiv & d \, A^\Lambda \, \nonumber\\ & \equiv & \,
{{1}\over{p !}} \, {\cal F}^\Lambda_{\mu_1 \dots\ \mu_{p}} \,
dx^{\mu_1} \, \wedge \, \dots \, \wedge dx^{\mu_{p}} \nonumber\\
{\cal F}^\Lambda_{\mu_1 \dots\ \mu_{p}} & \equiv & \partial_{\mu_1}
A^\Lambda_{\mu_2 \dots\ \mu_{p}} \, + \, \mbox{p-1 terms}
\nonumber\\ { F}^{\Lambda\star} & \equiv & {{1}\over{p !}} \, {\tilde
{\cal F}}^\Lambda_{\mu_1 \dots\ \mu_{p}} \, dx^{\mu_1} \, \wedge \,
\dots \, \wedge dx^{\mu_{p}} \nonumber\\
{\tilde {\cal
F}}^\Lambda_{\mu_1 \dots\ \mu_{p}} & \equiv &{{1}\over{p!}}
\varepsilon_{\mu_1\dots \mu_p \nu_1\dots \nu_p}\, {\cal F}^{\Lambda
\vert \nu_1 \dots \nu_p}
\label{campfort}
\end{eqnarray}
Defining
the space--time integration volume as it follows:
\begin{equation}
\mbox{d}^D x \, \equiv \, {{1}\over{D!}} \, \varepsilon_{\mu_1\dots
\mu_D} \, dx^{\mu_1} \, \wedge \, \dots \, \wedge dx^{\mu_{D}}
\label{volume}
\end{equation}
we obtain:
\begin{eqnarray}
& F^\Lambda \, \wedge \, F^\Sigma \, = &\nonumber\\ &
{{1}\over{(p!)^2}} \, \varepsilon^{\mu_1\dots \mu_p \nu_1\dots
\nu_p}\, {\cal F}^\Lambda_{\mu_1 \dots\ \mu_{p}} \, {\cal
F}^\Sigma_{\nu_1 \dots\ \nu_{p}}& \nonumber\\ & F^\Lambda \, \wedge
\, F^{\Sigma\star} \, = \nonumber\\ &(-)^p \, {{1}\over{(p!)}} \,
{\cal F}^\Lambda_{\mu_1 \dots\ \mu_{p}} \, {\cal F}^{\Sigma \vert
\mu_1 \dots\ \mu_{p}} &
\label{cinetici}
\end{eqnarray} In addition
to the $(p-1)$--forms let us also introduce a set of real scalar
fields $\phi^I$ ( $I=1,\dots , {\bar m}$) spanning an ${\bar
m}$--dimensional manifold ${\cal M}_{scalar}$ \footnotemark
\footnotetext{whether the $\phi^I$ can be arranged into complex
fields is not relevant at this level of the discussion } endowed
with a metric $g_{IJ}(\phi)$. Utilizing the above field
content we can write the following action functional:
\begin{eqnarray}
{\cal S}&=&{\cal S}_{tens} \, + \, {\cal
S}_{scal}\nonumber\\ {\cal S}_{tens}&=& \int \, \Bigl [ \,
\o{(-)^p}{2} \, \gamma_{\Lambda\Sigma}(\phi) \, F^\Lambda \, \wedge
\, F^{\Sigma\star} \, + \nonumber\\ &\null &\, \o{1}{2} \,
\theta_{\Lambda\Sigma}(\phi) \, F^\Lambda \, \wedge \, F^{\Sigma}
\Bigr ]\nonumber\\ {\cal S}_{scal}&=& \int \, \left [ \o{1}{2} \,
g_{IJ}(\phi) \, \partial_\mu \phi^I \, \partial^\mu \phi^J \right] \,
\mbox{d}^D x
\label{gaiazuma}
\end{eqnarray}
where the scalar field
dependent ${\bar m} \times {\bar m}$ matrix
$\gamma_{\Lambda\Sigma}(\phi)$ generalizes the inverse of the
squared coupling constant $\o{1}{g^2}$ appearing in ordinary
4D--gauge theories. The field dependent matrix
$\theta_{\Lambda\Sigma}(\phi)$ is instead a generalization of the
$theta$--angle of quantum chromodynamics. The matrix $\gamma$ is
symmetric in every space--time dimension, while $\theta$ is
symmetric or antisymmetric depending on whether $p=D/2$ is an even
or odd number. In view of this fact it is convenient to distinguish
between the two cases by setting:
\begin{equation}
D\, = \, \cases{4 \nu \quad
\qquad\nu \in \ZZ \, \vert \quad p=2\nu \cr 4 \nu + 2\quad \, \nu
\in \ZZ \, \vert \quad p=2\nu + 1 \cr}
\label{duecasi}
\end{equation}
Introducing a formal operator $j$ that maps a field
strength into its Hodge dual:
\begin{equation} \left ( j \, {\cal
F}^\Lambda \right )_{\mu_1 \dots \mu_p} \, \equiv \,
{{1}\over{(p!)}} \, \epsilon_{\mu_1 \dots\mu_p \nu_1 \dots \nu_p} \,
{\cal F}^{\Lambda \vert\nu_1 \dots \nu_p}
\label{opjei}
\end{equation} and a formal scalar product:
\begin{equation} \left (
G \, , \, K \right ) \equiv G^T K \, \equiv \, {{1}\over{(p!)}}
\sum_{\Lambda=1}^{\bar n} G^{\Lambda}_{\mu_1\dots\mu_p} K^{\Lambda
\vert \mu_1\dots\mu_p }
\label{formprod}
\end{equation} the total
lagrangian of eq.~\ref{gaiazuma} can be rewritten as
\begin{eqnarray}
&\cL^{(tot)}\,  =& \nonumber\\
& {\cal F}^T \, \left ( \gamma
\otimes \bfone + \theta \otimes j \right ) {\cal F} & \nonumber\\
&  + \o{1}{2} \, g_{IJ}(\phi) \, \partial_\mu \phi^I \,
\partial^\mu \phi^J &
\label{gaiazumadue}
\end{eqnarray}
and the essential distinction between the two cases of eq.~\ref{duecasi} is
given, besides the symmetry of $\theta$, by the involutive property
of $j$, namely we have:
\begin{equation} \matrix{ D=4 \nu & \vert &
\theta = \theta^T & j^2 = - \bfone \cr D=4 \nu + 2 & \vert & \theta
= - \theta^T & j^2 = \bfone \cr }
\label{splitta}
\end{equation}
Introducing dual and antiself--dual combinations:
\begin{equation}
\matrix{ D=4 \nu & \cases{ {\cal F}^{\pm} = {\cal F}\, \mp i \, j
{\cal F} \cr j \, {\cal F}^{\pm} = \pm \mbox{i} {\cal F}^{\pm} \cr}
\cr D=4 \nu+2 & \cases{ {\cal F}^{\pm} = {\cal F} \, \pm \, j {\cal
F}\cr j \, {\cal F}^{\pm} = \pm {\cal F}^{\pm} \cr} \cr}
\label{selfduals}
\end{equation}
and the field--dependent matrices:
\begin{equation} \matrix{D=4 \nu & \cases{ {\cal N} = \theta -
\mbox{i} \gamma \cr {\bar {\cal N}} = \theta + \mbox{i} \gamma \cr
}\cr D=4 \nu+2 & \cases{ {\cal N} = \theta + \gamma \cr -{\cal N}^T
= \theta - \gamma \cr }\cr }
\label{scripten}
\end{equation} the
vector part of the lagrangian ~\ref{gaiazumadue} can be rewritten in
the following way in the two cases:
\begin{eqnarray} D=4\nu &
: & \nonumber\\ {\cal L}_{vec} & = & {{\mbox{i}}\over{8}} \, \left [
{\cal F}^{+T} {\cal N} {\cal F}^{+} - {\cal F}^{-T} {\bar {\cal N}}
{\cal F}^{-} \right ] \nonumber\\ D=4\nu+2 & : & \nonumber\\ {\cal
L}_{vec} & = & {{\mbox{1}}\over{8}} \, \left [ {\cal F}^{+T} {\cal
N} {\cal F}^{+} + {\cal F}^{-T} { {\cal N}^T} {\cal F}^{-} \right
]\nonumber\\ \null&\null&\null
\label{lagrapm}
\end{eqnarray}
Introducing the new tensor:
\begin{equation} \matrix { {\tilde
G}^\Lambda_{\mu_1\dots\mu_p} &\equiv & -(p!)  { {\partial {\cal
L}}\over{\partial {\cal F}^\Lambda_{\mu_1\dots\mu_p}}} &
D=4\nu \cr {\tilde G}^\Lambda_{\mu_1\dots\mu_p} &\equiv & (p!)  {
{\partial {\cal L}}\over{\partial {\cal
F}^\Lambda_{\mu_1\dots\mu_p}}} &   D=4\nu+2 \cr}
\label{gtensor}
\end{equation}
which, in matrix notation, corresponds to:
\begin{eqnarray} &j \, G \, \equiv \, a \, {
{\partial {\cal L}}\over{\partial {\cal F}^T}} & \nonumber\\ & = \,
{\o{a}{p!}}\, \left ( \gamma\otimes\bfone + \theta\otimes j \right )
\, {\cal F}
\label{ggmatnot}
\end{eqnarray}
where $a=\mp$ depending
on whether $D=4\nu$ or $D=4\nu+2$, the Bianchi identities and field
equations associated with the lagrangian ~\ref{gaiazuma} can be
written as follows:
\begin{equation} \matrix {\partial^{\mu_1}
{\tilde {\cal F}}^{\Lambda}_{\mu_1\dots\mu_p} &=& 0 \cr \matrix
\partial^{\mu_1} {\tilde {\cal G}}^{\Lambda}_{\mu_1\dots\mu_p} &=&
0 \cr}
\label{biafieq}
\end{equation}
This suggests that we introduce the $2{\bar
n}$ column vector :
\begin{equation}
{\bf V} \, \equiv \, \left ( \matrix
{ j \, {\cal F}\cr
j \, {\cal G}\cr}\right )
\label{sympvec}
\end{equation}
and that we consider general linear transformations on such a vector:
\begin{equation}
\left ( \matrix
{ j \, {\cal F}\cr
j \, {\cal G}\cr}\right )^\prime \, =\,
\left (\matrix{ A & B \cr C & D \cr} \right )
\left ( \matrix
{ j \, {\cal F}\cr
j \, {\cal G}\cr}\right )
\label{dualrot}
\end{equation}
For any matrix $\left (\matrix{ A & B \cr C & D \cr} \right ) \, \in
\, GL(2{\bar n},\IR )$ the new vector ${\bf V}^\prime$ of {\it
magnetic and electric} field--strengths satisfies the same
equations ~\ref{biafieq} as the old one. In a condensed notation
we can write:
\begin{equation}
\partial \, {\bf V}^\prime \, = \, 0 \quad \Longleftrightarrow \quad
\partial \, {\bf V}^\prime \, = \, 0
\label{dualdue}
\end{equation}
Separating the self--dual and anti--self--dual parts
\begin{eqnarray}
{\cal F}={\o{1}{2}}\left ({\cal F}^+ +{\cal F}^- \right ) \nonumber\\
{\cal G}={\o{1}{2}}\left ({\cal G}^+ +{\cal G}^- \right )
\label{divorzio}
\end{eqnarray}
and taking into account that for $D=4\nu$ we have:
\begin{equation}
{\cal G}^+ \, = \, {\cal N}{\cal F}^+  \quad
{\cal G}^- \, = \, {\bar {\cal N}}{\cal F}^-
\label{gigiuno}
\end{equation}
while for $D=4\nu +2$ the same equation reads:
\begin{equation}
{\cal G}^+ \, = \, {\cal N}{\cal F}^+  \quad
{\cal G}^- \, = \, {- {\cal N}^T}{\cal F}^-
\label{gigidue}
\end{equation}
the duality
rotation of eq.~\ref{dualrot} can be rewritten as:
\begin{eqnarray}
D=4\nu: \qquad &\null & \nonumber\\
\left ( \matrix
{   {\cal F}^+ \cr
 {\cal G}^+\cr}\right )^\prime & = &
\left (\matrix{ A & B \cr C & D \cr} \right )
\left ( \matrix
{   {\cal F}^+\cr
{\cal N} {\cal F}^+\cr}\right ) \nonumber\\
\left ( \matrix
{   {\cal F}^- \cr
 {\cal G}^-\cr}\right )^\prime & = &
\left (\matrix{ A & B \cr C & D \cr} \right )
\left ( \matrix
{   {\cal F}^-\cr
{\bar {\cal N}} {\cal F}^-\cr}\right ) \nonumber\\
D=4\nu+2:&\null & \nonumber\\
\left ( \matrix
{   {\cal F}^+ \cr
 {\cal G}^+\cr}\right )^\prime & = &
\left (\matrix{ A & B \cr C & D \cr} \right )
\left ( \matrix
{   {\cal F}^+\cr
{\cal N} {\cal F}^+\cr}\right ) \nonumber\\
\left ( \matrix
{   {\cal F}^- \cr
 {\cal G}^-\cr}\right )^\prime & = &
\left (\matrix{ A & B \cr C & D \cr} \right )
\left ( \matrix
{   {\cal F}^-\cr
{ - {\cal N}^T} {\cal F}^-\cr}\right )
\label{trasform}
\end{eqnarray}
In both cases the problem is that the transformation rule
{}~\ref{trasform} of ${\cal G}^\pm$ must be consistent with the definition
of the latter
as variation of the Lagrangian with respect
to ${\cal F}^\pm$ (see eq.~\ref{gtensor}). This request
restricts the form of the matrix
$\Lambda =\left (\matrix{ A & B \cr C & D \cr} \right )$.
As we are just going to show, in the $D=4\nu$ case $\Lambda$ must belong
to the symplectic subgroup $Sp(2\bar n,\IR)$ of the special linear group,
while in the $D=4\nu +2$ case it must be in the pseudo--orthogonal
subgroup $SO(\bar n , \bar n)$:
\begin{eqnarray}
D=4\nu \quad :&\null& \nonumber\\
\left (\matrix{ A & B \cr C & D \cr} \right ) & \in &Sp(2\bar n,\IR)
\,\subset
\, GL(2\bar n ,\IR )\nonumber\\
D=4\nu +2 :&\null& \nonumber\\
\left (\matrix{ A & B \cr C & D \cr} \right ) & \in & SO(\bar n , \bar n)
\,\subset
\, GL(2\bar n ,\IR )\nonumber\\
\label{distinguo}
\end{eqnarray}
the above subgroups being defined as the set of $2\bar n \times
2\bar n$ matrices satisfying, respectively, the following conditions:
\begin{equation}
\matrix{
\Lambda \in Sp(2\bar n,\IR) \to &    \cr
 & \Lambda^T  \,
\left (\matrix{ {\bf 0}_{} &
\bfone_{} \cr
-\bfone_{} & {\bf 0}_{}
\cr }\right )
 \, \Lambda  = \cr
   \left (\matrix{ {\bf 0}_{} &
\bfone_{} \cr
-\bfone_{} & {\bf 0}_{}
\cr }\right ) & \null \cr
\null & \null  \cr
\Lambda \in SO(\bar n , \bar n)  \to &\cr
& \Lambda^T  \,
\left (\matrix{ {\bf 0}_{} &
\bfone_{} \cr
\bfone_{} & {\bf 0}_{}
\cr }\right )
\, \Lambda    = \cr
\left (\matrix{ {\bf 0}_{} &
\bfone_{} \cr
\bfone_{} & {\bf 0}_{}
\cr }\right ) &\null \cr }
\label{ortosymp}
\end{equation}
To prove the statement we just made, we calculate the transformed
lagrangian ${\cal L}^\prime$ and then we compare its variation
${\o{\partial {\cal L}^\prime}{\partial {\cal F}^{\prime T}}}$
with  ${\cal G}^{\pm\prime}$ as it follows from the postulated
transformation rule ~\ref{trasform}. To perform such a calculation
we rely on the following basic idea. While the
duality rotation~\ref{trasform} is performed on the field strengths
and on their duals, also the scalar fields are transformed by the action
of some diffeomorphism ${\xi }\,  \in \,
{\rm Diff}\left ( {\cal M}_{scalar}\right )$ of the scalar manifold
and, as a consequence of that, also the matrix ${\cal N}$ changes.
In other words given the scalar manifold ${\cal M}_{scalar}$ we
assume that in the two cases of interest there exists a
homomorphism of the following form :
\begin{equation}
\iota _{\delta} : \,  {\rm Diff}\left ( {\cal M}_{scalar}\right )
\, \longrightarrow \, GL(2\bar n,\IR)
\label{immersione}
\end{equation}
so that:
\begin{eqnarray}
\forall &  \xi   &\in \, {\rm Diff}\left ( {\cal M}_{scalar}\right ) \, :
\, \phi^I \, \stackrel{\xi}{\longrightarrow} \,  \phi^{I\prime}
\nonumber\\
\exists  & \iota _{\delta}(\xi) & = \left (\matrix{ A_\xi & B_\xi \cr
C_\xi & D_\xi \cr }\right ) \, \in \,  GL(2\bar n,\IR)
\label{apnea}
\end{eqnarray}
Using such a homomorphism
we can define the simultaneous action of $\xi$ on
all the fields of our theory by setting:
\begin{equation}
\xi \, : \, \cases{   \phi \,
\longrightarrow \, \xi (\phi) \cr
{\bf V} \,
\longrightarrow \, \iota _{\delta}(\xi) \, {\bf V} \cr
{\cal N}(\phi) \, \longrightarrow \, {\cal N}^\prime (\xi (\phi)) \cr }
\end{equation}
where the notation~\ref{sympvec} has been utilized.
In the tensor sector the transformed lagrangian, is
\begin{eqnarray}
& {\cal L}^{\prime}_{tens}  =  & \nonumber\\
& {\o{{\rm i}}{8}} \, \Bigl [ {\cal F}^{+T}
\, \bigl ( A + B {\cal N} \bigr )^T {\cal N}^\prime
( A + B {\cal N} \bigr ) {\cal F}^{+}  & \nonumber\\
& - \, {\cal F}^{-T}
\, \bigl ( A + B {\bar {\cal N}} \bigr )^T {\bar {\cal N}}^\prime
( A + B {\bar {\cal N}} \bigr ) {\cal F}^{-} \Bigr ]  &
\label{elleprima}
\end{eqnarray}
for the $D=4\nu$ case and
\begin{eqnarray}
&{\cal L}^{\prime}_{tens}  =  & \nonumber\\
& {\o{{\rm i}}{8}} \, \Bigl [ {\cal F}^{+T}
\, \bigl ( A + B {\cal N} \bigr )^T {\cal N}^\prime
( A + B {\cal N} \bigr ) {\cal F}^{+}  & \nonumber\\
&  - \, {\cal F}^{-T}
\, \bigl ( A - B {  {\cal N}^T} \bigr )^T {  {\cal N}^T}^\prime
( A - B {  {\cal N}^T} \bigr ) {\cal F}^{-} \Bigr ] &
\label{elleprimap}
\end{eqnarray}
Consistency with
the definition of ${\cal G}^+$ requires, in both cases that
\begin{eqnarray}
&{\cal N}^\prime \, \equiv \, {\cal N}^\prime (\xi(\phi)) \, =& \nonumber\\
&\left ( C_\xi + D_\xi  {\cal N}(\phi) \right )   \left ( A_\xi +
B_\xi {\cal N}(\phi)\right )^{-1} &
\label{Ntrasform}
\end{eqnarray}
while consistency with the definition of ${\cal G}^-$ imposes,
in the $D=4\nu$ case the transformation rule:
\begin{eqnarray}
&{\bar {\cal N}}^\prime  \, \equiv \, {\bar {\cal N}}^\prime (\xi(\phi)) \,
= &
\nonumber\\
&\left ( C_\xi + D_\xi  {\bar {\cal N}}(\phi ) \right )   \left ( A_\xi +
B_\xi {\bar {\cal N}}(\phi)\right )^{-1}&
\label{Nbtrasform}
\end{eqnarray}
and in the case $D=4\nu+2$ the other transformation rule:
\begin{eqnarray}
&{- {\cal N}^T}^\prime  \, \equiv \, {-{\cal N}^{T \prime}}(\xi(\phi)) \,
=&
\nonumber\\
&\left ( C_\xi - D_\xi  {  {\cal N}^T}(\phi) \right )   \left ( A_\xi -
B_\xi {  {\cal N}^T}(\phi )\right )^{-1}&
\label{Nttrasform}
\end{eqnarray}
It is from the transformation rules~\ref{Ntrasform},~\ref{Nbtrasform}
and ~\ref{Nttrasform} that we derive a restriction on the form of the
duality rotation matrix $\Lambda_\xi \equiv \iota_\delta(\xi)$.
Indeed, in the $D=4\nu$ case we have that by means of the fractional
linear transformation~\ref{Ntrasform} $\Lambda_\xi$ must map an
arbitrary {\it complex symmetric} matrix into another matrix of the
{\it same sort}. It is straightforward to verify that this condition is
the same as the first of conditions~\ref{ortosymp}, namely the
definition of the symplectic group $Sp(2\bar n ,\IR)$. Similarly
in the $D=4\nu +2$ case the matrix $\Lambda_\xi$ must obey the
property that taking the {\it negative of the transpose}
of an arbitrary
real matrix ${\cal N}$ {\it before or after} the fractional linear
transformation induced by $\Lambda_\xi$ {\it is immaterial}. Once again, it
is easy to verify that this condition is the same as the second
property in eq.~\ref{ortosymp}, namely the definition of the
pseudo--orthogonal group $SO(\bar n, \bar n)$. Consequently the
homomorphism of eq.~\ref{immersione} specializes
as follows in the two relevant cases
 \begin{equation}
\iota _{\delta} : \, \cases{ {\rm Diff}\left ( {\cal M}_{scalar}\right )
\, \longrightarrow \, Sp(2\bar n,\IR) \cr
{\rm Diff}\left ( {\cal M}_{scalar}\right )
\, \longrightarrow \, SO(\bar n , \bar n) \cr}
\label{spaccoindue}
\end{equation}
Clearly, since both $Sp(2\bar n,\IR)$ and $SO(\bar n , \bar n)$ are
finite dimensional Lie groups, while
${\rm Diff}\left ( {\cal M}_{scalar}\right )$ is
infinite--dimensional, the homomorphism  $\iota _{\delta}$ can
never be an isomorphism. Defining the Torelli group of the
scalar manifold as:
\begin{equation}
{\rm Diff}\left ( {\cal M}_{scalar}\right ) \, \supset \,
\mbox{Tor} \left ({\cal M}_{scalar} \right ) \, \equiv \,
\mbox{ker} \, \iota_\delta
\label{torellus}
\end{equation}
we always have:
\begin{equation}
\mbox{dim} \, \mbox{Tor} \left ({\cal M}_{scalar} \right ) \, = \,
\infty
\label{infitor}
\end{equation}
The reason why we have given the name of Torelli to the group defined
by eq.~\ref{torellus} is because of its similarity with the
Torelli group that occurs in algebraic geometry. There one deals
with the {\it moduli space} ${\it M}_{moduli}$
of complex structures of a $(p+1)$--fold ${\cal M}_{p+1}$ and
considers the action of the diffeomorphism group
$\mbox{Diff}\left ( {\it M}_{moduli} \right )$ on canonical homology
bases of $(p+1)$--cycles.  Since this action must be linear and must
respect the intersection matrix, which is either  symmetric or
antisymmetric depending on the odd or even parity of $p$, it follows that
one obtains a homomorphism similar to that in eq.~\ref{spaccoindue}:
\begin{equation}
\iota _h : \, \cases{ {\rm Diff}\left ( {\cal M}_{moduli}\right )
\, \longrightarrow \, Sp(2\bar n,\IR) \cr
{\rm Diff}\left ( {\cal M}_{moduli}\right )
\, \longrightarrow \, SO(\bar n , \bar n) \cr}
\label{spaccointre}
\end{equation}
The Torelli group is usually defined as the kernel of such
a homomorphism. When cohomology with real coefficients is
replaced by cohomology with integer coefficients the homomorphism
of eq.~\ref{spaccointre} reduces to
\begin{equation}
\iota _h : \, \cases{ {\rm Diff}\left ( {\cal M}_{moduli}\right )
\, \longrightarrow \, Sp(2\bar n,\ZZ) \cr
{\rm Diff}\left ( {\cal M}_{moduli}\right )
\, \longrightarrow \, SO(\bar n , \bar n,\ZZ) \cr}
\label{spaccointrez}
\end{equation}
and the Torelli group becomes even larger \footnotemark
\footnotetext{To be precise one considers the mapping class
group $\mbox{Diff}/\mbox{Diff}_0$ and defines the Torelli
group as the kernel of the homomorphism $\iota_h \,:\,
\mbox{Diff}/\mbox{Diff}_0 \to Sp(2\bar n ,\ZZ)$ or
$\to SO(\bar n , \bar n)$. Yet since $\mbox{Diff}_0$ is
certainly in the kernel of $\iota_h$ we have slightly extended
the notion. In other words, the Torelli group of algebraic geometry
is $\mbox{Tor}/\mbox{Diff}_0$ with respect to the Torelli
group defined $\mbox{Tor}$ here}
\par
This similarity between two problems that are, at first sight,
totally disconnected is by no means accidental. In the theories
recently considered in the literature and largely discussed
at this Spring School, where electro--magnetic duality
rotations are indeed realized as a quantum symmetry,
the scalar manifold ${\it M}_{scalar}$ is identified
with the moduli--space of  complex structures for suitable
complex $(p+1)$--folds and the duality rotations are related
with changes of integer homology basis. From the physical
point of view what requires the restriction from the
continuous duality groups $Sp(2\bar n,\IR)$, $SO(\bar n , \bar n,\ZZ)$
to their discrete counterparts $Sp(2\bar n,\ZZ)$,
$SO(\bar n , \bar n,\ZZ)$ is the Dirac quantization condition
of electric and magnetic charges eq.~\ref{dirquant}, which
obviously occurs when electric and magnetic currents are introduced.
Indeed the lattice spanned by electric and magnetic charges is
eventually identified with the integer homology lattice of
the corresponding  $(p+1)$--fold.
\par
In view of this analogy, the natural question which arises is
the following: what is the counterpart in algebraic geometry
of the matrix ${\cal N}$ that appears in the kinetic terms of
the gauge fields? In view of its transformation property
(see eq.~\ref{Ntrasform}) the answer is very simple: it is the
{\it period matrix}. Consider for instance the situation,
occurring in Calabi--Yau three--folds, where the middle
cohomology group $H^{3}_{DR}\left ({\cal M}_{3}\right )$ admits
a Hodge--decomposition of the type:
\begin{eqnarray}
 H^{(3)}_{DR}({\cal M}_n)& =& H^{(3,0)}\, \oplus
 \, H^{(2,1)} \nonumber\\
 & &  \oplus \, H^{(1,2)} \, \oplus \, H^{(0,3) }
\label{residue20}
\end{eqnarray}
and where the canonical bundle is trivial:
\begin{equation}
c_1\left( T{\cal M}\right) \, =\, 0 \,
\longleftrightarrow \, \mbox{dim} H^{(3,0)} \, = \, 1
\end{equation}
naming $\Omega^{(3,0)}$ the unique (up to a multiplicative constant)
holomorphic $3$--form, and choosing a canonical homology basis of
$3$--cycles $(A^\Lambda , B_ \Sigma)$ satisfying :
\begin{equation}
\matrix{
 A^\Lambda \, \cap \, A^\Sigma \, 0 \, & A^\Lambda \, \cap \, B_ \Delta \,
 = \, \delta^\Lambda_\Delta \cr
 B_ \Gamma \, \cap \, A^\Sigma  \,
 = - \, \delta^\Lambda_\Sigma  &  B_\Gamma \, \cap \, B_\Delta \, = \,
 0 \cr }
 \label{interseca}
\end{equation}
where
\begin{equation}
\Lambda , \Sigma \, \dots \, = \, 1,\dots\, {\bar n}=1+h^{(2,1)}
\end{equation}
we can define the periods:
\begin{eqnarray}
X^\Lambda (\phi) & = &  \int_{A^\Lambda} \, \Omega^{(3,0)}(\phi)
\nonumber\\
F_ \Sigma (\phi) & = &  \int_{B_\Sigma} \, \Omega^{(3,0)}(\phi)
\label{periodando}
\end{eqnarray}
where $\phi^i$ ($i=1,\dots\, h^{(2,1)}$)
are the moduli of the complex structures and we can implicitly define the
{\it period matrix} by the relation:
\begin{equation}
{\bar F}_ \Lambda \,  =  \, {\bar {\cal N}}_{\Lambda \Sigma} \,
{\bar X}^\Sigma ~,~{\o{\partial F_\Lambda}{\partial \phi^i}}
\, = \, {\bar {\cal N}}_{\Lambda \Sigma} \,
{\o{\partial X^\Sigma}{\partial \phi^i}}
\label{matper}
\end{equation}
Under a diffeomorphism $\xi$ of the manifold of complex structures
the period vector
 \begin{equation}
   V (\phi) \, = \, \left ( \matrix{ X^\Lambda (\phi) \cr
 F_ \Sigma (\phi) \cr } \right )
 \label{pervec}
\end{equation}
will transform linearly through the $Sp(2\bar n, \IR)$ matrix
$\iota_h (\xi)$ defined by the homomorphism in eq.~\ref{spaccointre}
and the period matrix ${\cal N}$ will obey the linear fractional
transformation rule of eq.~\ref{Ntrasform}. Indeed the
intersection relations in eq.~\ref{interseca} define the symplectic
invariant metric $\left (\matrix { {\bf 0} & \bfone \cr -\bfone &
{\bf 0} \cr } \right )$.
\par
Alternatively we can consider the more familiar example of a
Riemann surface of genus $g$.  Introducing a basis of homology
one--cycles $(A^\alpha , B_\beta)$ that satisfy  the analogue
of eq.~\ref{interseca}:
\begin{equation}
\matrix{
 A^\alpha \, \cap \, A^\beta \, = \, 0 & A^\alpha \, \cap \, B_ \gamma \,
 = \, \delta^\alpha_\gamma \cr
 B_ \mu \, \cap \, A^\beta  \,
 = - \, \delta^\beta_\mu  &  B_\mu \, \cap \, B_\gamma \, = \,
 0 \cr }
 \label{intsecrie}
\end{equation}
where
\begin{equation}
\alpha , \beta \, \dots \, = \, 1,\dots\, ,{\bar n}=g
\end{equation}
and a basis of holomorphic differentials $\omega_i$ ($i=1,\dots,g$)
we can set:
\begin{eqnarray}
& f_i^\alpha \, = \, \int_{A^\alpha} \, \omega_i   \quad ; \quad
h_{j \vert \beta} \, = \, \int_{B^\beta} \, \omega _j & \nonumber\\
& h_{i \vert \alpha} \, = \,  {\bar {\cal N}}_{\alpha\beta} \, f_i^\beta &
\label{rieperiod}
\end{eqnarray}
which provides the standard definition of the {\it period matrix}
${\cal N}$ in
Riemann surface theory \footnotemark
\footnotetext{usually one normalizes the choice of holomorphic differentials
in such a way that $f^\alpha_i = \delta^\alpha_i$ }
\par
The two examples we have recalled here from algebraic geometry will
be relevant for the issue of exact quantum duality symmetries in
the context of {\it local}, respectively {\it global } N=2 theories
and we shall have more to say about them in the sequel. What should
be clear from the above discussion is that a family of Lagrangians
as in eq.~\ref{gaiazuma} will admit a group of
duality--rotations/field--redefinitions that will map one into the
other member of the family, as long as a {\it kinetic matrix}
${\cal N}_{\Lambda\Sigma}$ can be constructed  that transforms as
in eq.~\ref{Ntrasform}. A way to obtain such an object is to identify
it with the {\it period matrix} occurring in problems of algebraic
geometry. At the level of the present discussion, however, this
identification is by no means essential: any construction of
${\cal N}_{\Lambda\Sigma}$ with the appropriate transformation
properties is acceptable.
\par
Note also that so far we have used the
words {\it duality--rotations/field--redefinitions} and not the word
duality symmetry. Indeed the diffeomorphisms of the scalar manifold
we have considered were quite general and, as such had no pretension
to be symmetries of the action, or of the theory. Indeed the question
we have answered is the following: what are the appropriate
transformation properties of the tensor gauge fields and of the generalized
coupling constants under diffeomorphisms of the scalar manifold?
The next question is obviously that of duality symmetries. Suppose
that a certain diffeomorphism $\xi \in \mbox{Diff}\left ( {\cal
M}_{scalar} \right )$ is actually an {\it isometry} of the scalar metric
$g_{IJ}$. Naming
$\xi^\star : \, T{\cal M}_{scalar} \, \rightarrow \, T{\cal M}_{scalar}$
the push--forward of $\xi$, this means that
\begin{eqnarray}
&\forall\,  X,Y \, \in \, T{\cal M}_{scalar}& \nonumber\\
& g\left ( X, Y \right
)\, = \, g \left ( \xi^\star X, \xi^\star Y \right )&
\label{isom}
\end{eqnarray}
and $\xi$ is an exact global symmetry of the scalar  part of the
Lagrangian in eq~\ref{gaiazuma}. The obvious question is:
{\it " can this symmetry be extended to a symmetry of the complete
action?}  Clearly the answer  is that, in general, this is not possible.
The best we can do is to extend it to a symmetry of the field
equations plus Bianchi identities letting it act as a duality
rotation on the field--strengths plus their duals. This requires
that the group of isometries of the scalar metric
${\cal I} ({\cal M}_{scalar})$ be suitably embedded into
the duality group (either $Sp(2\bar n,\IR)$ or $SO(\bar n , \bar n)$
depending on the case) and that the kinetic matrix ${\cal
N}_{\Lambda\Sigma}$ satisfies the covariance law:
\begin{equation}
{\cal N}\left ( \xi (\phi)\right ) \, = \,
\left ( C_\xi + D_\xi {\cal N}(\phi) \right )
\left ( A_\xi + B_\xi {\cal N}( \phi )\right )
\label{covarianza}
\end{equation}
A general class of solutions to this programme can be derived in the
case where the scalar manifold is taken to be a homogeneous space
${\cal G}/{\cal H}$. This is the subject of next section.

\section{Symplectic embeddings of homogenous spaces ${\cal G}/{\cal
H}$, the period matrix ${\cal N}$
and the structure of extended supergravities}
\label{LL2}
As remarked in the last section, the problem of constructing
duality--symmetric lagrangians of type~\ref{gaiazuma} admits
general solutions  when the scalar manifold is a homogeneous
space ${\cal G}/{\cal H}$. This is what happens in all
extended supergravities for $N \ge 3$ and also in specific
instances of N=2 theories. For this reason I devote the present
section to a review of the construction of the {\it kinetic
period matrix} ${\cal N}$ in the case of homogeneous spaces.
{}From now on I also concentrate on the case of $D=4$ Minkowski
space, so that the relevant homomorphism $\iota_\delta$ (see
eq.~\ref{spaccoindue}) becomes:
\begin{equation}
\iota_\delta : \, \mbox{Diff}\left ({{\cal G}\over{\cal
H}} \right ) \, \longrightarrow \, Sp(2\bar n, \IR)
\label{embeddif}
\end{equation}
In particular, focusing on the isometry group of the canonical metric
defined on ${{\cal G}\over{\cal H}}$\footnotemark
\footnotetext{Actually, in order to be true, eq.~\ref{isogroup} requires
that that the normaliser of ${\cal H}$ in ${\cal G}$ be the
identity group, a condition that is verified in all the relevant examples}:
\begin{equation}
 {\cal I} \left ({{\cal G}\over{\cal H}}\right ) \, = \, {\cal G}
 \label{isogroup}
\end{equation}
we must consider the embedding:
\begin{equation}
\iota_\delta : \,  {\cal G}  \, \longrightarrow \, Sp(2\bar n, \IR)
\label{embediso}
\end{equation}
That in eq.~\ref{embeddif} is a homomorphism of finite dimensional
Lie groups and as such it constitutes a problem that can be solved
in explicit form. What we just need to know is the dimension of the
symplectic group, namely the number $\bar n$ of gauge fields appearing
in the theory. Without supersymmetry the dimension $m$ of the scalar
manifold (namely the possible choices of ${{\cal G}\over{\cal H}}$) and
the number of vectors $\bar n$ are unrelated so that the
possibilities covered by eq.~\ref{embediso} are infinitely many.
In supersymmetric theories, instead, the two numbers $m$ and $\bar n$
are related, so that there are finitely many cases to be studied
corresponding to the possible  embeddings of given groups
${\cal G}$ into a symplectic group $Sp(2\bar n, \IR)$ of fixed dimension
$\bar n$. Actually taking into account further conditions on
the holonomy of the scalar manifold that are also imposed by
supersymmetry, the solution for the symplectic embedding problem
is unique for all extended supergravities with $N \ge 3$ as we
have already remarked. This yields the unique scalar manifold
choice displayed in Table~\ref{topotable}.
\par
\begin{table*}
\begin{center}
\caption{\sl Scalar Manifolds of Extended Supergravities}
\label{topotable}
\begin{tabular}{|c||c|c|c||c|c||c||c| }
\hline
\hline
{}~ & $\#$ scal. & $\#$ scal. & $\#$ scal. & $\#$ vect. &
 $\#$ vect. &~ & $~ $ \\
N & in & in & in & in  &
 in  &$\Gamma_{cont}$ & ${\cal M}_{scalar}$   \\
 ~ & scal.m. & vec. m. & grav. m. & vec. m. & grav. m. & ~ &~
\\
\hline
\hline
{}~    &~    &~   &~   &~  &~  & ~ & ~ \\
$1$  & 2 m &~   & ~  & n &~  &  ${\cal I}$  & ~   \\
{}~    &~    &~   &~   &~  &~  &  $\subset Sp(2n,\IR)$ & K\"ahler \\
{}~    &~    &~   &~   &~  &~  & ~ & ~ \\
\hline
{}~    &~    &~   &~   &~  &~  & ~ & ~ \\
$2$  & 4 m & 2 n& ~  & n & 1 &  ${\cal I}$ & Quaternionic $\otimes$
\\
{}~    &~    &~   &~   &~  &~  &  $\subset Sp(2n+2,\IR)$ & Special K\"ahler \\
{}~    &~    &~   &~   &~  &~  & ~ & ~ \\
\hline
{}~    &~    &~   &~   &~  &~  & ~ & ~ \\
$3$  & ~   & 6 n& ~  & n & 3 &  $SU(3,n)$ &~  \\
{}~    &~    &~   &~   &~  &~  & $\subset Sp(2n+6,\IR)$ & $\o{SU(3,n)}
{S(U(3)\times U(n))}$ \\
{}~    &~    & ~  &~   &~  &~  & ~ & ~ \\
\hline
{}~    &~    &~   &~   &~  &~  & ~ & ~ \\
$4$  & ~   & 6 n& 2  & n & 6 &  $SU(1,1)\otimes SO(6,n)$ &
$\o{SU(1,1)}{U(1)} \otimes $ \\
{}~    &~    &~   &~   &~  &~  & $\subset Sp(2n+12,\IR)$ &
$\o{SO(6,n)}{SO(6)\times SO(n)}$ \\
{}~    &~    &~   &~   &~  &~  & ~ & ~ \\
\hline
{}~    &~    &~   &~   &~  &~  & ~ & ~ \\
$5$  & ~   & ~  & 10 & ~ & 10 & $SU(1,5)$ & ~  \\
{}~    &~    &~   &~   &~  &~  & $\subset Sp(20,\IR)$ & $\o{SU(1,5)}
{S(U(1)\times U(5))}$ \\
{}~    &~    &~   &~   &~  &~  & ~ & ~ \\
\hline
{}~    &~    &~   &~   &~  &~  & ~ & ~ \\
$6$  & ~   & ~  & 30 & ~ & 16 & $SO^\star(12)$ & ~ \\
{}~    &~    &~   &~   &~  &~  & $\subset Sp(32,\IR)$ &
$\o{SO^\star(12)}{U(1)\times SU(6)}$ \\
{}~    &~    &~   &~   &~  &~  & ~ & ~ \\
\hline
{}~    &~    &~   &~   &~  &~  & ~ & ~ \\
$7,8$& ~   & ~  & 70 & ~ & 56 & $E_{7(-7)}$  & ~ \\
{}~    &~    &~   &~   &~  &~  & $\subset Sp(128,\IR)$ &
$\o{ E_{7(-7)} }{SU(8)}$ \\
{}~    &~    &~   &~   &~  &~  & ~ & ~ \\
\hline
 \hline
\end{tabular}
\end{center}
\end{table*}
Apart from the details of the specific case considered
once a symplectic embedding is given there is a general
formula one can write down for the {\it period matrix}
${\cal N}$ that guarantees symmetry (${\cal N}^T = {\cal N}$)
and the required transformation property~\ref{covarianza}.
This is the first result I want to present.
\par
The real symplectic group $Sp(2\bar n ,\IR)$ is defined as the set
of all {\it real} $2\bar n \times 2\bar n$ matrices
\begin{equation}
\Lambda \, = \, \left ( \matrix{ A & B \cr C & D \cr } \right )
\label{matriciana}
\end{equation}
satisfying the first of equations~\ref{ortosymp}, namely
\begin{equation}
\Lambda^T \, \IC \, \Lambda \, = \, \IC
\label{condiziona}
\end{equation}
 where
\begin{equation}
\IC  \, \equiv \, \left ( \matrix{ {\bf 0} & \bfone \cr -\bfone &
{\bf 0} \cr } \right )
\label{definizia}
\end{equation}
If we relax the condition that the matrix should be real but we
still impose eq.~\ref{condiziona} we obtain the definition
of the complex symplectic group $Sp(2\bar n, \IC)$. It is
a well known fact that the following isomorphism is true:
\begin{equation}
Sp(2\bar n, \IR)  \sim  Usp(\bar n , \bar n)   \equiv
Sp(2\bar n, \IC)   \cap   U(\bar n , \bar n)
\label{usplet}
\end{equation}
By definition an element ${\cal S}\,\in \, Usp(\bar n , \bar n)$
is a complex matrix that satisfies simultaneously eq.~\ref{condiziona}
and a pseudo--unitarity condition, that is:
\begin{eqnarray}
{\cal S}^T \, \IC \, {\cal S} &=& \IC \nonumber\\
{\cal S}^\dagger \, \IH \, {\cal S} &=& \IH \nonumber\\
\IH & \equiv & \left ( \matrix{ \bfone & {\bf 0} \cr {\bf 0} & -\bfone
 \cr } \right )
\label{uspcondo}
\end{eqnarray}
The general block form of the matrix ${\cal S}$ is:
\begin{equation}
{\cal S}\, = \, \left ( \matrix{ T & V^\star \cr V & T^\star \cr } \right )
\label{blocusplet}
\end{equation}
and eq.s~\ref{uspcondo} are equivalent to:
\begin{eqnarray}
T^\dagger \, T \, - \, V^\dagger \, V &=& \bfone \nonumber\\
T^\dagger \, V^\star  \, - \,  V^\dagger \, T^\star &=& {\bf 0}
\label{relazie}
\end{eqnarray}
The isomorphism of eq.~\ref{usplet} is explicitly realized by
the so called Cayley matrix:
\begin{equation}
{\cal C} \, \equiv \, {\o{1}{\sqrt{2}}} \,
\left ( \matrix{ \bfone & {\rm i}\bfone \cr \bfone & -{\rm i}\bfone
 \cr } \right )
\label{cayley}
\end{equation}
via the relation:
\begin{equation}
{\cal S}\, = \, {\cal C} \, \Lambda \, {\cal C}^{-1}
\label{isomorfo}
\end{equation}
which yields:
\begin{eqnarray}
T &=& {\o{1}{2}}\, \left ( A - {\rm i} B \right ) +
{\o{1}{2}}\, \left ( C + {\rm i} D \right ) \nonumber\\
V &=& {\o{1}{2}}\, \left ( A - {\rm i} B \right ) -
{\o{1}{2}}\, \left ( C + {\rm i} D \right ) \nonumber\\
\label{mappetta}
\end{eqnarray}
When we set $V=0$ we obtain the subgroup $U(\bar n) \subset Usp (\bar
n , \bar n)$, that in the real basis is given by the subset of
symplectic matrices of the form
$\left ( \matrix{ A & B \cr -B & A
 \cr } \right )$. The basic idea, to obtain the
general formula for the period matrix, is that the symplectic embedding
of the isometry group ${\cal G}$ will be such that the isotropy
subgroup ${\cal H}\subset {\cal G}$ gets embedded into the maximal
compact subgroup $U(\bar n)$, namely:
\begin{eqnarray}
{\cal G} & {\stackrel{\iota_\delta}{\longrightarrow}} & Usp (\bar
n , \bar n) \nonumber\\
{\cal G} \supset {\cal H} & {\stackrel{\iota_\delta}{\longrightarrow}} &
U(\bar n) \subset Usp (\bar n , \bar n)
\label{gruppino}
\end{eqnarray}
If this condition is realized let $L(\phi)$ be a parametrization of
the coset ${\cal G}/{\cal H}$ by means of coset representatives.
By this we mean the following. Let $\phi^I$ be local coordinates on the
manifold ${\cal G}/{\cal H}$: to each point $\phi   \in
{\cal G}/{\cal H}$ we assign an element $L(\phi) \in {\cal G}$ in
such a way that if $\phi^\prime \ne \phi$, then no $h \in {\cal H}$
can exist such that $L(\phi^\prime)=L(\phi)\cdot h$.
In other words for each equivalence class
of the coset (labelled by the coordinate $\phi$) we choose one
representative element $L(\phi)$ of the class.
Relying on the symplectic embedding of eq.~\ref{gruppino} we obtain
a map:
\begin{eqnarray}
& L(\phi)  \, \longrightarrow  {\cal O}(\phi)\, =  \, &\nonumber\\
&\left ( \matrix{ U_0(\phi) & U^\star_1(\phi) \cr U_1(\phi)
& U^\star_0(\phi) \cr } \right )\,  \in  \, Usp(\bar n , \bar n)&
\label{darstel}
\end{eqnarray}
that associates to $L(\phi)$ a coset representative of $Usp(\bar n ,
\bar n)/U(\bar n)$. By construction if $\phi^\prime \ne \phi$
{\it no} unitary $\bar n \times \bar n$ matrix $W$ {\it can exist}
such that:
\begin{equation}
 {\cal O}(\phi^\prime)  =  {\cal O}(\phi) \,
 \left ( \matrix{ W & {\bf 0} \cr {\bf 0}
& W^\star \cr } \right )
\end{equation}
On the other hand let $\xi \in {\cal G}$ be an element of the
isometry group of ${{\cal G}/{\cal H}}$. Via the symplectic embedding
of eq.~\ref{gruppino} we obtain a $Usp(\bar n, \bar n)$ matrix
\begin{equation}
{\cal S}_ \xi \, = \,
\left ( \matrix{ T_\xi & V^\star_\xi \cr V_\xi & T^\star_\xi \cr } \right )
\label{uspimag}
\end{equation}
such that
\begin{equation}
{\cal S}_ \xi \,{\cal O}(\phi) \, = \, {\cal O}(\xi(\phi)) \,
\left ( \matrix{ W(\xi,\phi) & {\bf 0} \cr {\bf 0}
& W^\star(\xi,\phi) \cr } \right )
\label{cosettone}
\end{equation}
where $\xi(\phi)$ denotes the image of the point
$\phi \in  {{\cal G}/{\cal H}}$ through $\xi$ and $W(\xi,\phi)$ is
a suitable $U(\bar n)$ compensator depending both on $\xi$ and
$\phi$.
Combining eq.s~\ref{cosettone},~\ref{darstel}, with eq.s~\ref{mappetta}
we immediately obtain:
\begin{eqnarray}
&U_0^\dagger \left( \xi(\phi) \right ) +
U^\dagger_1 \left (\xi(\phi) \right)   = & \nonumber\\
& W^\dagger   \left [ U_0^\dagger \left( \phi \right )   \left (
A^T + {\rm i}B^T \right ) + U_1^\dagger \left( \phi \right )   \left (
A^T - {\rm i}B^T \right ) \right ]  &\nonumber\\
&U_0^\dagger \left( \xi(\phi) \right ) -
U^\dagger_1 \left (\xi(\phi) \right)   = & \nonumber\\
& W^\dagger \, \left [ U_0^\dagger \left( \phi \right )   \left (
D^T - {\rm i}C^T \right ) - U_1^\dagger \left( \phi \right )   \left (
D^T + {\rm i}C^T \right ) \right ]  &\nonumber\\
\label{semitrasform}
\end{eqnarray}
Setting:
\begin{equation}
{\cal N} \, \equiv \, {\rm i} \left [ U_0^\dagger + U_1^\dagger \right
]^{-1} \, \left [ U_0^\dagger - U_1^\dagger \right ]
\label{masterformula}
\end{equation}
and using the result of eq.~\ref{semitrasform} one verifies
that the transformation rule~\ref{covarianza} is verified.
It is also an immediate consequence of the analogue of
eq.s~\ref{relazie} satisfied by $U_0$ and $U_1$ that the matrix
in eq.~\ref{masterformula} is symmetric
\begin{equation}
{\cal N}^T \, = \, {\cal N}
\label{massi}
\end{equation}
Eq.~\ref{masterformula} is the masterformula derived in 1981 by
Gaillard and Zumino \cite{gaizum}.
It explains the structure of the gauge field
kinetic terms in all $N\ge 3$ extended supergravity theories and
also in those $N=2$ theories where, using the parlance of my later
lectures,
the {\it Special K\"ahler manifold} ${\cal SM}$
is a homogeneous manifold ${\cal G}/{\cal H}$. In particular, using
eq.~\ref{masterformula} we can easily retrieve the structure of
$N=4$ supergravity, extensively discussed in Sen's lectures
\cite{senlecture} as the basic example of theory where {\it
Strong--Weak duality} is realized through {\it Electric--Magnetic
duality} rotations. In Sen's approach the structure of the $N=4$ bosonic
Lagrangian is derived by means of dimensional reduction from ten
dimensions. Obviously the same structure can be directly derived in $D=4$
by supersymmetry, since the $N=4$ theory is unique. Actually, given the
information (following from $N=4$ supersymmetry)
that the scalar manifold is the following coset manifold (see
table~\ref{topotable}):
\begin{eqnarray}
{\cal M}^{N=4}_{scalar} & = & {\cal ST}\left [ 6,n \right ]
\nonumber\\
{\cal ST}\left [ m,n \right ] & \equiv &
{\o{SU(1,1)}{U(1)}} \, \otimes \, {\o{SO(m,n)}{SO(m)\otimes SO(n)}}
\nonumber\\
\null & \null & \null
\label{stmanif}
\end{eqnarray}
what we just need to study is the symplectic embedding of
the coset manifolds ${\cal ST}\left [ 6,n \right ]$ where $n$ is
the number vector multiplets in the theory. This is what I do
in the next subsection where I actually consider the
general case of  ${\cal ST}\left [ m,n \right ]$ manifolds.
\subsection{Symplectic embedding of the ${\cal ST}\left [ m,n \right ]$
homogeneous manifolds}
The first thing I should do is to justify the name I have given to
the particular class of coset manifolds I propose to study. The
letters ${\cal ST}$ stand for space--time and target space duality.
Indeed, the isometry group of the ${\cal ST}\left [ m,n \right ]$
manifolds defined in eq.~\ref{stmanif} contains a factor ($SU(1,1)$)
whose transformations act as non--perturbative $S$--dualities and
another factor $(SO(m,n)$ whose transformations act as
$T$--dualities,
holding true at each order in string perturbation theory. Furthermore
$S$ is the traditional name given, in superstring theory, to the
complex field obtained by combining together the {\it dilaton} $D$ and
{\it axion} ${\cal A}$:
\begin{eqnarray}
S & = & {\cal A} - {\rm i} \mbox{exp}[D] \nonumber\\
\partial^\mu {\cal A} & \equiv & \varepsilon^{\mu\nu\rho\sigma} \,
\partial_\nu \, B_{\rho\sigma}
\label{scampo}
\end{eqnarray}
while $t^i$ is the name usually given to the moduli--fields of the
compactified target space. Now in string and supergravity
applications $S$ will be identified with the complex coordinate
on the manifold ${\o{SU(1,1)}{U(1)}}$, while  $t^i$  will be
the coordinates of the coset space  ${\o{SO(m,n)}{SO(m)\otimes SO(n)}}$.
Although as differentiable and metric manifolds
the spaces ${\cal ST}\left [ m,n \right ]$ are just direct products
of two factors (corresponding to the above mentioned different
physical interpretation of the coordinates $S$ and $t^i$), from the
point of view of the symplectic embedding and duality rotations
they have to be regarded as a single entity. This is even more
evident in the case $m=2,n=\mbox{arbitrary}$,
where the following theorem has been proven by
Ferrara and Van Proeyen \cite{ferratoine}:
${\cal ST}\left [ 2,n \right ]$ are the only special K\"ahler
manifolds with a direct product structure. For the definition
of special K\"ahler manifolds I ask the audience to be patient
and wait until my further lectures, yet the anticipation of this result
should make
clear that the special K\"ahler structure (encoding the
duality rotations in the $N=2$ case)
is not a property of the individual factors
but of the product as a whole. Neither factor
is by itself a special manifold although the product is.
\par
{}From this anticipation it also appears that the proposed study
of symplectic embedding is relevant not only for the unique $N=4$
theory, but also for a class of $N=2$ theories. Are they physically
relevant? Very much so, since the ${\cal ST}\left [ 2,r \right ]$
special manifold is what emerges, at tree level, as moduli space of
compactified N=2 string theories with a rank $r$ gauge group.
The basic question addressed in the recent literature and leading
to the most exciting results is: {\it How is the
${\cal ST}\left [ 2,r \right ]$ manifold deformed by quantum corrections?}
Or, in different words,  {\it How does the quantum moduli space
${\hat {\cal ST}}\left [ 2,r \right ]$ look like?}
The same question has a simple answer in the $N=4$ case. There the
tree--level moduli space of a heterotic string compactified on a $T^6$
torus is
\begin{equation}
{\cal M}^{N=4}_{moduli} \, = \, {  {\cal ST}}\left [ 6,22 \right ]
\label{n4moduli}
\end{equation}
the number of abelian gauge fields being
$22= 6\, \mbox{(moduli of $T^6$) }
\oplus    16 \, \mbox{ (rank of $E_8 \times E_8$ )}$.
Because of the uniqueness of $N=4$
supergravity the quantum moduli--space
${\hat  {\cal ST}}\left [ 6,22 \right ]$
cannot be anything else but a manifold with the same covering space
as ${  {\cal ST}}\left [ 6,22 \right ]$, namely a manifold with the
same local structure. Indeed the only thing which is not fixed
by $N=4$ supersymmetry is the global structure of the
scalar manifold.  What actually comes out is the following result
\begin{equation}
{\hat  {\cal ST}}\left [ 6,22 \right ] \, = \,
{\o{{  {\cal ST}}\left [ 6,22 \right ] }{SL(2,\ZZ)\otimes
SO(6,22,\ZZ)}}
\label{qn4mod}
\end{equation}
The homotopy group of the quantum moduli space:
\begin{equation}
 \pi_1 \left ( {\hat  {\cal ST}}\left [ 6,22 \right ] \right ) \,= \,
 {SL(2,\ZZ)\otimes SO(6,22,\ZZ)}
\label{homotop}
\end{equation}
is just the restriction to the integers $\ZZ$ of the original continuous
duality group $SL(2,\IR)\otimes SO(6,22,\IR)$ associated with the
manifold ${  {\cal ST}}\left [ 6,22 \right ]$. After modding by
this discrete group the only duality--rotations that survive as
exact duality symmetries of the quantum theory are those contained
in $\pi_1 \left ( {\hat  {\cal ST}}\left [ 6,22 \right ] \right )$
itself. This happens because of the Dirac quantization condition in
eq.~\ref{dirquant} of electric and magnetic charges, the lattice spanned
by these charges being invariant under the discrete group of
eq.~\ref{homotop}.
At this junction the relevance, in the quantum theory, of the symplectic
embedding should appear. What does  restriction to the integers
exactly, mean? It means that the image in $Sp(56,\IR)$ of those
matrices of $SL(2,\IR) \times SO(6,22,\IR)$ that are retained as
elements of $\pi_1 \left ( {\hat  {\cal ST}}\left [ 6,22 \right ] \right )$
should be integer valued. In other words we define:
\begin{eqnarray}
 & SL(2,\ZZ) \times SO(6,22,\ZZ) \, \equiv &  \nonumber\\
 & \iota_\delta \left ( SL(2,\IR) \times SO(6,22,\IR) \right ) \cap
 Sp(56,\ZZ) &
\label{astrazeta}
\end{eqnarray}
As we see the statement in eq.~\ref{astrazeta} is dependent on the
symplectic embedding. What is integer valued in one embedding is
not integer valued in another embedding. This raises the question
of the correct symplectic embedding. Such a question has two aspects:
\begin{enumerate}
\item{Intrinsically inequivalent embeddings}
\item{Symplectically equivalent embeddings that become inequivalent
after gauging}
\end{enumerate}
The first issue in the above list is group--theoretical in nature.
When we say that the group ${\cal G}$ is embedded into $Sp(2\bar
n,\IR)$ we must specify how this is done from the point of view
of irreducible representations. Group--theoretically the matter is
settled by specifying how the fundamental representation of
$Sp(2\bar n)$ splits into irreducible representations of ${\cal G}$:
\begin{eqnarray}
& {\bf {2 \bar n}} \, {\stackrel{{\cal G}}{\longrightarrow}}
\oplus_{i=1}^{\ell} \, {\bf D}_i &
\label{splitsplit}
\end{eqnarray}
Once eq.~\ref{splitsplit} is given (in supersymmetric theories
such information is provided by supersymmetry ) the only arbitrariness
which is left is that of conjugation by arbitrary $Sp(2\bar n,\IR)$
matrices. Suppose we have determined an embedding $\iota_\delta$ that
obeys the law in eq.~\ref{splitsplit}, then:
\begin{equation}
\forall \, {\cal S} \, \in \, Sp(2\bar n,\IR) \, : \,
\iota_\delta^\prime \, \equiv \, {\cal S} \circ  \iota_\delta \circ
{\cal S}^{-1}
\label{matrim}
\end{equation}
will obey the same law. That in eq.~\ref{matrim} is a symplectic
transformation that corresponds to an allowed
duality--rotation/field--redefinition in the abelian theory of
type in eq.~\ref{gaiazuma} discussed in the previous subsection. Therefore
all abelian lagrangians related by such transformations are physically
equivalent.
\par
The matter changes in presence of {\it gauging}. When we switch
on the gauge coupling constant and the electric charges, symplectic
transformations cease to yield physically equivalent theories. This
is the second issue in the above list. The choice of a symplectic
gauge becomes physically significant. As I have emphasized in the
introduction, the construction of supergravity theories proceeds in
two steps. In the first step, which is the most extensive and complicated,
one constructs the abelian theory: at that level the only relevant
constraint is that encoded in eq.~\ref{splitsplit} and the choice of
a symplectic gauge is immaterial. Actually one can write the entire
theory in such a way that {\it symplectic covariance} is manifest.
In the second step one {\it gauges} the theory. This {\it breaks
symplectic covariance} and the choice of the correct symplectic gauge
becomes a physical issue. This issue has been recently emphasized
by the results in \cite{porfergir} where it has been shown that
whether N=2 supersymmetry can be spontaneously broken to N=1 or
not depends on the symplectic gauge.
\par
These facts being cleared I proceed to discuss the symplectic
embedding of the ${\cal ST}\left [ m,n \right ]$ manifolds.
\par
Let $\eta$ be the symmetric flat metric with signature
$(m,n)$ that defines the $SO(m,n)$ group, via the relation
\begin{equation}
L \, \in \, SO(m,n) \, \Longleftrightarrow \, L^T \, \eta L \, = \,
\eta
\label{ortogruppo}
\end{equation}
Both in the $N=4$ and in the $N=2$ theory, the number of gauge fields
in the theory is given by:
\begin{equation}
\# \mbox{vector fields} \, = \, m \oplus n
\label{vectornum}
\end{equation}
$m$ being the number of {\it graviphotons} and $n$ the number of
{\it vector multiplets}. Hence we have to embed $SO(m,n)$ into
$Sp(2m+2n,\IR)$ and the explicit form of the decomposition in
eq.~\ref{splitsplit} required by supersymmetry is:
\begin{equation}
{\bf {2m+2n}} \, {\stackrel{SO(m,n)}{\longrightarrow}} \, {\bf { m+n}}
\oplus  {\bf { m+n}}
\label{ortosplitsplit}
\end{equation}
where ${\bf { m+n}}$ denotes the fundamental representation of
$SO(m,n)$. Eq.~\ref{ortosplitsplit} is easily understood in physical
terms. $SO(m,n)$ must be a T--duality group, namely a symmetry
 holding true order by order in perturbation theory. As such it must
 rotate electric  field strengths into electric field strengths and
 magnetic field strengths into magnetic field field strengths. The
 two irreducible representations into which the the fundamental
 representation of the symplectic group decomposes when reduced to
 $SO(m,n)$ correspond precisely to electric and magnetic sectors,
 respectively.
In the {\it simplest  gauge} the symplectic embedding satisfying
eq.~\ref{ortosplitsplit} is block--diagonal and takes the form:
\begin{eqnarray}
& \forall \,  L \, \in \, SO(m,n) \quad {\stackrel{\iota_\delta}
{\hookrightarrow}} & \nonumber\\
&\left ( \matrix{ L & {\bf 0}\cr {\bf 0} & (L^T)^{-1} \cr } \right )
\, \in \, Sp(2m+2n,\IR) &
\label{ortoletto}
\end{eqnarray}
Consider instead the group $SU(1,1) \sim SL(2,\IR)$. This is the
factor in the isometry group of ${\cal ST}[m,n]$
that is going to act by means of S--duality non perturbative
rotations. Typically it will rotate each electric field strength into
its homologous magnetic one. Correspondingly supersymmetry implies
that its embedding into the symplectic group must satisfy the
following condition:
\begin{equation}
{\bf {2m+2n}} \, {\stackrel{SL(2,\IR)}{\longrightarrow}} \,
\oplus_{i=1}^{m+n} \, {\bf 2}
\label{simposplisplit}
\end{equation}
where  ${\bf 2}$ denotes the fundamental representation of
$SL(2,\IR)$. In addition it must commute with the embedding
of $SO(m,n)$ in eq.~\ref{ortoletto} . Both conditions are fulfilled
by setting:
\begin{eqnarray}
& \forall \,   \left ( \matrix{a & b \cr  c &d \cr }\right )
\, \in \, SL(2,\IR) \quad {\stackrel{\iota_\delta}
{\hookrightarrow}} & \nonumber\\
&\left ( \matrix{ a \, \bfone & b \, \eta \cr c \, \eta &
d \, \bfone \cr } \right )
\, \in \, Sp(2m+2n,\IR) &
\label{ortolettodue}
\end{eqnarray}
Utilizing eq.s~\ref{isomorfo} the corresponding  embeddings into
the group $Usp(m+n,m+n)$ are immediately derived:
\begin{eqnarray}
& \forall \,  L \, \in \, SO(m,n) \quad {\stackrel{\iota_\delta}
{\hookrightarrow}} & \nonumber\\
&\left ( \matrix{ {\o{1}{2}}  \left ( L+
\eta L \eta \right ) & {\o{1}{2}}  \left ( L-
\eta L \eta \right )\cr {\o{1}{2}}  \left ( L -
\eta L \eta \right ) & {\o{1}{2}}  \left ( L+
\eta L \eta \right ) \cr } \right ) & \nonumber\\
& \, \in \, Usp(m+n,m+n) & \nonumber\\
& \forall \,   \left ( \matrix{t & v^\star \cr  v &t^\star \cr }\right )
\, \in \, SU(1,1) \quad {\stackrel{\iota_\delta}
{\hookrightarrow}} & \nonumber\\
&\left ( \matrix{ {\rm Re}t \bfone +{\rm i}{\rm Im}t\eta &
{\rm Re}v \bfone -{\rm i}{\rm Im}v \eta   \cr
{\rm Re}v \bfone +{\rm i}{\rm Im}v\eta &
{\rm Re}t \bfone - {\rm i}{\rm Im}t\eta \cr } \right )& \nonumber\\
&\, \in \, Usp(m+n,m+n) &
\label{uspembed}
\end{eqnarray}
where the relation between the entries of the $SU(1,1)$ matrix
and those of the corresponding $SL(2,\IR)$ matrix are provided
by the relation in eq.~\ref{mappetta}.
\par
Equipped with these relations we can proceed to derive the explicit
form of the {\it period matrix} ${\cal N}$.
\par
The homogeneous manifold $SU(1,1)/U(1)$ can be conveniently
parametrized in terms of a single complex coordinate $S$, whose
physical interpretation will be that of {\it axion--dilaton},
according to eq.~\ref{scampo}. The coset parametrization appropriate
for comparison with other constructions (Special Geometry (see later
sections) or dimensional reduction (see~\cite{senlecture})) is given
by the matrices:
\begin{eqnarray}
& M(S) \, \equiv \, {\o{1}{n(S)} } \, \left (
\matrix{ \bfone & { \o{{\rm i} -S }{ {\rm i} + S } }\cr
{\o{ {\rm i} + {\bar S} }{ {\rm i} -{\bar S} } } & \bfone \cr}
\right )& \nonumber \\
& n(S) \, \equiv \, \sqrt{ {\o{4 {\rm Im}S } {
 1+\vert S \vert^2 +2 {\rm Im}S } } }&
 \label{su11coset}
\end{eqnarray}
To parametrize the coset $SO(m,n)/SO(m)\times SO(n)$ we can instead
take the usual coset representatives
(see for instance~\cite{castdauriafre}):
\begin{equation}
L(X) \, \equiv \, \left (\matrix{ \left ( \bfone + XX^T \right )^{1/2}
& X \cr X^T & \left ( \bfone + X^T X \right )^{1/2}\cr } \right )
\label{somncoset}
\end{equation}
where the $m \times n $ real matrix $X$ provides a set of independent
coordinates. Inserting these matrices into the embedding formulae of
eq.s~\ref{uspembed} we obtain a matrix:
\begin{eqnarray}
& Usp(n+m , n+m) \,
\ni  \, \iota_\delta \left ( M (S) \right ) \circ  \iota_\delta
\left ( L(X) \right ) & \nonumber\\
& = \, \left ( \matrix{ U_0(S,X) & U^\star_1(S,X) \cr
U_1(S,X)
& U^\star_0(S,X) \cr } \right )& \nonumber\\
\label{uspuspusp}
\end{eqnarray}
that inserted into the master formula of eq.~\ref{masterformula}
yields the following result:
\begin{equation}
{\cal N}\, = \, {\rm i} {\rm Im}S \, \eta L(X) L^T(X) \eta
+ {\rm Re}S \, \eta
\label{maestrina}
\end{equation}
Alternatively, remarking that if $L(X)$ is an $SO(m,n)$ matrix
also $L(X)^\prime =\eta L(X) \eta$ is such a matrix and represents
the same equivalence class, we can rewrite ~\ref{maestrina} in the
simpler form:
\begin{equation}
{\cal N}\, = \, {\rm i} {\rm Im}S \,   L(X)^\prime L^{T\prime}
(X)
+ {\rm Re}S \, \eta
\label{maestrino}
\end{equation}
\subsection{The bosonic lagrangian of $N=4$ supergravity}
As anticipated, eq.s~\ref{maestrino} and ~\ref{maestrina} are
the group--theoretical explanation of the form taken by $N=2$
supergravity when the special manifold ${\cal ST}[2,n]$ is used
as scalar manifold in the vector multiplet sector.
To this point I will come back
after discussing the notion of special K\"ahler geometry. But
eq.~\ref{maestrina} is also the group theoretical explanation
of the $N=4$ lagrangian in the specific form discussed by Sen in
his lectures at this school \cite{senlecture} and best suited
to study $S$ and $T$ duality. Let me then show how such a lagrangian
is retrieved from the previous discussion focusing on the
case ${\cal ST}[6,n]$.
\par
Since we deal with the bosonic sector of $N=4$ supergravity
we have, in addition to the scalar and vector fields, also the
graviton. Correspondingly we write:
\begin{eqnarray}
{\cal L}^{N=4}_{bose} &=& \sqrt{-g} \, \Biggl [ R[g] \nonumber\\
& & + {\o{1}{4({\rm Im}S)^2}} \, \partial_\mu S \partial^\mu {\bar
S}\nonumber\\
& & -{\o{1}{4}} \, {\rm Im}S \, {\cal F}^{\Lambda}_{\mu\nu} \,
\left ( \eta M \eta \right )_{\Lambda\Sigma} \,
{\cal F}^{\Sigma \vert \mu \nu} \nonumber\\
& & +{\o{1}{8\sqrt{-g}}} \, {\rm Re}S \, {\cal F}^{\Lambda}_{\mu\nu} \,
 \eta_{\Lambda\Sigma} \,
{\cal F}^{\Sigma}_{\rho\sigma} \,
\varepsilon^{\mu\nu\rho\sigma}\nonumber\\
& & - {\o{1}{4}} \, {\rm Tr} \left ( \partial_\mu M \, \partial^\mu M
\, \right ) \, \Biggr ]
\label{n4lagra}
\end{eqnarray}
where, by definition I have set:
\begin{equation}
M \, = \, L(X) \, L^T(X)
\label{mdefi}
\end{equation}
By construction the matrix $M(X)$ is symmetric. Comparison of
eq.s~\ref{n4lagra} with eq.s~\ref{gaiazuma} and
{}~\ref{scripten} shows that in writing the kinetic terms for the
vector fields, we have indeed utilized the formula of eq.~\ref{maestrina}
for the matrix ${\cal N}_{\Lambda\Sigma}$. The expression
used in eq.~\ref{n4lagra} for the
kinetic term of the $SO(6,n)/SO(6)\times SO(n)$ scalars
is explained as follows.
Let
\begin{equation}
\Theta \, \equiv \, L^{-1}(X) \, d L(X)
\label{lif}
\end{equation}
be the left invariant 1--form on the coset manifold
$SO(m,n)/SO(m)\times SO(n)$.
Suppose that $n=m+r, \, r\ge 0$ and choose for the
pseudo--orthogonal metric $\eta$ the following form:
\begin{equation}
\eta \,= \, \left ( \matrix{ {\bf O}_{m \times m}& \bfone_{m \times
m} & {\bf O}_{m \times r} \cr
\bfone_{m \times m}& {\bf 0}_{m \times
m} & {\bf O}_{m \times r} \cr
 {\bf O}_{r \times m} & {\bf 0}_{r \times
m} & \bfone_{r \times r} \cr }\right )
\end{equation}
In this basis the subalgebra
\begin{equation}
SO(m)\otimes SO(m+r) \subset SO(m,m+r)
\end{equation}
is made by those matrices $\Lambda$ that in addition to satisfying the
condition:
\begin{equation}
\eta \, \L^T + L \, \eta \, = \, 0
\end{equation}
are also antisymmetric:
\begin{equation}
\Lambda = - \Lambda^T
\end{equation}
Relying on this fact the decomposition of the left invariant
1--form along the subalgebra and coset directions becomes very
simple. The $\IH$--connection is just the antisymmetric part of
$\Theta$, while the vielbein on the coset manifold is given by
its symmetric part:
\begin{equation}
E \, = \, {\o{1}{2}} \, \left ( \Theta + \Theta^T \right)
\label{quattrozamponi}
\end{equation}
It is then a matter of straightforward algebra to verify that
the coset manifold metric
\begin{equation}
ds^2 \, \equiv \, {\rm Tr} \left ( E \, \otimes E \right )
\end{equation}
can be rewritten as
\begin{equation}
 ds^2 \, = -\,{\o{1}{4}} {\rm Tr} \left ( dM\, \otimes dM \right )
\end{equation}
the matrix $M(X)$ being defined in eq.~\ref{mdefi}. This proves
why $- {\o{1}{4}} \, {\rm Tr} \left ( \partial_\mu M \, \partial^\mu M
\, \right ) \,$ is a permissible way of writing the sigma--model
lagrangian when the target manifold is $SO(m,n)/SO(m)\times SO(n)$.
Actually the entries of the symmetric matrix $M$ can be taken as
coordinates on the coset manifold.
\par
In Sen's lectures the lagrangian of eq.~\ref{n4lagra} is derived
by dimensional reduction from D=10 supergravity and it is the
starting point of all duality considerations. In my lectures I
have derived it from duality--covariance and from the symplectic
embedding of the sigma--model target manifold ${\cal ST}[m,n]$.
For $m=6$ it is indeed the bosonic lagrangian of $N=4$ supergravity,
but for $m=2$ it is the bosonic action of a specific $N=2$
supergravity: that with no hypermultiplets and with ${\cal ST}[2,n]$
as special manifold in the vector multiplet sector. It is then
appropriate to turn to the general definition of special K\"ahler
manifolds.
\section{Special K\"ahler Geometry}
\label{LL4}
Let me begin by reviewing the notions of K\"ahler and Hodge--K\"ahler
manifolds that are the prerequisites to introduce the notion of
Special K\"ahler manifolds.
\subsection{K\"ahler manifolds}
Let
${\cal M}$ be a 2n-dimensional manifold with a complex
structure $J: {\cal TM} \longrightarrow {\cal TM}$, $J^2=-\bfone$.
A metric $g$ on ${\cal M}$ is  hermitian with respect to $J$ if
\begin{equation}
g( J \vec u  , J \vec w )= g (\vec u , \vec w)
\end{equation}
Given the metric $g$ and the complex structure
$J$ we introduce the following differential 2-form $K$:
\begin{equation}
K ( \vec u, \vec w)=\o{1}{2\pi} g (J \vec u , \vec w)
\end{equation}
The components $K_{\alpha \beta}$ of $K$ are given by
\be
K_{\alpha \beta} = g_{\gamma \beta} J^\gamma_{\phantom{\gamma}\alpha}
\end{equation}
and by direct computation we can easily verify that
$g$ is hermitian if and only if $K$ is anti-symmetric.
By definition a hermitian  complex manifold is a complex manifold
endowed with a hermitian metric $g$.
In a well-adapted basis:
\begin{equation}
J \o{\partial}{\partial z^{i}} = {\rm i} \o{\partial}{\partial z^{i}}
\end{equation}
we can write
$g (u,w) = g_{ij}u^i w^j + g_{{i^\star}{j^\star}}u^{i^\star}
w^{j^\star} +g_{\i{j^\star}}u^i w^{j^\star}
+g_{{i^\star} j}u^i w^{j^\star}
$.
Reality of $g(u,w)$ implies
$ g_{ij}  =  (g_{{i^\star} {j^\star}})^\star $,
$ g_{{i^\star} j}  = \left ( g_{i {j^\star}}\right )^\star$,
symmetry  yields
$ g_{ij} =g_{ji} $,
$ g_{{j^\star} i}=g_{i {j^\star}}$
while the hermiticity condition gives
$g_{ij}= g_{{i^\star} {j^\star}}=0$.
Hence in a well-adapted basis the 2-form
$K$ associated to the hermitian metric $g$ can be written as
follows:
\begin{equation}
K= \o{i}{2\pi} g_{ij^\star} dz^i \wedge d {\bar z}^{j^\star}
\label{kahlerclass}
\end{equation}
{\it
A hermitian metric on a complex manifold ${\cal M}$ is named a
K\"ahler metric if the associated 2-form $K$ is closed:}
\begin{equation}
d K=0 \label{formach}
\end{equation}
A hermitian complex manifold endowed with a K\"ahler metric is called
a {\it K\"ahler manifold} and $K$ is named its {\it K\"ahler 2--form}.
The Dolbeault cohomology class $[K] \in H^{(1,1)}$
of the K\"ahler 2--form is named the {\it K\"ahler
class} of the metric.
Equation \ref{formach}  is a differential equation
for $g_{i {j^\star}}$ whose
general solution, in any local chart, is given by the following
expression:
\begin{equation}
g_{i {j^\star}}= \partial_i \partial_{j^\star} {\cal K}
\label{popov}
\end{equation}
where ${\cal K}={\cal K}^\star = {\cal K}( z, z^*)$
is a real function of $z^i, z^{i^\star}$.
The function $\cal K$ is named the {\it K\"ahler potential}
and it is defined
only up to the real part of a holomorphic function $f(z)$. Indeed one
sees that
\begin{equation}
{\cal K}^\prime (z, z^{i^\star} )={\cal K} (z, z^{i^\star} ) +
{\rm Re}f(z)
\label{041}
\end{equation}
gives rise to the same metric $g_{i {j^\star}}$ as ${\cal K}$.
The transformation in eq.~\ref{041} is called a
{\it K\"ahler transformation}.
\par
To fix our notations we write the formulae for the Levi--Civita
connection 1--form  and Riemann curvature 2--form on a K\"ahler manifold:
\begin{eqnarray}
\Gamma^i_j &=& \Gamma^i_{kj} d z^k \nonumber\\
\Gamma^i_{kj} &=& - g^{i\ell^*}(\partial_j g_{k\ell^*})\nonumber\\
\Gamma^{i^*}_{j^*} &=& \Gamma^{i^*}_{k^*j^*} d \bar
z^{k^*}\nonumber\\
\Gamma^{i^*}_{k^*j^*}&=& -g^{i^*\ell}(\partial_{j^*} g_{k^*\ell})
\nonumber\\
{\cal R}^i_j &=& {\cal R}^i_{jk^*\ell} d \bar{z}^{k^*} \wedge d
z^\ell\nonumber \\
{\cal R}^i_{jk^*\ell} &=&\partial_{k^*} \Gamma^i_{j\ell}\nonumber\\
{\cal R}^{i^*}_{j^*}&=& {\cal R}^{i^*}_{j^*k \ell^*}d z^k \wedge d\bar
z^{\ell^*}\nonumber\\
{\cal R}^{i^*}_{j^* k \ell^*}&=& \partial_k \Gamma^{i^*}_{j^*\ell^*}
\label{curvelievi}
\end{eqnarray}
The Ricci tensor has a remarkable simple expression:
\begin{equation}
R_{m^*}^n= R_{m^* n\, i}^i= \partial_{m^*} \Gamma_{ni}^i=
\partial_{m^*} \partial_n \, \mbox{log} \,(\sqrt{g} )
\end{equation}
where $g= det | g_{\alpha \beta} |= (det |g_{i{j^\star}}|)^2$.
\subsection{Hodge--K\"ahler manifolds}
\def\mom{{M(k, \IC)}}
Consider next a {\sl line bundle}
${\cal L} {\stackrel{\pi}{\longrightarrow}} {\cal M}$ over the K\"ahler
manifold. By definition this is a holomorphic
vector bundle of rank $r=1$. For such bundles the only available
Chern class is the first:
\begin{eqnarray}
c_1 ( {\cal L} ) &=&\o{i}{2\pi}
\, {\bar \partial} \,
\left ( \, h^{-1} \, \partial \, h \, \right )\nonumber\\
&=&\o{i}{2\pi} \,
{\bar \partial} \,\partial \, \mbox{log} \,  h
\label{chernclass23}
\end{eqnarray}
where the 1-component real function $h(z,{\bar z})$ is some hermitian
fibre metric on ${\cal L}$. Let $\xi (z)$ be a holomorphic section of
the line bundle
${\cal L}$: noting that  under the action of the operator ${\bar
\partial} \,\partial \, $ the term $\mbox{log} \left ({\bar \xi}({\bar z})
\, \xi (z) \right )$ yields a vanishing contribution, we conclude that
the formula in eq.~\ref{chernclass23}  for the first Chern class can be
re-expressed as follows:
\begin{equation}
c_1 ( {\cal L} ) ~=~\o{i}{2\pi} \,
{\bar \partial} \,\partial \, \mbox{log} \,\parallel \, \xi(z) \, \parallel^2
\label{chernclass24}
\end{equation}
where $\parallel \, \xi(z) \, \parallel^2 ~=~h(z,{\bar z}) \,
{\bar \xi}({\bar z}) \,
\xi (z) $ denotes
the norm of the holomorphic section $\xi (z) $.
\par
Eq.~\ref{chernclass24} is the starting point for the definition
of Hodge K\"ahler manifolds,
an essential notion in supergravity theory.
\par
A K\"ahler manifold ${\cal M}$ is a Hodge manifold if and
only if there exists
a line
bundle ${\cal L} \, \longrightarrow \, {\cal M}$ such that its
first Chern class equals
the cohomology class of the K\"ahler 2-form K:
\begin{equation}
c_1({\cal L} )~=~\left [ \, K \, \right ]
\label{chernclass25}
\end{equation}
\par
In local terms this means that there is a holomorphic section
$W(z)$ such that we can write
\begin{eqnarray}
K&=&\o{i}{2\pi} \, g_{ij^{\star}} \, dz^{i} \, \wedge \,
d{\bar z}^{j^{\star}} \nonumber\\
&=&
\o{i}{2\pi} \, {\bar \partial} \,\partial \, \mbox{log} \,\parallel \, W(z) \,
\parallel^2
\label{chernclass26}
\end{eqnarray}
Recalling the local expression of the K\"ahler metric
in terms of the K\"ahler potential
$ g_{ij^{\star}} ~=~{\partial}_i \, {\partial}_{j^{\star}}
{\cal K} (z,{\bar z})$,
it follows from eq.~\ref{chernclass26} that if the
manifold ${\cal M}$ is a Hodge manifold,
then the exponential of the K\"ahler potential
can be interpreted as the metric
$h(z,{\bar z})~=~\exp \left ( {\cal K} (z,{\bar z})\right )$
on an appropriate line bundle ${\cal L}$.
\par
This structure is precisely that advocated by the lagrangian of
$N=1$ matter coupled supergravity:
the holomorphic section $W(z)$ of the line bundle
${\cal L}$ is what, in N=1 supergravity theory is
named the superpotential and the logarithm of
its norm  $\mbox{log} \,\parallel \, W(z) \, \parallel^2\, = \,
{\cal K} (z,{\bar z})\, + \, \mbox{log} \, | \, W(z) \, |^2  ~=~ G(z,{\bar z})$
is precisely the  invariant function in terms of which one writes the
potential and Yukawa coupling terms of the supergravity action.
\par
\subsection{Special K\"ahler Manifolds: general discussion}
\par
As I have emphasized several times there are in fact two kinds
of special K\"ahler geometry: the local and the rigid one.
The former describes the scalar field sector of vector multiplets
in $N=2$ supergravity while the latter describes the same sector
in rigid $N=2$  Yang--Mills theories. Since $N=2$
includes $N=1$ supersymmetry, local and rigid special
K\"ahler manifolds must be compatible with the geometric structures that are
respectively enforced by local and rigid $N=1$ supersymmetry in the
scalar sector. What is the distinction between the two cases in the
$N=1$ theory? It deals with the first Chern--class of the line--bundle
${\cal L} {\stackrel{\pi}{\longrightarrow}} {\cal M}$, whose sections
are the possible superpotentials.  In the local theory $c_1({\cal L})
=[K]$ and this restricts ${\cal M}$ to be a Hodge--K\"ahler manifold.
In the rigid theory, instead, we have $c_1({\cal L})=0$. At the level
of the lagrangian this reflects into a different behaviour of the
fermion fields. These latter are sections of ${\cal L}^{1/2}$ and
couple to the canonical hermitian connection defined on ${\cal L}$:
\begin{eqnarray}
{\theta}& \equiv & h^{-1} \, \partial  \, h = {\o{1}{h}}\, \partial_i h \,
dz^{i} \nonumber\\
{\bar \theta}& \equiv & h^{-1} \, {\bar \partial}  \, h = {\o{1}{h}} \,
\partial_{i^\star} h  \,
d{\bar z}^{i^\star}
\label{canconline}
\end{eqnarray}
In the local case where
\begin{equation}
\left  [ \, {\bar \partial}\,\theta \,  \right ] \, = \,
c_1({\cal L}) \, = \, [K]
\label{curvc1}
\end{equation}
the fibre metric $h$ can be identified with the exponential of the
K\"ahler potential and we obtain:
\begin{eqnarray}
{\theta}& = &  \partial  \,{\cal K} =  \partial_i {\cal K}
dz^{i} \nonumber\\
{\bar \theta}& = &   \partial  \, {\cal K} =
\partial_{i^\star} {\cal K}
d{\bar z}^{i^\star}
\label{curvconline}
\end{eqnarray}
In the rigid case,  ${\cal L}$ is instead a flat bundle and its
metric is unrelated to the K\"ahler potential. Actually one can choose
a vanishing connection:
\begin{equation}
\theta \,= \, {\bar \theta} \, = \, 0
\label{rigconline}
\end{equation}
The distinction between rigid and local special manifolds is the
$N=2$ generalization of this difference occurring at the
$N=1$ level. In the $N=2$ case, in addition to the line--bundle
${\cal L}$ we need a flat holomorphic vector bundle ${\cal SV}
\, \longrightarrow \, {\cal M}$ whose sections can be identified
with the {\it fermi--fermi} components of electric and magnetic
field--strengths. In this way, according to the discussion of
previous sections the diffeomorphisms of the scalar manifolds
will be lifted to produce an action on the gauge--field strengths
as well. In a supersymmetric theory where scalars and gauge fields
belong to the same multiplet this is a mandatory condition.
However  this symplectic bundle structure must be made
compatible with the line--bundle structure already requested
by $N=1$ supersymmetry. This leads to the existence of
two kinds of special geometry. Another essential distinction
between the two kind of geometries arises from the different
number of vector fields in the theory. In the rigid case
this number equals that of the vector multiplets so that
\begin{eqnarray}
\# \, \mbox{vector fields}\, \equiv \, {\bar n} & = & n
\nonumber\\
\# \, \mbox{vector multiplets}\equiv n & = &
\mbox{dim}_{\bf C} \, {\cal M}\nonumber\\
\mbox{rank} \, {\cal SV}   \, \equiv \, 2\bar n & = & 2 n
 \label{rigrank}
\end{eqnarray}
On the other hand,
in the local case, in addition to the vector fields arising
from the vector multiplets we have also the graviphoton coming from
the graviton multiplet. Hence we conclude:
\begin{eqnarray}
\# \, \mbox{vector fields}\, \equiv \, {\bar n} & = & n+1
\nonumber\\
\# \, \mbox{vector multiplets}\equiv n & = &
\mbox{dim}_{\bf C} \, {\cal M}\nonumber\\
\mbox{rank} \, {\cal SV}   \, \equiv \, 2\bar n & = & 2 n+2
 \label{locrank}
\end{eqnarray}
In the sequel we make extensive use of covariant derivatives with
respect to the canonical connection of the line--bundle ${\cal L}$.
Let us review its normalization. As it is well known there exists
a correspondence between line--bundles and
$U(1)$--bundles. If $\mbox{exp}[f_{\alpha\beta}(z)]$ is the transition
function between two local trivializations of the line--bundle
${\cal L} \, \longrightarrow \, {\cal M}$, the transition function
in the corresponding principal $U(1)$--bundle ${\cal U} \,
\longrightarrow {\cal M}$ is just
$\mbox{exp}[{\rm i}{\rm Im}f_{\alpha\beta}(z)]$.
At the level of connections this correspondence is formulated by
setting:
\begin{eqnarray}
\mbox{ $U(1)$--connection}   \equiv   {\cal Q} & = &  \mbox{Im}
\theta = -{\o{\rm i}{2}}   \left ( \theta - {\bar \theta}
\right) \nonumber\\
\label{qcon}
\end{eqnarray}
If we apply the above formula to the case of the $U(1)$--bundle
${\cal U} \, \longrightarrow \, {\cal M}$
associated with the line--bundle ${\cal L}$ whose first Chern class equals
the K\"ahler class, we get:
\begin{equation}
{\cal Q}  =    -{\o{\rm i}{2}} \left ( \partial_i {\cal K}
dz^{i} -
\partial_{i^\star} {\cal K}
d{\bar z}^{i^\star} \right )
\label{u1conect}
\end{equation}
 Let now
 $\Phi (z, \bar z)$ be a section of ${\cal U}^p$.  By definition its
covariant derivative is
\begin{equation}
\nabla \Phi = (d + i p {\cal Q}) \Phi
\end{equation}
or, in components,
\begin{eqnarray}
\nabla_i \Phi &=&
 (\partial_i + {1\over 2} p \partial_i {\cal K}) \Phi \label{scrivo1}\\
\nabla_{i^*}\Phi &=&(\partial_{i^*}-{1\over 2} p \partial_{i^*} {\cal K})
\Phi \label{scrivo2}
\end{eqnarray}
A covariantly holomorphic section of ${\cal U}$ is defined by the equation:
\begin{equation}
\nabla_{i^*} \Phi = 0 \label{scrivo3}
\end{equation}
We can easily map each  section $\Phi (z, \bar z)$
of ${\cal U}^p$
into a  section of the line--bundle ${\cal L}$ by setting:
\begin{equation}
\tilde{\Phi} = e^{-p {\cal K}/2} \Phi
\label{mappuccia}
\end{equation}
With this position we obtain:
\begin{eqnarray}
\nabla_i    \tilde{\Phi}&    =&   (\partial_i   +   p   \partial_i  {\cal K})
\tilde{\Phi}\\
\nabla_{i^*}\tilde{\Phi}&=& \partial_{i^*} \tilde{\Phi}\quad
\end{eqnarray}
Under the map of eq.~\ref{mappuccia} covariantly holomorphic sections
of ${\cal U}$ flow into holomorphic sections of ${\cal L}$
and viceversa.
\subsection{Special K\"ahler manifolds: the local case}
We are now ready to give the definition of local special K\"ahler
manifolds and illustrate their properties.
A first definition that does not  make direct reference to the
symplectic bundle is the following:
\bd
A Hodge K\"ahler manifold is {\bf Special K\"ahler (of the local type)}
if there exists a completely symmetric holomorphic 3-index section $W_{i
j k}$ of $(T^\star{\cal M})^3 \otimes {\cal L}^2$ (and its
antiholomorphic conjugate $W_{i^* j^* k^*}$) such that the following
identity is satisfied by the Riemann tensor of the Levi--Civita
connection:
\begin{eqnarray}
\partial_{m^*}   W_{ijk}& =& 0   \quad   \partial_m  W_{i^*  j^*  k^*}
=0 \nonumber \\
\nabla_{[m}      W_{i]jk}& =&  0
\quad \nabla_{[m}W_{i^*]j^*k^*}= 0 \nonumber \\
{\cal R}_{i^*j\ell^*k}& =&  g_{\ell^*j}g_{ki^*}
+g_{\ell^*k}g_{j i^*}   \nonumber\\
& & + e^{2 {\cal K}}
W_{i^* \ell^* s^*} W_{t k j} g^{s^*t}
\label{specialone}
\end{eqnarray}
\label{defspecial}
\ed
In the above equations $\nabla$ denotes the covariant derivative with
respect to both the Levi--Civita and the $U(1)$ holomorphic connection
of eq.~\ref{u1conect}.
In the case of $W_{ijk}$, the $U(1)$ weight is $p = 2$.
\par
The holomorphic sections $W_{ijk}$ have two different physical
interpretations in the case that the special manifold is utilized
as scalar manifold in an N=1 or N=2 theory. In the first case
they correspond to the Yukawa couplings of Fermi families
\cite{petropaolo}. In the second case they provide the coefficients
for the anomalous magnetic moments of the gauginos.
Out of the $W_{ijk}$ we can construct covariantly holomorphic
sections by setting:
\begin{equation}
C_{ijk}\,=\,W_{ijk}\,e^{  {\cal K}}  \quad ; \quad
C_{i^\star j^\star k^\star}\,=\,W_{i^\star j^\star k^\star}\,e^{  {\cal K}}
\label{specialissimo}
\end{equation}
Next we can give the second more intrinsic definition that relies
on the notion of the flat symplectic bundle.
Let ${\cal L}\, \longrightarrow \,{\cal  M}$ denote the complex
line bundle whose first Chern class equals
the K\"ahler form $K$ of an $n$-dimensional Hodge--K\"ahler
manifold ${\cal M}$. Let ${\cal SV} \, \longrightarrow \,{\cal  M}$
denote a holomorphic flat vector bundle of rank $2n+2$ with structural
group $Sp(2n+2,\IR)$. Consider   tensor bundles of the type
${\cal H}\,=\,{\cal SV} \otimes {\cal L}$.
A typical holomorphic section of such a bundle will be
denoted by ${\Omega}$ and will have the following structure:
\begin{equation}
{\Omega} \, = \, {\twovec{{X}^\Lambda}{{F}_ \Sigma} } \quad
\Lambda,\Sigma =0,1,\dots,n
\label{ololo}
\end{equation}
By definition
the transition functions between two local trivializations
$U_i \subset {\cal M}$ and $U_j \subset {\cal M}$
of the bundle ${\cal H}$ have the following form:
\begin{equation}
{\twovec{X}{ F}}_i=e^{f_{ij}} M_{ij}{\twovec{X}{F}}_j
\end{equation}
where   $f_{ij}$ are holomorphic maps $U_i \cap U_j \, \rightarrow
\,\IC $
while $M_{ij}$ is a constant $Sp(2n+2,\IR)$ matrix. For a consistent
definition of the bundle the transition functions are obviously
subject to the cocycle condition on a triple overlap:
\begin{eqnarray}
e^{f_{ij}+f_{jk}+f_{ki}} &=&1 \ \nn\\
M_{ij} M_{jk} M_{ki} &=& 1 \
\end{eqnarray}
Let $i\langle\ \vert\ \rangle$ be the compatible
hermitian metric on $\cal H$
\begin{equation}
i\langle \Omega \, \vert \, \bar \Omega \rangle= -
i \Omega^\dagger \twomat {0} {\bfone} {-\bfone}{0} \Omega
\label{compati}
\end{equation}
\bd
We say that a Hodge--K\"ahler manifold ${\cal M}$
is {\bf special K\"ahler of the local type} if there exists
a bundle ${\cal H}$ of the type described above such that
for some section $\Omega \, \in \, \Gamma({\cal H},{\cal M})$
the K\"ahler two form is given by:
\begin{equation}
K= \o{i}{2\pi}
 \partial \bar \partial \mbox{log}\left ({\rm i}\langle \Omega \,
 \vert \, \bar \Omega
\rangle \right )
\label{compati1} .
\end{equation}
\ed
{}From the point of view of local properties, eq.~\ref{compati1}
implies that we have an expression for the K\"ahler potential
in terms of the holomorphic section $\Omega$:
\begin{eqnarray}
{\cal K}& = & -\mbox{log}\left ({\rm i}\langle \Omega \,
 \vert \, \bar \Omega
\rangle \right )\nonumber\\
&=&-\mbox{log}\left [ {\rm i} \left ({\bar X}^\Lambda F_\Lambda -
{\bar F}_\Sigma X^\Sigma \right ) \right ]
\label{specpot}
\end{eqnarray}
The relation between the two definitions of special manifolds is
obtained by introducing a non--holomorphic section of the bundle
${\cal H}$ according to:
\begin{equation}
V \, = \, \twovec{L^{\Lambda}}{M_\Sigma} \, \equiv \, e^{{\cal K}/2}\Omega
\,= \, e^{{\cal K}/2} \twovec{X^{\Lambda}}{F_\Sigma}
\label{covholsec}
\end{equation}
so that eq.~\ref{specpot} becomes:
\begin{eqnarray}
1 & = &  {\rm i}\langle V  \,
 \vert \, \bar V
\rangle  \nonumber\\
&=&  {\rm i} \left ({\bar L}^\Lambda M_\Lambda -
{\bar M}_\Sigma L^\Sigma \right )
\label{specpotuno}
\end{eqnarray}
Since $V$ is related to a holomorphic section by eq.~\ref{covholsec}
it immediately follows that:
\begin{equation}
\nabla_{i^\star} V \, = \, \left ( \partial_{i^\star} - {\o{1}{2}}
\partial_{i^\star}{\cal K} \right ) \, V \, = \, 0
\label{nonsabeo}
\end{equation}
On the other hand, from eq.~\ref{specpotuno}, defining:
\begin{equation}
U_i   =   \nabla_i V  =   \left ( \partial_{i} + {\o{1}{2}}
\partial_{i}{\cal K} \right ) \, V   \equiv
\twovec{f^{\Lambda}_{i} }{h_{\Sigma\vert i}}
\label{uvector}
\end{equation}
it follows that:
\begin{equation}
\nabla_i U_j  = {\rm i} C_{ijk} \, g^{k\ell^\star} \, {\bar U}_{\ell^\star}
\label{ctensor}
\end{equation}
where $\nabla_i$ denotes the covariant derivative containing both
the Levi--Civita connection on the bundle ${\cal TM}$ and the
canonical connection $\theta$ on the line bundle ${\cal L}$.
In eq.~\ref{ctensor} the symbol $C_{ijk}$ denotes a covariantly
holomorphic (
$\nabla_{\ell^\star}C_{ijk}=0$) section of the bundle
${\cal TM}^3\otimes{\cal L}^2$ that is totally symmetric in its indices.
This tensor can be identified with the tensor of eq.~\ref{specialissimo}
appearing in eq.~\ref{specialone}. Indeed eq.~\ref{specialone} is
just the integrability condition of eq.~\ref{uvector}.
The {\it period matrix} is now introduced via the relations:
\begin{equation}
{\bar M}_\Lambda = {\bar {\cal N}}_{\Lambda\Sigma}{\bar L}^\Sigma \quad ;
\quad
h_{\Sigma\vert i} = {\bar {\cal N}}_{\Lambda\Sigma} f^\Sigma_i
\label{etamedia}
\end{equation}
which can be solved introducing the two $(n+1)\times (n+1)$ vectors
\begin{equation}
f^\Lambda_I = \twovec{f^\Lambda_i}{{\bar L}^\Lambda} \quad ; \quad
h_{\Lambda \vert I} =  \twovec{h_{\Lambda \vert i}}{{\bar M}_\Lambda}
\label{nuovivec}
\end{equation}
and setting
\begin{equation}
{\bar {\cal N}}_{\Lambda\Sigma}= h_{\Lambda \vert I} \circ \left (
f^{-1} \right )^I_{\phantom{I} \Sigma}
\label{intriscripen}
\end{equation}
As a consequence of its definition the matrix ${\cal N}$ transforms,
under diffeomorphisms of the base K\"ahler manifold exactly as it
is requested by the rule in eq.~\ref{covarianza}.
Indeed this is the very reason
why the structure of special geometry has been introduced. The
existence of the symplectic bundle ${\cal H} \, \longrightarrow \,
{\cal M}$ is required in order to be able to pull--back the action
of diffeomorphisms on the field
strengths and to construct the kinetic matrix ${\cal N}$.
\par
It is clear from our discussion that nowhere we have
assumed the base K\"ahler manifold to be a homogeneous space. So,
in general, special manifolds are not homogeneous spaces. Yet
there is a subclass of homogenous special manifolds. The homogeneous
symmetric ones were classified by Cremmer and Van Proeyen in
\cite{cremvanp} and are displayed in table~\ref{homospectable}.
\begin{table*}
\begin{center}
\caption{\sl Homogeneous Symmetric Special Manifolds}
\label{homospectable}
\begin{tabular}{|c||c||c||c|}
\hline
n  & $G/H$  & $Sp(2n+2)$  & symp rep of G
\\
\hline
{}~~&~~&~~& ~~\\
$1$ & $\o{SU(1,1)}{U(1)}$ & $Sp(4)$ &
${\underline {\bf 4}}$ \\
{}~~&~~&~~& ~~\\
\hline
{}~~&~~&~~& ~~\\
{}~~&~~&~~& ~~\\
$n$ & $\o{SU(1,n)}{SU(n)\times U(1)}$ & $Sp(2n+2)$ &
${\underline {\bf n+1}}\oplus{\underline {\bf n+1}}$ \\
{}~~&~~&~~& ~~\\
\hline
{}~~&~~&~~& ~~\\
$n+1$ & $\o{SU(1,1)}{U(1)}\otimes \o{SO(2,n)}{SO(2)\times SO(n)}$ & $Sp(2n+4)$
& $
{\underline {\bf 2}}\otimes \left ({\underline {\bf n+2}}
\oplus{\underline {\bf n+2}}\right )  $ \\
{}~~&~~&~~& ~~\\
\hline
{}~~&~~&~~& ~~\\
$6$ & $\o{Sp(6,\IR)}{SU(3)\times U(1)}$ & $Sp(14)$ & $
{\underline {\bf 14}}$ \\
{}~~&~~&~~& ~~\\
\hline
\hline
{}~~&~~&~~& ~~\\
$9$ & $\o{SU(3,3)}{S(U(3)\times U(3))}$& $Sp(20)$ & $
{\underline {\bf 20}}$ \\
{}~~&~~&~~& ~~\\
\hline
{}~~&~~&~~& ~~\\
$15$ & $\o{SO^\star(12)}{SU(6)\times U(1)}$& $Sp(32)$ & $
{\underline {\bf 32}}$ \\
{}~~&~~&~~& ~~\\
\hline
{}~~&~~&~~& ~~\\
$27$ & $\o{E_{7(-6)}}{E_6\times SO(2)}$& $Sp(56)$ & $
{\underline {\bf 56}}$ \\
{}~~&~~&~~& ~~\\
\hline
\end{tabular}
\end{center}
\end{table*}
It goes without saying that for homogeneous special manifolds the two
constructions of the period matrix, that provided by the master
formula in eq.~\ref{masterformula} and that given by eq.~\ref{intriscripen}
must agree. We shall shortly verify it in the case of the manifolds
${\cal ST}[2,n]$ that correspond to the second infinite family of
homogeneous special manifolds displayed in table
{}~\ref{homospectable}.
\par
Anyhow, since special geometry guarantees the existence of a kinetic
period matrix with the correct covariance property it is evident that
to each special manifold we can associate a duality covariant bosonic
lagrangian of the type considered in eq.~\ref{gaiazuma}. However special
geometry contains more structures than just the period matrix ${\cal
N}$ and the scalar metric $g_{ij^\star}$.
All the other items of the construction do have
a place and play an essential role in the supergravity lagrangian and
the supersymmetry transformation rules. I will briefly discuss this in
later lectures when the geometry of the hypermultiplet sector has
also been introduced.
\par
To complete the comparison between the two equivalent definitions of
special geometry let me write down the formulae that express the
K\"ahler metric and the magnetic moments/Yukawa couplings in terms
of symplectic invariants:
\begin{eqnarray}
g_{ij^\star} &=& -{\rm i} \langle \, U_{i} \, \vert \, {\bar U}_{j^\star}
\, \rangle \nonumber\\
C_{ijk} &=& \langle \, \nabla_i U_{j} \, \vert \, {  U}_{k} \,
\rangle
\label{sympinvloc}
\end{eqnarray}
Eq.s~\ref{sympinvloc} are a straightforward consequence of the very
definition of $U_i$ as it is given in eq.~\ref{uvector}, yet their
virtue is that of showing how the hermitian geometry of the
base--manifold is determined in terms of holomorphic data on the
symplectic vector bundle. Furthermore eq.s~\ref{sympinvloc} have
a natural counterpart in rigid special geometry which is my next
topic. Before turning to it, let me mention another very useful
relation that plays a role in the structure of the scalar potential
of $N=2$ supergravity, as we are going to see:
\begin{eqnarray}
U^{\Lambda\Sigma} &=& f^\Lambda_i \, f^\Sigma_{j^\star} \,
g^{ij^\star} \nonumber\\
&=& -{\o{1}{2}} \, \left ( {\rm Im}{\cal N} \right )^{-1 \vert
\Lambda\Sigma} \, -\, {\bar L}^\Lambda L^\Sigma
\label{utillima}
\end{eqnarray}
\subsection{Special K\"ahler manifolds: the rigid case}
Let ${\cal M}$ be a K\"ahler manifold  with $\mbox{dim}_{\bf C} \,
{\cal M} \, = \, n$ and  let
${\cal L}\, \longrightarrow \,{\cal  M}$ be a {\bf flat}
line bundle $c_1({\cal L})=0$\footnotemark
\footnotetext{the holomorphic sections of ${\cal L}$ would be
the possible superpotentials if ${\cal M}$ were used as scalar
manifold in an $N=1$ globally supersymmetric theory.}.
Let ${\cal SV} \, \longrightarrow \,{\cal  M}$
denote a holomorphic {\bf flat} vector bundle of rank $2n$ with structural
group $Sp(2n,\IR)$. Consider   tensor bundles of the type
${\cal H}\,=\,{\cal SV} \otimes {\cal L}$.
A typical holomorphic section of such a bundle will be
denoted by ${\Omega}$ and will have the following structure:
\begin{equation}
{\Omega} \, = \, {\twovec{{Y}^\alpha}{{F}_ \beta} } \quad
\alpha,\beta =1,\dots,n
\label{ololorig}
\end{equation}
By definition
the transition functions between two local trivializations
$U_i \subset {\cal M}$ and $U_j \subset {\cal M}$
of the bundle ${\cal H}$ have the following form:
\begin{equation}
{\twovec{Y}{ F}}_i=e^{{\hat f}_{ij}} {\hat M}_{ij}{\twovec{Y}{F}}_j
\end{equation}
where   ${\hat f}_{ij}\in \IC$ are constant complex numbers
while $M_{ij}$ is a constant $Sp(2n,\IR)$ matrix. For a consistent
definition of the bundle the transition functions are obviously
subject to the cocycle condition on a triple overlap:
\begin{eqnarray}
e^{{\hat f}_{ij}+{\hat f}_{jk}+{\hat f}_{ki}} &=& 1 \ \nonumber\\
{\hat M}_{ij} {\hat M}_{jk} {\hat M}_{ki} &=& \bfone
\end{eqnarray}
Let $i\langle\ \vert\ \rangle$ be the compatible
hermitian metric on $\cal H$
\begin{equation}
i\langle \Omega \, \vert \, \bar \Omega \rangle=
i \Omega^\dagger \twomat {0} {-\bfone} {\bfone}{0} \Omega
\label{compatirig}
\end{equation}
\bd
We say that a Hodge--K\"ahler manifold ${\cal M}$
is {\bf special K\"ahler of the rigid type} if there exists
a bundle ${\cal H}$ of the type described above such that
for some section ${\hat \Omega} \, \in \, \Gamma({\cal H},{\cal M})$
the K\"ahler two form is given by:
\begin{equation}
K= \o{i}{2\pi}
 \partial \bar \partial \left ({\rm i}\langle {\hat \Omega} \,
 \vert \, {\hat {\bar \Omega}}
\rangle \right )
\label{compati1rig} .
\end{equation}
\ed
Just as in the local case eq.~\ref{compati1rig}
yields an expression for the K\"ahler potential
in terms of the holomorphic section ${\hat \Omega}$:
\begin{eqnarray}
{\cal K}& = &  \left ({\rm i}\langle {\hat \Omega} \,
 \vert \, {\hat {\bar \Omega}}
\rangle \right )\nonumber\\
&=& \left [ {\rm i} \left ({\bar Y}^\alpha F_\alpha -
{\bar F}_\beta Y^\beta \right ) \right ]
\label{specpotrig}
\end{eqnarray}
Similarly defining
\begin{equation}
{\hat U}_i   =   \partial_i {\hat \Omega}   \equiv
\twovec{f^{\alpha}_{i} }{h_{\beta\vert i}}
\label{uvectorrig}
\end{equation}
one finds:
\begin{equation}
D_i {\hat U}_j  = {\rm i} C_{ijk} \, g^{k\ell^\star} \,
{\hat {\bar U}}_{\ell^\star}
\label{ctensorrig}
\end{equation}
where $D_i$ is the covariant derivative with respect to the
Levi--Civita connection on ${\cal TM}$ and
where $C_{ijk}$ is a totally symmetric holomorphic
section of the bundle ${\cal TM}^3\otimes{\cal L}^2$:
$\partial_{\ell^\star}C_{ijk}=0$.
Just as in the local case we obtain formulae that express
the metric and the magnetic moments in terms of the symplectic
sections:
\begin{eqnarray}
g_{ij^\star} &=& -{\rm i} \langle \,{\hat U}_{i} \, \vert \,
{\hat {\bar U}}_{j^\star}
\, \rangle \nonumber\\
C_{ijk} &=& \langle \, \partial_i {\hat U}_{j} \, \vert \, {\hat  U}_{k} \,
\rangle
\label{sympinvrig}
\end{eqnarray}
 The integrability condition of eq.~\ref{ctensorrig}  is similar
but  different from eq.~\ref{specialone} due to the replacement of
the
covariant derivative on ${\cal TM}\times{\cal L}$ by that
on ${\cal TM}$, due to the flatness of ${\cal L}$. We get
\begin{eqnarray}
\partial_{m^*}   C_{ijk}& =& 0   \quad   \partial_m  C_{i^*  j^*  k^*}
=0 \nonumber \\
\nabla_{[m}      C_{i]jk}& =&  0
\quad D_{[m}C_{i^*]j^*k^*}= 0 \nonumber \\
{\cal R}_{i^*j\ell^*k}& =&  C_{i^* \ell^* s^*} C_{t k j} g^{s^*t}
\label{specialonerig}
\end{eqnarray}
In a way similar to the local case, conditions~\ref{specialonerig}
can be taken as an alternative definition of  special geometry
of the rigid type. The definition of the {\it period matrix} is
obtained in full analogy to eq.~\ref{etamedia}:
\begin{equation}
h_{\alpha\vert i} = {\bar {\cal N}}_{\alpha\beta} f^\Sigma_i
\label{etaalta}
\end{equation}
that yields:
\begin{equation}
{\bar {\cal N}}_{\alpha\beta}= h_{\alpha \vert i} \circ \left (
f^{-1} \right )^i_{\phantom{i} \beta}
\label{intriscripenrig}
\end{equation}
\subsection{Special K\"ahler manifolds: the issue of special
coordinates}
 So far no privileged coordinate system has been chosen on the base
 K\"ahler manifold ${\cal M}$ and no mention has been made
 of the holomorphic prepotential $F(X)$ that is ubiquitous in the $N=2$
 literature and recently has been made famous, by Seiberg--Witten results,
 also to non supersymmetry experts. The simultaneous avoidance
 of privileged coordinates and of the prepotential is not
 accidental. Indeed, when the definition of special K\"ahler
 manifolds  is given in intrinsic terms, as we did in the previous
 subsection, the holomorphic prepotential $F(X)$ can be dispensed
 of. Although the structure of $N=2$ supersymmetric theories is very often
 summarized by saying that the lagrangian is completely determined in
 terms of a single holomorphic function, the very existence of this
 function is not even guaranteed by  special geometry which,
 alone, is the necessary and sufficient structure required
 to write down a supersymmetric
 lagrangian. Actually it appears that some of the physically most
 interesting cases are precisely instances where $F(X)$ does not
 exist. Let us then see how the notion of  $F(X)$
 emerges if we resort to special coordinate systems.
 \par
 Note that under a K\"ahler transformation ${\cal K} \, \to \,
 {\cal K} + \mbox{Re}f(z)$ the holomorphic section transforms,
 in the local case as $\Omega \, \to \, \Omega \, e^{-f}$, so that
 we have $X^\Lambda \, \to \, X^\Lambda \, e^{-f}$. This means that,
 at least locally, the upper half of $\Omega$ (associated with
 the electric field strengths) can be regarded as a set $X^\Lambda$
 of homogeneous coordinates on ${\cal M}$, provided that the
 jacobian matrix
 \begin{equation}
e^a_{i}(z) = \partial_{i} \left ( {\o{X^a}{X^0}}\right ) \quad ;
\quad a=1,\dots ,n
\label{nonsingcoord}
\end{equation}
is invertible. In this case, for the lower part of the symplectic
section $\Omega$ we obtain $F_\Lambda = F_\Lambda(X)$. Recalling
eq.s~\ref{specpotuno},~\ref{covholsec} and ~\ref{uvector}
from which it follows
\begin{eqnarray}
0 &=& \langle\,  V \, \vert \, U_i \, \rangle \nonumber\\
&=& X^\Lambda \, \partial_{i} F_\Lambda - \partial_{i} X^\Lambda \,
F_\Lambda
\label{ortogonalcosa}
\end{eqnarray}
we obtain:
\begin{equation}
X^\Sigma \, \partial_ \Sigma F_ \Lambda (x) \, = \, F_ \Lambda (X)
\label{nipiolbuit}
\end{equation}
so that we can conclude:
\begin{equation}
F_ \Lambda (X) \, = \, {\o{\partial}{\partial X^\Lambda} } F(X)
\label{lattedisoia}
\end{equation}
where $F(X)$ is a homogeneous function of degree 2 of the homogeneous
coordinates $X^\Lambda$. Therefore, when the condition
in eq.~\ref{nonsingcoord}
is verified we can use the {\it special coordinates}:
\begin{equation}
t^a \, \equiv \, {\o{X^a}{X^0}}
\label{speccoord}
\end{equation}
and the whole geometric structure can be derived by a single
holomorphic prepotential:
\begin{equation}
{\cal F}(t) \, \equiv \,  (X^0)^{-2} F(X)
\label{gianduia}
\end{equation}
In particular, eq.~\ref{specpot} for
the K\"ahler potential becomes
\begin{eqnarray}
{\cal K}(t, \bar t)&=&-\mbox{log} \,\,  {\rm i}\Bigl [
2 \left ( {\cal F} - \bar {\cal F} \right )\nonumber\\
&& -
\left ( \partial_a {\cal F}
+\partial_{{a^\star}} \bar {\cal F} \right )\left  ( t^a -\bar t^{a^\star}
\right ) \Bigr ]
\label{vecchiaspec}
\end{eqnarray}
while eq.~\ref{sympinvloc} for the magnetic moments simplifies into
\begin{equation}
W_{abc}= \partial_a \partial_b \partial_c {\cal F}(t)
\label{yuktre}
\end{equation}
\section{Examples of Special Manifolds of the local type}
\label{LL5}
In this section I consider explicit examples of Special
K\"ahler manifolds of the local type. In particular my
goal is to let the audience appreciate the difference
between the continuous duality groups that occur before
the gauging of the theory and the discrete duality groups
that occur in the effective quantum theory of the massless
modes after symmetry breaking.
On the concept of gauging I will make retour once I have explained
also the other geometrical item needed to construct
an $N=2$ theory, namely {\it hypergeometry}. At this level of
the discussion I just want to compare special manifolds with continuous or
discrete isometry groups from a purely geometrical viewpoint.
\par
The special manifolds with continuous dualities
are, as it is obvious, homogeneous manifolds ${\cal G}/{\cal H}$.
Hence their special geometry is completely determined by the choice
of the symplectic embedding. In particular the {\it period matrix}
${\cal N}$ can be calculated by means of the master formula
in eq.~\ref{masterformula}. As promised I want to show that such a
calculation agrees with the definition of ${\cal N}$ inside
special geometry (see eq.~\ref{etamedia}). I will do that in the
particular case of the ${\cal ST}[2,n]$ manifolds whose relevance
for string theory effective actions has already been pointed out.
\subsection{One--dimensional special manifolds}
We begin with the simplest possible choice of special K\"ahler
manifolds, those in complex dimension one.
Let then
\begin{equation}
     \mbox{dim}_{\bf C} \, {\cal SM} \, = \, 1
     \label{altobasso}
\end{equation}
 A generic complex coordinate describing such a manifold will be denoted
 by $z$ while a special coordinate will be denoted by $t$.
 In  an arbitrary symplectic basis
 the symplectic section $\Omega$ has the following form:
 \begin{equation}
\Omega \, = \, \left ( \matrix { X^0 \cr X^1 \cr F_0 \cr F_1 \cr
}\right )
\label{arbitrio}
\end{equation}
 In a symplectic basis where the holomorphic prepotential $F(X)$
 exists, setting, as in eq.~\ref{gianduia}:
\begin{equation}
{\cal F}(t) \, \equiv \, (X^0)^{-2} \, F(X) \, \quad \, t \, \equiv
\, {\o{X^1}{X^0}}
\label{brighella}
\end{equation}
we obtain :
\par
\begin{equation}
\o{1}{X^0} \, \Omega \,{\stackrel{\rm def}{=}}\, {\tilde \Omega }(t)
\, = \,
\left ( \matrix {
1 \cr t \cr 2{\cal F}(t) \, - \, t \, {\cal F}^{\prime} (t)\cr
{\cal F}^{\prime} (t)\cr }\right )
\end{equation}
I will discuss three examples of one--dimensional special
manifolds characterized by three different choices of the
prepotential ${\cal F}(t)$. The first two cases are homogeneous
manifolds and hence have a continuous duality group. The third
case has only a discrete duality group. It is the moduli space
of complex structures of the mirror quintic three--fold.
Let us begin with the homogeneous manifolds. If
we look at the classification of homogeneous symmetric
special manifolds displayed in Table~\ref{homospectable}
we see that there are just two one--dimensional cases.
They correspond to two different symplectic embeddings:
\begin{eqnarray}
a)&\o{SU(1,1)}{U(1)}\,\sim\, \o{SL(2,\IR)}{SO(2)}& \quad
   {\cal F}_2(t)~=~{\rm i}\, ( 1 - \o{1}{2} t^2) \nonumber\\
b)&\o{SU(1,1)}{U(1)}\,\sim\, \o{SL(2,\IR)}{SO(2)}& \quad
   {\cal F}_3(t)~=~{\rm i}\,\o{1}{3!}\, d_0 \, t^3 \nonumber\\
\label{pasodoble}
\end{eqnarray}
{}From the {\it metric viewpoint} the two manifolds are the same,
but the {\it Yukawa couplings}/{\it anomalous magnetic moments}
are different. In the first case one has $W_{ttt}=0$, while
in the second case one obtains $W_{ttt} = d_0 = \mbox{const}$.
Hence as special manifolds they are different. To prove what
we have just stated I utilize the following strategy.
Recalling eq.~\ref{su11coset} that provides a coset parametrization
for the manifold $SU(1,1)/U(1)$, calculating the left--invariant
one--form
$M^{-1}(S) \, dM(S)$ and extracting the
vielbein component, we easily obtain the form
of the invariant metric on such a coset:
\begin{equation}
ds^2 \, = \,- \, \mbox{ p } \,
{\o{ 1}{({\bar S} - S)^2}}\, d{\bar S}\,
\otimes \, dS
\label{invsumetric}
\end{equation}
As far as the isometry group is concerned the real positive
constant $p \, > \, 0$ appearing
in eq.~\ref{invsumetric} is arbitrary since it just corresponds
to an overall rescaling of the vielbein with a factor $\sqrt{p}$.
For any choice of this
number $p \, \in \, \IR_{+}$
the corresponding metric manifold is a copy of the homogeneous
space $SU(1,1)/U(1)$.
\par
a) Upon the field identification
\begin{equation}
t={\o{S +{\rm i}}{S-{\rm i}}}
\end{equation}
 the metric of eq.~\ref{invsumetric}
can be obtained from the K\"ahler potential:
\begin{equation}
{\cal K}(t) \, =\, - \mbox{log} \left [ 1 - t{\bar t}\right ]
\label{qformpot}
\end{equation}
which  follows from the choice a) of the holomorphic superpotential
in eq.~\ref{pasodoble}:
\begin{eqnarray}
{\cal F}(t)&=&={\cal F}_2(t)\, =  \,
\,( X^0 )^{-2} F_2(X)\nonumber\\
F_2(X)&=&{\rm i}(X_0^2  -   X_1^2)
\label{quadrica}
\end{eqnarray}
via the general formula in eq.~\ref{vecchiaspec}
\par
b) Alternatively, identifying
\begin{equation}
S=    t
\end{equation}
the metric~\ref{invsumetric} can be retrieved from the following
K\"ahler potential:
\begin{equation}
{\cal K}(t)= - \, p \, \mbox{ log } \left (t -{\bar t} \right )
\label{pformkapot}
\end{equation}
For $p=3$ eq.~\ref{pformkapot}
is obtained from the general
eq.~\ref{vecchiaspec} if the prepotential ${\cal F}(t)$ is
chosen according to option b) of eq.~\ref{pasodoble}:
\begin{eqnarray}
{\cal F}(t)&=&{\cal F}_3(t)  \, =
\,( X^0 )^{-2} F_3(X)
\nonumber\\
F_3(X)& \equiv &\o{1}{3!}\, d_0 \,\o{ X_1^3}{X0}
\label{cubica}
\end{eqnarray}
{}From eq.s~\ref{cubica} and ~\ref{quadrica}, utilizing the general
formula in eq.~\ref{yuktre} one easily verifies that in   case b)
the anomalous magnetic moment $W_{ttt}$ is indeed equal
to the non--zero constant $d_0$, while in the   case a) it
vanishes.
\par
Group--theoretically  case a) and case b) correspond
to two different symplectic embeddings:
\begin{equation}
M \, : \, SL(2,\IR) \, \longrightarrow \, Sp(4,\IR)\nonumber\\
\end{equation}
We have:
\begin{eqnarray}
a)~~  {\underline {\bf 4}} \, &\stackrel{SL(2,\IR)}{\longrightarrow}&\,
{\underline {\bf 2}}\oplus{\underline {\bf 2}}\nonumber\\
b)~~  {\underline {\bf 4}} \, &\stackrel{SL(2,\IR)}{\longrightarrow}&\,
{\underline {\bf 4}}\,\sim\, \mbox{three-times symm.}
\end{eqnarray}
Explicitly the two embedding maps are displayed in Table~\ref{symbedtable}:
\begin{table*}
\begin{center}
\caption{\sl Symplectic embeddings of $SL(2,\IR)$ in $Sp(4,\IR)$}
\label{symbedtable}
\begin{tabular}{ccc}
\hline
\hline
$\null$ & $\null$ & $\null$ \\
$\forall \, g $&=& $\, \left  ( \matrix { a & b\cr c & d\cr }\right )
\, \in \, SL(2, \IR) \quad\quad M_2(g), M_3(g) \in  \, Sp(4,\IR )$ \\
$\null$ & $\null$ & $\null$ \\
$M_2(g)$&=&$\left ( \matrix{ a & b & 0 & 0\cr
c & d & 0 & 0\cr 0 & 0 & a & b\cr 0 & 0 & c & d \cr}\right )$ \\
$\null$ & $\null$ & $\null$ \\
$M_3(g)$&=&$
\left ( \matrix{ {a^3} & 3\,{a^2}\,b & {{-6\,{b^3}}\over {d_0}} &
  {{6\,a\,{b^2}}\over {d_0}} \cr {a^2}\,c &
  2\,a\,b\,c + {a^2}\,d & {{-6\,{b^2}\,d}\over {d_0}} &
  {{2\,{b^2}\,c + 4\,a\,b\,d}\over {d_0}} \cr
  {{-\left( {c^3}\,d_0 \right) }\over 6} &
  {{-\left( {c^2}\,d\,d_0 \right) }\over 2} & {d^3} &
  -\left( c\,{d^2} \right)  \cr {{a\,{c^2}\,d_0}\over 2} &
  \left( {{b\,{c^2}}\over 2} + a\,c\,d \right) \,d_0 &
  -3\,b\,{d^2} & 2\,b\,c\,d + a\,{d^2} \cr  } \right ) $\\
  $\null$ & $\null$ & $\null$ \\
\hline
\hline
\end{tabular}
\end{center}
\end{table*}
The proof that these two embeddings correspond to the two
choices in eq.~\ref{quadrica} and eq.~\ref{cubica} of the superpotential
is done by checking that, for some overall rescaling
 $\exp[\varphi_i(g,t)]$
the following transformation rule is true:
\begin{equation}
M_i(g) \, {\tilde \Omega}_i \left ( t \right ) ~=~\exp [\varphi_i(g,t)] \,
{\tilde \Omega}_i \left ( \o{ d \, t + c}{ b \, t + a}\right ) \,
\end{equation}
\begin{equation}
\forall g\, =\left  ( \matrix { a & b\cr c & d\cr }\right )
\, \in \, SL(2, \IR)  \quad i=\cases{2\cr 3\cr}
\end{equation}
When one deals with string compactifications on Calabi--Yau
threefolds $M^{CY}_3$, the case b) of one--dimensional
special manifolds
emerges as the moduli space of {\it K\"ahler class deformations}
\footnotemark \footnotetext{For an exhaustive review see, for
instance  the book by the present author ~\cite{petropaolo} }
whenever we have:
\begin{equation}
h^{(1,1)} \, \equiv \, {\rm dim}_{\bf C} \, H^{(1,1)}
\left (M^{CY}_3 \right ) \, = \, 1
\end{equation}
 Yet the choice ${\cal F}_3(t)$ is correct only in
the {\bf large radius limit} $\Im m\, t \, \longrightarrow \, \infty$.
In this limit $d_0 \, \sim \, W_{ttt}$, is geometrically interpreted
as the {\it intersection number} of three
2--cycles, (Poincar\'e duals to the $(1,1)$--form $\omega^{(1,1)}$):
\begin{equation}
d_0~=~W_{ttt}^{(0)}~=~\int_{{M}_3^{CY}} \, \omega^{(1,1)} \,
\wedge \, \omega^{(1,1)} \,
\wedge \, \omega^{(1,1)}
\label{lucy}
\end{equation}
On the other hand
  at the quantum level  we have infinitely many corrections due
to world--sheet instantons \cite{candelinas}. Eq.~\ref{lucy} is replaced by:
\begin{equation}
W_{ttt}~=~W_{ttt}^{(0)}\, + \, \sum_{k=1}^{\infty}
\o{n_k \, k^3 \, q^k}{1\, - \, q} \quad q=\exp[2i\pi t]
\label{treporcellini}
\end{equation}
where
 $n_k$ = {\it number of rational curves} of degree $k$ embedded in
$M_3^{CY}$ \cite{candelinas}.
This result is obtained from topological field--theories
and for a review of the derivation I refer the audience
to \cite{petropaolo}.
For the    quintic threefold, the determination of the
special manifold, with the   values
of $n_k$ appearing in the Yukawa coupling/anomalous magnetic
moment   explicitly evaluated, has been obtained by means
of Picard--Fuchs equations on the  mirror manifold \cite{candelinas}.
A review of this procedure can also be found in \cite{petropaolo}.
Here we are not interested in these details. What we want to
discuss is the form of the outcoming special geometry.
We have:
\begin{eqnarray}
{\cal F}(t)&=& {\cal F}_\infty (t) \, + \, \Delta {\cal F}(t)
\nonumber\\
{\cal F}_\infty(t) &=& \o{1}{3!} \,5 \, t^3 \, + \,
\, \o{11}{4} \, t^2 \,\nonumber\\
&& - \, \o{25}{12} \, t \, + \, i \,
\o{25 \, \zeta (3)}{\pi^3} \nonumber\\
\Delta {\cal F}(t)&=& - \, i \, \sum_{N=1}^{\infty} \,
\o{d_N}{(2\pi N)^3} \, e^{2\pi {\rm i} N t}
\end{eqnarray}
the coefficients $d_N$ being related to $n_k$ by:
\begin{eqnarray}
\sum_{N=0}^{\infty} \, d_N \, q^N&=&5\, + \, \sum_{k=1}^{\infty}
\o{n_k \, k^3 \, q^k}{1\, - \, q}\nonumber\\
&=& 5 \,  + \,  n_1 \,  q  \,  +  \,  (8 n_2 +n_1 ) \,  q^2 \,
\nonumber\\
& & +  \dots
\end{eqnarray}
The quantum symplectic section
\begin{equation}
{\tilde \Omega}_{quantum} \left ( t \right) ~=~
\left ( \matrix{
1 \cr    t \cr    2{\cal F}(t) \, - \, t \,
{\cal F}^{\prime} (t)\cr
{\cal F}^{\prime} (t)\cr } \right )
\end{equation}
has {  no longer} a {\it  continuous group} of duality rotations,
but it rather admits a {\it discrete, infinite,  group} of such rotations:
$\Gamma_{duality} \, \subset \, Sp(4,\ZZ)$
generated by the following two {\it  integer--valued} symplectic
matrices:
\begin{eqnarray}
{\cal A}_{q}&=&\left ( \matrix{ 1 & 0 & -1 & 0 \cr -1 & 1 & 1 & 0 \cr 5 & -8 &
-4 & 1 \cr -3
   & -5 & 3 & 1 \cr  } \right ) \nonumber \\
{\cal T}_{q} & = &
\left ( \matrix
{ 1      &  0      & 1       &  0      \cr
 0       &1        &0        &0        \cr
0        & 0       & 1       & 0       \cr
0        &0        &0        &1        \cr } \right )
\end{eqnarray}
The interpretation of these generators of the duality group $\Gamma$
is in terms of  symmetries of the polynomial constraint
${\cal W}(X)=0$ defining the Calabi--Yau threefold  as a vanishing
locus in $\IC\IP_4$
\begin{equation}
W(X,\psi)~=~\o{1}{5}\,\sum_{i = 1}^{5} \, X_i^5 \, - \,\psi\,
\prod_{i=1}^{5} \, X_i
\label{polcostretto}
\end{equation}
and monodromies around the singular points
of the Picard--Fuchs differential system. We have:
\begin{eqnarray}
{\cal A}_{q}^5 &=& \bfone \nonumber\\
  \ZZ_5 & \sim &\Gamma_W ~\mbox{symm of
def. polyn.}\nonumber\\
{\cal T}&=&{\cal T}_0 ~~ \nonumber\\
\mbox{ monodromy }  & \mbox{at}&
{\psi}=\exp[2i\pi]=1 \nonumber\\
{\cal T}_k&=&{\cal A}^{-k}{\cal T}{\cal A}^{k}\nonumber\\
\mbox{ monodromy}
&\mbox{at}&
{\psi}=\exp[2i\pi k/5] \nonumber\\
{\cal T}_\infty&=&\left ({\cal A}{\cal T}\right)^5 \nonumber\\
 \mbox{monodromy} & \mbox{at}&
{\psi}=\infty\nonumber\\
\bfone &=&{\cal T}_0 \, {\cal T}_1 \, {\cal T}_2 \, {\cal T}_3 \,
{\cal T}_4 \, {\cal T}_{\infty}\nonumber\\
\Gamma_W&=&\o{\Gamma_{duality}}{\Gamma_{monodromy}}
\end{eqnarray}
Monodromies, in particular
refer to the singularities of the Picard--Fuchs
equation in the non--special coordinate
$\psi$ that appears in eq.~\ref{polcostretto}.
The Picard--Fuchs equations precisely determine the relation
between the special
coordinate $t$ and the non--special coordinate $\psi$. Among the
group elements of the duality group one determines the translation
\begin{equation}
t \, \longrightarrow \, t + 1
\end{equation}
of special coordinate. It is
\begin{equation}
{\cal AT}^{-1}_{q}~=~\left (
\matrix{ 1 & 0 & 0 & 0 \cr 1 & 1 & 0 & 0 \cr -5 & 3 & 1 & -1 \cr 8 &
  5 & 0 & 1 \cr  }\right )
\end{equation}
 For comparison in the classical coset case the $t \longrightarrow t+1$
translation is effected by
\begin{equation}
{\cal AT}_{class}~=~\left (\matrix{ 1 & 0 & 0 & 0
\cr 1 & 1 & 0 & 0 \cr -{5\over 6} &
  -{5\over 2} & 1 & -1 \cr {5\over 2} & 5 & 0 & 1 \cr  } \right )
\end{equation}
which is  symplectic but not integer!
Similarly the matrix that performs a $\ZZ_5$ rotation in the classical
  coset manifold case is simply  the $Sp(4,\IR)$--image of the $O(2)$
  rotation of an angle $\theta = 2\pi/5$ contained in $SL(2,\IR)$.
  This is displayed in Table~\ref{z5table} and it is far
  from being integer valued.
\begin{table*}
\begin{center}
\caption{\sl The $Z_5$ generator in the $M_3$ embedding of $SL(2,\IR)$ }
\label{z5table}
\begin{tabular}{ccc}
\hline
\hline
$\null$ & $\null$ & $\null$ \\
$\left (\matrix {Cos [2\pi/5]&Sin [2\pi/5] \cr -Sin[2\pi/5]
  & Cos [2\pi/5]\cr } \right ) $& $\hookrightarrow$ & $
  \left (\matrix{ 0.0295085 & 0.272453 & -1.03229 & 0.33541 \cr -0.0908178 &
  -0.529508 & -0.33541 & -0.271441 \cr 0.716866 & -0.698771 &
  0.0295085 & 0.0908178 \cr 0.698771 & 1.69651 & -0.272453 &
  -0.529508 \cr  }\right ) $ \\
$\null$ & $\null$ & $\null$ \\
\hline
\hline
\end{tabular}
\end{center}
\end{table*}
What is the lesson learned from this example? The operations in
the duality group have a geometrical origin in the cohomology
of the target manifold $M^{CY}_3$ and are integer valued
because they must map {\it integer homology cycles} into
{\it integer homology cycles}. Yet if the Calabi--Yau threefold
is utilized as a compactification space for a type II superstring
all the operations of the duality group act as electric--magnetic
duality in the sense discussed in previous lectures. Hence the
integer--valuedness of the transformations can be alternatively
traced back to the charge--lattice and to Dirac quantization
condition in eq.~\ref{dirquant}. In the classical limit the
theory has continuous duality symmetries that can be understood in
terms of symplectic embeddings of Lie groups. The quantum theory
has a discrete group of duality symmetries that, abstractly, is
isomorphic to a discrete subgroup of the continuous one appearing
in the classical case. Yet what changes from the classical to the
quantum case is the symplectic embedding of this subgroup. This
change of embedding is where all the non--perturbative effects of
strong coupling physics reside.
\par
I shall illustrate these ideas in my last lecture, presenting
in minute details the case of Seiberg Witten solution of the
N=2 SU(2) theory in terms of an auxiliary dynamical Riemann
surface. Such a surface plays for rigid special geometry
the role played
by the Calabi--Yau threefolds in the case of local special geometry.
\par
Let me now turn to another multidimensional example
of special K\"ahler manifold of the local type.
\subsection{The ${\cal ST}[2,n]$ special manifolds and the Calabi
Visentini coordinates}
When I studied the symplectic embeddings of the ${\cal ST}[m,n]$
manifolds,
defined by eq.~\ref{stmanif}, a study that lead me to
the general formula in eq.~\ref{maestrino}, I remarked that the
subclass ${\cal ST}[2,n]$ constitutes a family of special K\"ahler
manifolds: actually a quite relevant one. Here I survey the
special geometry of this class.
\par
Besides their applications in the large radius limit of superstring
compactifications, the ${\cal ST}[2,n]$ manifolds are
interesting under another respect. They provide an example
where the holomorphic prepotential  can be non--existing.
\par
Consider a standard parametrization of the $SO(2,n)/SO(2)\times
SO(n)$ manifold, like for instance that in eq.~\ref{somncoset}.
In the $m=2$ case we can introduce a canonical complex structure
on the manifold by setting:
\begin{eqnarray}
\Phi^\Lambda (X)& \equiv &{\o{1}{\sqrt{2}}} \, \left (
L^\Lambda_{\phantom{\Lambda}0 } + {\rm i} \,
L^\Lambda_{\phantom{\Lambda}0 } \right )\nonumber\\
&& \Lambda=0,1,\alpha \quad \alpha=2,\dots,n+1
\label{cicciophi}
\end{eqnarray}
The relations satisfied by the upper two rows of the coset
representative (consequence of $L(X)$ being pseudo--orthogonal with
respect to metric $\eta_{\Lambda\Sigma}={\rm diag}(+,+,-,\dots,-)$):
\begin{eqnarray}
L^\Lambda_{\phantom{\Lambda}0 } \, L^\Sigma_{\phantom{\Lambda}0 } \,
\eta_{\Lambda\Sigma} &=& 1  \nonumber\\
L^\Lambda_{\phantom{\Lambda}0 } \, L^\Sigma_{\phantom{\Lambda}1 } \,
\eta_{\Lambda\Sigma} &=& 0  \nonumber\\
L^\Lambda_{\phantom{\Lambda}1 } \, L^\Sigma_{\phantom{\Lambda}1 } \,
\eta_{\Lambda\Sigma} &=& 1
\label{pseudodionigi}
\end{eqnarray}
can be summarized into the complex equations:
\begin{eqnarray}
{\bar \Phi}^\Lambda  \,
\Phi^\Sigma
\eta_{\Lambda\Sigma} &=& 1  \nonumber\\
{\Phi}^\Lambda  \,
\Phi^\Sigma
\eta_{\Lambda\Sigma} &=& 0
 \label{pseudomichele}
\end{eqnarray}
Eq.s ~\ref{pseudomichele} are solved by posing:
\begin{equation}
\Phi^\Lambda \, = \,{\o{X^\Lambda}{\sqrt{{\bar X}^\Lambda \,
X^\Sigma \, \eta_{\Lambda\Sigma}}}}
\label{fungomarcio}
\end{equation}
where $X^\Lambda$ denotes any set of complex parameters, determined
up to an overall multiplicative constant and satisfying the
constraint:
\begin{equation}
 X^\Lambda  \,
X^\Sigma
\eta_{\Lambda\Sigma} \, = \, 0
\label{ciliegia}
\end{equation}
In this way we have proved the identification, as differentiable
manifolds, of the coset space $SO(2,n)/SO(2)\times
SO(n)$ with the vanishing locus of the quadric in
eq.~\ref{ciliegia}. Taking
any holomorphic solution of eq.~\ref{ciliegia}, for instance:
\begin{equation}
 X^{\Lambda}(y)  \, \equiv \,
\left(\begin{array}{c} 1/2\hskip 2pt (1 + y^2) \\
{\rm i}/2\hskip 2pt (1 -
y^2)\\ y^{\alpha}\end{array}\right)
\label{calabivise}
\end{equation}
where $y^\alpha$ is a set of $n$ independent complex coordinates,
inserting it into eq.~\ref{fungomarcio} and comparing with
eq.~\ref{cicciophi} we obtain the relation between whatever
coordinates we had previously used to write the coset representative
$L(X)$ and the complex coordinates $y^\alpha$. In other words
we can regard the matrix $L$ as a function of the $y^\alpha$ that
are named the Calabi Visentini coordinates.
\par
Consider in addition the {\it axion--dilaton} field  $S$
that parametrizes the $SU(1,1)/U(1)$ coset according with
eq.~\ref{su11coset}. The special geometry of the manifold
${\cal ST}[2,n]$ is completely specified by writing the
holomorphic symplectic section $\Omega$ as follows
(\cite{CDFVP}):
\par
\begin{equation}
\label{so2nssec}
\Omega (y,S) \, = \, \left (\matrix{
X^{\Lambda}  \cr F_{\Lambda}\cr }\right)
\, = \, \left ( \matrix { X^{\Lambda}(y) \cr
{\cal  S }\, \eta_{\Lambda\Sigma}
X^{\Sigma}(y) \cr } \right  )
\end{equation}
Notice that  with the above choice, it is
not possible to describe $F_\Lambda$ as derivatives of any
prepotential. Yet everything else can be calculated utilizing
the formulae I presented in previous lectures.
The K\"ahler potential is:
\begin{eqnarray} \label{so2nkpotskn}
\cK &=&\cK_1(S)+\cK_2(y)\nonumber \\
&=& -\mbox{log} {\rm i} (\bar S - S) -\log X^T\eta X
\end{eqnarray}
The K\"ahler metric  is  block diagonal:
\begin{equation}
\label{so2n5}
g_{ij^\star} \, = \,
\left(\begin{array}{cc}g_{S\bar S} & {\bf
0}\\ {\bf 0} & g_{\alpha \bar\beta}\end{array}\right)
\end{equation}
\begin{equation}
\left\{\begin{array}{l} g_{S\bar S} = \partial_S \partial_{\bar S} \cK_1 =
{-1\over (\bar S - S)^2}\\ g_{\alpha\bar\beta}(y)=
\partial_{\alpha}\partial_{\bar\beta} \cK_2 \end{array}\right.
\end{equation}
as expected.
The anomalous magnetic moments-Yukawa
couplings $C_{ijk}$ ($i=S, \alpha$) have a very simple expression
in the chosen coordinates:
\begin{equation}
\label{so2n6bis} C_{S\alpha\beta} = -{\rm exp}[{\cal K}] \,
\delta_{\alpha\beta},
\end{equation}
all the other components being
zero.
\par
Using the definition of the {\it period matrix} given
in eq.~\ref{intriscripen} we obtain
\begin{equation} \label{so2n10} \cN_{\Lambda\Sigma} = (S -
\bar S) {X_{\Lambda} {\bar X}_{\Sigma} + {\bar X}_{\Lambda} X_{\Sigma}
\over X^T\eta X} + \bar S\eta_{\Lambda\Sigma}.
\label{discepolo}
\end{equation}
In order to see that eq.~\ref{discepolo} just coincides
with eq.~\ref{maestrino} it suffices to note that as a
consequence of its definition~\ref{cicciophi} and of the
pseudo--orthogonality of the coset representative $L(X)$,
the vector $\Phi^\Lambda$ satisfies the following
identity:
\begin{eqnarray}
&\Phi^\Lambda \, {\bar \Phi}^\Sigma + \Phi^\Sigma \, {\bar
\Phi}^\Lambda   = & \nonumber\\
& {\o{1}{2}} \,  L^{\Lambda}_{\phantom{\Lambda}\Gamma} \,
L^{\Sigma}_{\phantom{\Sigma}\Delta} \left ( \delta^{\Gamma\Delta} +
\eta^{\Gamma\Delta} \, \right )   &
\label{meraviglia}
\end{eqnarray}
Inserting eq.~\ref{meraviglia} into eq.~\ref{discepolo},
formula ~\ref{maestrino} is retrieved.
\par
This completes the proof that the choice~\ref{so2nssec} of
the special geometry holomorphic section corresponds to the
symplectic embedding~\ref{ortoletto} and ~\ref{ortolettodue}
of the coset manifold ${\cal ST}[2,n]$. In this symplectic
gauge the symplectic transformations of the isometry group
are the simplest possible ones and,  anticipating
the language of later lectures, the entire group $SO(2,n)$
is represented by means of {\it classical} transformations
that do not mix electric fields with magnetic fields.
The disadvantage of this basis, if any,
is that there is no holomorphic prepotential. To find
an $F(X)$ it suffices to make a symplectic rotation to
a different basis.
\par
 If we set:
\begin{eqnarray}
X^1 = {{1}\over{2}} (1 + y^2) &=& -{{1}\over{2}} (1 - \eta_{ij} t^{i}
 t^{j})\nonumber\\
X^2 = i {{1}\over{2}} (1 - y^2) &=& t^2\nonumber\\
X^\alpha = y^\alpha &=& t^{2+\alpha} \quad
{\alpha=1,\dots,n-1}\nonumber\\
X^{\alpha=n} = y^{n} &=& {{1}\over{2}} (1+\eta_{ij} t^{i} t^{j})
\end{eqnarray}
where
\begin{equation}
\eta_{ij} = {\rm diag} \left ( + , -, \dots , -\right ) ~
i,j=2,\dots , n +1
\end{equation}
Then we can show that $\exists \, {\cal C}\, \in Sp(2 n+2,\IR)$ such that:
\begin{eqnarray}
&{\cal C} \,
\left (
       \matrix {
                 X^\Lambda \cr
                 S\eta_{\Lambda\Sigma} \, X^\Lambda \cr
                }
\right ) ~=~& \nonumber\\
& \exp[\varphi(t)]\,
\left (
\matrix{
         1 \cr
         S \cr
           t^{i} \cr
          2 \, {\cal F} - t^{i}
         {{\partial}\over{\partial t^{i}}}{\cal F}
          - S{{\partial}\over S}{\cal F} \cr
           S{{\partial}\over S}{\cal F}\cr
       {{\partial}\over{\partial t^{i}}}{\cal F}\cr
      }
\right )&
\end{eqnarray}
with
\begin{eqnarray}
{\cal F}(S,t) &=&{{1}\over{2}} \,
S\,  \eta_{ij} t^{i} t^{j}
= {{1}\over{2}} \, d_{IJK} t^{I} t^{J} t^{K}
\nonumber\\
t^{1} &=&S\nonumber\\
d_{IJK}&=&\cases{
d_{1jk}=\eta_{ij}\cr
0 ~~\mbox{otherwise}\cr}
\label{vecchiume}
\end{eqnarray}
and
\begin{equation}
W_{IJK} = d_{IJK} = {{\partial^3{\cal F}(S, t^{i})}
\over{\partial t^{I}\partial
t^{J} \partial t^{K} } }
\end{equation}
This means that in the new basis
the symplectic holomorphic section ${\cal C}\Omega$ can be derived
from the following cubic prepotential:
\begin{equation}
F(X) \, =\,{\o{1}{3!}}\, {\o{ d_{IJK} \, X^I \, X^J \, X^K}{X_0}}
\label{duplocubetto}
\end{equation}
For instance in the case $n=1$  the matrix which does such a job
is:
\begin{equation}
{\cal C}=\left (\matrix{ 1 & 0 & -1 & 0 & 0 & 0 \cr 0 & 0 & 0 & 1
& 0 & 1 \cr 0 & -1
   & 0 & 0 & 0 & 0 \cr 0 & 0 & 0 & {1\over 2} & 0 & -{1\over 2} \cr
  -{1\over 2} & 0 & -{1\over 2} & 0 & 0 & 0 \cr 0 & 0 & 0 & 0 & -1
   & 0 \cr  }\right )
\end{equation}
\subsection{Comments on the ${\cal ST}[2,2]$ case: S--duality and
R--symmetry}
To conclude let us focus  on the case ${\cal ST}[2,2]$.
This manifold has two coordinates that we can either call $S$ and
$t$, in the parametrization of eq.~\ref{vecchiume} or  $S$
and $y$ in the Calabi Visentini basis. The relation between $t$ and
$y$ simplifies enormously in this case:
 \begin{equation}
t ~=~i {{y+1}\over{y-1}}
\end{equation}
It is then a matter of choice to regard the holomorphic section
in whatever basis as a function of $y$ or of $t$, in addition to
$S$. Independently from this choice the manifold
${\cal ST}[2,2]$ emerges as {\it moduli space (at tree--level)}
in a locally N=2
supersymmetric gauge theory of a rank one gauge group, namely
$SU(2)$. The two fields spanning the manifold have very different
interpretations. The field $y$ is the scalar partner of the
gauge field that remains massless after Higgs mechanism. Its vacuum
expectation value is the modulus of the gauge theory. It is the
same field that occurs also in a globally supersymmetric theory
and which I shall amply discuss in my last lecture.
On the other hand the field $S$ is the dilaton--axion.
It plays the role of generalized coupling constant and generalized
$theta$--angle.
There are two $SL(2,\IR)$ groups embedded in $SP(6,\IR)$, they
act as standard fractional linear transformations on the
{\it dilaton--axion} $S$ and on the special coordinate $t$
for the gauge modulus.
Using the Calabi--Visentini section of eq.~\ref{so2nssec} and the
embedding eq.s~\ref{ortoletto} and ~\ref{ortolettodue},
we have that
\par
{\bf S--duality} $ S \longrightarrow \, - 1/S$ is generated
by the symplectic matrix:
\begin{equation}
S_{duality}~=~\left ( \matrix{ 0 & 0 & 0 & 1 & 0 & 0 \cr 0 & 0 & 0 & 0 & 1 & 0
\cr 0 & 0 &
  0 & 0 & 0 & -1 \cr -1 & 0 & 0 & 0 & 0 & 0 \cr 0 & -1 & 0 & 0 & 0 &
  0 \cr 0 & 0 & 1 & 0 & 0 & 0 \cr  }\right )
\label{sduality}
\end{equation}
while {\bf T--duality} $ t \longrightarrow \, - 1/t$ is generated
by the symplectic matrix:
\begin{equation}
R_{symmetry}~=~\left (\matrix{ -1 & 0 & 0 & 0 & 0 & 0 \cr 0 & -1 & 0 & 0 & 0 &
0 \cr 0 & 0
   & 1 & 0 & 0 & 0 \cr 0 & 0 & 0 & -1 & 0 & 0 \cr 0 & 0 & 0 & 0 & -1
   & 0 \cr 0 & 0 & 0 & 0 & 0 & 1 \cr  } \right )
   \label{tduality}
\end{equation}
If we think of the $t$--field as the {\it modulus} of some compact
internal manifold then T--duality is just the transformation from
small to large compactification radius. Looking at the same
transformation in terms of the $y$ variable its meaning
becomes more clear. It is
$R$--symmetry $ y \longrightarrow - y$, an exact global
symmetry of the {\it microscopic} lagrangian. The fact that
the matrix generating T--duality or R--symmetry is block--diagonal
agrees with the fact that this is a perturbative symmetry, holding
at each order in perturbation theory and never exchanging electric
with magnetic states. Very different is the nature of
S--duality. Since it inverts the coupling constant it is by definition
non--perturbative. It exchanges strong and weak coupling regimes and
because of that it is supposed to exchange elementary states
with soliton states. For this reason it must mix electric with magnetic
field strengths and it is off--diagonal. These symmetries exist in the
microscopic theory which is derived by {\it gauging} (see later lectures)
the abelian theories possessing continuous duality symmetries (in
this case the two $SL(2,\IR)$ groups). After gauging the continuous
duality symmetries will be broken. The question is will the integer
valued symplectic generators of S--duality and R--symmetry survive
given that they respect the Dirac quantization condition? The answer
is yes, but in the effective quantum theory they will be
represented by new integer valued elements of $Sp(6,\ZZ)$ not
derivable from the classical embedding. This is the same phenomenon
already observed in the Calabi--Yau case with one modulus. The $\ZZ_5$
generator is not the image through the classical embedding map of
the $\ZZ_5$ subgroup in the classical $O(2)$. Since the special
geometry in the effective theory is corrected by the instanton
contributions and has a new complicated transcendental structure,
the duality generators must change basis to adapt themselves to the new
situation and be integer valued in the new non--perturbative geometry.
Alternatively one can turn matters around. If we know the new quantum
symplectic embedding of the discrete duality group we have
essentially
determined the non perturbative geometry. It is this point of view
that has proven very
fruitful in the very recent literature.
\section{Hypergeometry}
\label{LL6}
Next we turn to the hypermultiplet sector of an $N=2$ theory.
Here there are $4$ real scalar fields for each hypermultiplet
and, at least locally, they can be regarded as the four
components of a quaternion. The locality caveat is, in this case,
very substantial because global quaternionic coordinates can
be constructed only occasionally even on those manifolds
that are denominated quaternionic in the mathematical literature.
Anyhow, what is important  is that,
in the hypermultiplet sector, the scalar manifold ${\cal HM}$
has dimension multiple of four:
\begin{eqnarray}
\mbox{dim}_{\bf R} \, {\cal HM} & = & 4 \, m \nonumber\\
m & \equiv & \# \, \mbox{of hypermultiplets}
\label{quatdim}
\end{eqnarray}
and, in some appropriate
sense, it has a quaternionic structure.
\par
As {\it Special K\"ahler} is the collective name given
to the vector multiplet geometry both in the rigid and
in the local case, in the same way  we name {\it Hypergeometry}
that pertaining to the hypermultiplet sector, irrespectively whether
we deal with global or local N=2 theories. Yet in the very same
way as there are two kinds of special geometries, there are also
two kind of hypergeometries and for a very similar reason.
Supersymmetry requires the existence of a principal
$SU(2)$--bundle
\begin{equation}
{\cal SU} \, \longrightarrow \, {\cal HM}
\label{su2bundle}
\end{equation}
that plays for hypermultiplets the same role played by the
the line--bundle ${\cal L} \, \longrightarrow \, {\cal SM}$
in the case of vector multiplets. As it happens there the
bundle ${\cal SU}$ is {\bf flat} in the {\it rigid case} while its
curvature is proportional to the K\"ahler forms in the
{\it local case}.
\par
The difference with the case of vector multiplets
is that rigid and local hypergeometries were already known in
mathematics prior to their use in the context of $N=2$ supersymmetry
and had the following names:
\begin{eqnarray}
\mbox{rigid hypergeometry} & \equiv & \mbox{HyperK\"ahler geom.}
\nonumber\\
\mbox{local hypergeometry} & \equiv & \mbox{Quaternionic
geom.}\nonumber\\
\label{picchio}
\end{eqnarray}
\subsection{Quaternionic, versus HyperK\"ahler manifolds}
Both a quaternionic or a HyperK\"ahler manifold ${\cal HM}$
  is a $4 m$-dimensional real manifold
endowed with a metric $h$:
\begin{equation}
d s^2 = h_{u v} (q) d q^u \otimes d q^v   \quad ; \quad u,v=1,\dots,
4  m
\label{qmetrica}
\end{equation}
and three complex structures
\begin{eqnarray}
(J^x) &:& T({\cal HM}) \, \longrightarrow \, T({\cal HM}) \nonumber\\
(x=1,2,3)& &
\end{eqnarray}
that satisfy the quaternionic algebra
\begin{equation}
J^x J^y = - \delta^{xy} \, \bfone \,  +  \, \epsilon^{xyz} J^z
\label{quatalgebra}
\end{equation}
and respect to which the metric is hermitian:
\begin{eqnarray}
\forall   \mbox{\bf X} ,\mbox{\bf Y}  \in   T{\cal HM}   &:&
h \left( J^x \mbox{\bf X}, J^x \mbox{\bf Y} \right )   =
h \left( \mbox{\bf X}, \mbox{\bf Y} \right ) \nonumber\\
  (x=1,2,3) &\null &\null
\label{hermit}
\end{eqnarray}
{}From eq.~\ref{hermit} it follows that one can introduce
a triplet of  2-forms
\begin{eqnarray}
K^x& = &K^x_{u v} d q^u \wedge d q^v \nonumber\\
K^x_{uv} &=&   h_{uw} (J^x)^w_v
\label{iperforme}
\end{eqnarray}
that provides the generalization of the concept of K\"ahler form
occurring in  the complex case. The triplet $K^x$ is named
the {\it HyperK\"ahler} form. It is an $SU(2)$ Lie--algebra valued
2--form  in the same way as the K\"ahler form is a $U(1)$ Lie--algebra
valued 2--form. In the complex case the definition of K\"ahler manifold
involves the statement
that the K\"ahler 2--form is closed. At the same time in
Hodge--K\"ahler manifolds (those appropriate to local supersymmetry)
the K\"ahler 2--form can be identified with the curvature of
a line--bundle which in the case of rigid supersymmetry
is flat. Similar steps can be taken also here and lead to two possibilities:
either HyperK\"ahler or Quaternionic manifolds.
\par
Let us  introduce a principal $SU(2)$--bundle
${\cal SU}$ as defined in eq.~\ref{su2bundle}. Let $\omega^x$ denote a
connection on such a bundle.
To obtain either a HyperK\"ahler or a quaternionic manifold
we must impose the condition that the HyperK\"ahler 2--form
is covariantly closed with respect to the connection $\omega^x$:
\begin{equation}
\nabla K^x \equiv d K^x + \epsilon^{x y z} \omega^y \wedge
K^z    \, = \, 0
\label{closkform}
\end{equation}
The only difference between the two kinds of geometries resides
in the structure of the ${\cal SU}$--bundle.
\bd
A HyperK\"ahler manifold is a $4 m$--dimensional manifold with
the structure described above and such that the ${\cal SU}$--bundle
is {\bf flat}
\ed
 Defining the ${\cal SU}$--curvature by:
\begin{equation}
\Omega^x \, \equiv \, d \omega^x +
{1\over 2} \epsilon^{x y z} \omega^y \wedge \omega^z
\label{su2curv}
\end{equation}
in the HyperK\"ahler case we have:
\begin{equation}
\Omega^x \, = \, 0
\label{piattello}
\end{equation}
Viceversa
\bd
A quaternionic manifold is a $4 m$--dimensional manifold with
the structure described above and such that the curvature
of the ${\cal SU}$--bundle
is proportional to the HyperK\"ahler 2--form
\ed
Hence, in the quaternionic case we can write:
\begin{equation}
\Omega^x \, = \, {\o{1}{\lambda}}\, K^x
\label{piegatello}
\end{equation}
where $\lambda$ is a non vanishing real number. Actually the
limit of rigid supersymmetry can be identified with $\lambda \to
\infty$.
\par
As a consequence of the above structure the manifold ${\cal HM}$ has a
holonomy group of the following type:
\begin{eqnarray}
{\rm Hol}({\cal HM})&=& SU(2)\otimes {\cal H} \quad \mbox{quater.}
\nonumber \\
{\rm Hol}({\cal HM})&=& \bfone \otimes {\cal H} \quad \mbox{HyperK\"ahler}
\nonumber \\
{\cal H} & \subset & Sp (2m,\IR)
\label{olonomia}
\end{eqnarray}
In both cases, introducing flat
indices $\{A,B,C= 1,2\} \{\alpha,\beta,\gamma = 1,.., 2m\}$  that run,
respectively, in the fundamental representations of $SU(2)$ and
$Sp(2m,\IR)$, we can find a vielbein 1-form
\begin{equation}
{\cal U}^{A\alpha} = {\cal U}^{A\alpha}_u (q) d q^u
\label{quatvielbein}
\end{equation}
such that
\begin{equation}
h_{uv} = {\cal U}^{A\alpha}_u {\cal U}^{B\beta}_v
\IC_{\alpha\beta}\epsilon_{AB}
\label{quatmet}
\end{equation}
where $\IC_{\alpha \beta} = - \IC_{\beta \alpha}$ and
$\epsilon_{AB} = - \epsilon_{BA}$ are, respectively, the flat $Sp(2m)$
and $Sp(2) \sim SU(2)$ invariant metrics.
The vielbein ${\cal U}^{A\alpha}$ is covariantly closed with respect
to the $SU(2)$-connection $\omega^z$ and to some $Sp(2m,\IR)$-Lie Algebra
valued connection $\Delta^{\alpha\beta} = \Delta^{\beta \alpha}$:
\begin{eqnarray}
\nabla {\cal U}^{A\alpha}& \equiv & d{\cal U}^{A\alpha}
+{i\over 2} \omega^x (\epsilon \sigma_x\epsilon^{-1})^A_{\phantom{A}B}
\wedge{\cal U}^{B\alpha} \nonumber\\
&+& \Delta^{\alpha\beta} \wedge {\cal U}^{A\gamma} \IC_{\beta\gamma}
=0
\label{quattorsion}
\end{eqnarray}
\noindent
where $(\sigma^x)_A^{\phantom{A}B}$ are the standard Pauli matrices.
Furthermore ${ \cal U}^{A\alpha}$ satisfies  the reality condition:
\begin{equation}
{\cal U}_{A\alpha} \equiv ({\cal U}^{A\alpha})^* = \epsilon_{AB}
\IC_{\alpha\beta} {\cal U}^{B\beta}
\label{quatreality}
\end{equation}
Eq.\ref{quatreality}  defines  the  rule to lower the symplectic indices by
means   of  the  flat  symplectic   metrics $\epsilon_{AB}$   and
$\IC_{\alpha \beta}$.
More specifically we can write a stronger version of eq.~\ref{quatmet}
\begin{eqnarray}
({\cal U}^{A\alpha}_u {\cal U}^{B\beta}_v + {\cal U}^{A\alpha}_v {\cal
 U}^{B\beta}_u)\IC_{\alpha\beta}&=& h_{uv} \epsilon^{AB}\nonumber\\
({\cal U}^{A\alpha}_u {\cal U}^{B\beta}_v + {\cal U}^{A\alpha}_v {\cal
U}^{B\beta}_u) \epsilon_{AB} &=& h_{uv} {1\over m} \IC^{\alpha
\beta}
\label{piuforte}
\end{eqnarray}
\noindent
We have also the inverse vielbein ${\cal U}^u_{A\alpha}$ defined by the
equation
\begin{equation}
{\cal U}^u_{A\alpha} {\cal U}^{A\alpha}_v = \delta^u_v
\label{2.64}
\end{equation}
Flattening a pair of indices of the Riemann
tensor ${\cal R}^{uv}_{\phantom{uv}{ts}}$
we obtain
\begin{equation}
R^{uv}_{\phantom{uv}{ts}} {\cal U}^{\alpha A}_u {\cal U}^{\beta B}_v =
\Omega^x_{ts}
 {i\over 2} (\epsilon^{-1} \sigma_x)^{AB} \IC^{\alpha \beta}+
 \IR^{\alpha\beta}_{ts}
\label{2.65}
\end{equation}
\noindent
where $\IR^{\alpha\beta}_{ts}$ is the field strength of the $Sp(2m)
$ connection:
\begin{equation}
d \Delta^{\alpha\beta} + \Delta^{\alpha \gamma} \wedge \Delta^{\delta \beta}
\IC_{\gamma \delta} \equiv \IR^{\alpha\beta} = \IR^{\alpha \beta}_{ts}
dq^t \wedge dq^s
\label{2.66}
\end{equation}
Eq. ~\ref{2.65} is the explicit statement that the Levi Civita connection
associated with the metric $h$ has a holonomy group contained in
$SU(2) \otimes Sp(2m)$. Consider now eq.s~\ref{quatalgebra},
\ref{iperforme} and~\ref{piegatello}.
We easily deduce the following relation:
\begin{equation}
h^{st} K^x_{us} K^y_{tw} = -   \delta^{xy} h_{uw} +
  \epsilon^{xyz} K^z_{uw}
\label{universala}
\end{equation}
that holds true both in the HyperK\"ahler and in the Quaternionic case.
In the latter case, where the $\lambda$ parameter is finite,
eq.~\ref{universala} can be rewritten as follows:
\begin{equation}
h^{st} \Omega^x_{us} \Omega^y_{tw} = - \lambda^2 \delta^{xy} h_{uw} +
\lambda \epsilon^{xyz} \Omega^z_{uw}
\label{2.67}
\end{equation}
Eq.~\ref{2.67} implies that the intrinsic components of the curvature
 2-form $\Omega^x$ yield a representation of the quaternion algebra.
In the HyperK\"ahler case such a representation is provided only
by the HyperK\"ahler form.
In the quaternionic case we can write:
\begin{equation}
\Omega^x_{A\alpha, B \beta} \equiv \Omega^x_{uv} {\cal U}^u_{A\alpha}
{\cal U}^v_{B\beta} = - i \lambda \IC_{\alpha\beta} (\sigma^x
\epsilon)_{AB}
\label{2.68}
\end{equation}
\noindent
Alternatively eq.\ref{2.68} can be rewritten in an intrinsic form as
\begin{equation}
\Omega^x = i \lambda \IC_{\alpha\beta} (\sigma^x \epsilon^{-1})
_{AB}{\cal U}^{\alpha A} \wedge {\cal U}^{\beta B}
\label{2.69}
\end{equation}
\noindent
where from we also get:
\begin{equation}
{i\over 2} \Omega^x (\sigma_x)_A^{\phantom{A}B} =
\lambda{\cal U}_{A\alpha} \wedge {\cal
U}^{B\alpha}
\label{2.70}
\end{equation}
Homogeneous symmetric quaternionic spaces are displayed in Table
{}~\ref{quatotable}.
\begin{table*}
\begin{center}
\caption{\sl Homogeneous symmetric quaternionic manifolds}
\label{quatotable}
\begin{tabular}{|c||c|}
\hline
m  & $G/H$
\\
\hline
{}~~&~~\\
$m$ & $\o{Sp(2m+2)}{Sp(2)\times SP(2m)}$ \\
{}~~&~~\\
\hline
{}~~&~~\\
$m$ & $\o{SU(m,2)}{SU(m)\times SU(2)\times U(1)}$  \\
{}~~&~~\\
\hline
{}~~&~~\\
$m$ & $\o{SO(4,m)}{SO(4)\times SO(m)}$  \\
{}~~&~~\\
\hline
\hline
{}~~&~~\\
$2$ & $\o{G_2}{S0(4)}$ \\
{}~~&~~\\
\hline
{}~~&~~\\
$7$ & $\o{F_4}{Sp(6)\times Sp(2)}$\\
{}~~&~~~\\
\hline
{}~~&~~\\
$10$ & $\o{E_6}{SU(6)\times U(1)}$\\
{}~~&~~\\
\hline
{}~~&~~\\
$16$ & $\o{E_7}{S0(12)\times SU(2)}$\\
{}~~&~~\\
\hline
{}~~&~~\\
$28$ & $\o{E_8}{E_7\times SU(2)}$\\
{}~~&~~\\
\hline
\end{tabular}
\end{center}
\end{table*}
\section{The Gauging}
\label{LL7}
With the above discussion of HyperK\"ahler and Quaternionic manifolds
I have  completed my review of the geometric structures
involved in the construction of {\it abelian, ungauged} $N=2$
supergravity or of $N=2$ rigid gauge theory.
The relation of supersymmetry with duality--rotations
has been repeatedly emphasized.  Sticking to my promise of
restricting my own attention to the bosonic sector of all
lagrangians, the situation we have so far reached is the following.
The bosonic Lagrangian of $N=2$ supergravity coupled to $n$
abelian vector multiplets and $m$ hypermultiplets is the
following:
\begin{eqnarray}
&{\cal L}_{ungauged}^{SUGRA} \, =& \nonumber\\
& \sqrt{-g}\Bigl  [ \,  R[g]
\, + \, g_{i {j^\star}}(z,\bar z )\, \partial^{\mu} z^i \,
\partial _{\mu} \bar z^{j^\star} & \nonumber \\
& - \, 2 \,\lambda\, h_{uv}(q) \, \partial^{\mu} q^u \,
 \partial _{\mu} q^v
\nonumber\\
&  + \,{\rm i} \,\left(
\bar {\cal N}_{\Lambda \Sigma} {\cal F}^{- \Lambda}_{\mu \nu}
{\cal F}^{- \Sigma \vert {\mu \nu}}
\, - \,
{\cal N}_{\Lambda \Sigma}
{\cal F}^{+ \Lambda}_{\mu \nu} {\cal F}^{+ \Sigma \vert {\mu \nu}} \right )
\, \Bigr ] & \nonumber \\
\label{ungausugra}
\end{eqnarray}
where the  $n$ complex fields $z^{i}$ span some {\it special K\"ahler
manifold of the local type} ${\cal SM}$ and the $4 m$ real fields
$q^u$ span a quaternionic manifold ${\cal HM}$. By $g_{ij^\star}$
and $h_{uv}$ we have denoted the metrics on these two manifolds.
The proportionality constant between the $SU(2)$ curvature and the
HyperK\"ahler form  appearing in the lagrangian is fixed by
to the value $\lambda=-1$ if we want canonical kinetic terms
for the hypermultiplet scalars. The period matrix
${\cal N}_{\Lambda\Sigma}$ depends only on the special manifold
coordinates $z^{i}, z^{j^\star}$ and it is expressed in terms
of the symplectic sections of the flat symplectic bundle by
eq.~\ref{intriscripen}.
On the other hand the bosonic Lagrangian of an $N=2$ abelian gauge
theory containing $n$ vector multiplets and coupled to
$m$ hypermultiplets is the following one:
\begin{eqnarray}
&{\cal L}_{ungauged}^{YM} \, =& \nonumber\\
&  g_{i {j^\star}}(z,\bar z )\, \partial^{\mu} z^i \,
\partial _{\mu} \bar z^{j^\star} \,
+ \, 2 \,  h_{uv}(q) \, \partial^{\mu} q^u \,
 \partial _{\mu} q^v
\nonumber\\
&  + \,{\rm i} \,\left(
\bar {\cal N}_{\alpha\beta} {\cal F}^{- \alpha}_{\mu \nu}
{\cal F}^{- \beta \vert {\mu \nu}}
\, - \,
{\cal N}_{\alpha\beta}
{\cal F}^{+ \alpha}_{\mu \nu} {\cal F}^{+ \beta \vert {\mu \nu}} \right )
\,   & \nonumber \\
\label{ungaugauga}
\end{eqnarray}
where the  $n$ complex fields $z^{i}$ span some {\it special K\"ahler
manifold of the rigid type} ${\cal SM}$ and the $4 m$ real fields
$q^u$ span a HyperK\"ahler manifold ${\cal HM}$. By $g_{ij^\star}$
and $h_{uv}$ we have denoted the metrics on these two manifolds.
   The period matrix
${\cal N}_{\alpha\beta}$ depends only on the special manifold
coordinates $z^{i}, z^{j^\star}$ and it is expressed in terms
of the symplectic sections of the flat symplectic bundle by
eq.~\ref{intriscripenrig}.
In both theories there are no electric or magnetic currents and
we have {\it symplectic covariance}. By means of the
the first homomorphism in eq.~\ref{spaccoindue} any diffeomorphism
of the scalar manifold can be lifted to a symplectic transformation
on the electric--magnetic field strengths, the {\it period} matrix
transforming, by construction covariantly as required by
eq.~\ref{covarianza}. Under this lifting any isometry of the
scalar manifold becomes a symmetry not just of the lagrangian
but of the differential system made by the equations of motions
plus Bianchi identities. There are in fact three type of these
isometries:
\begin{enumerate}
\item{{\it The classical symmetries}, namely those isometries
$\xi \, \in \, {\cal I} \left ( {\cal M}_{scalar} \right )$
whose image in the symplectic group is block--diagonal:
\begin{equation}
\iota_\delta ( \xi ) \, = \, \twomat{A_ \xi}{  0}{  0}{(A^T_ \xi)^{-1}}
\end{equation}
 These transformations are exact ordinary symmetries of the lagrangian.
They clearly form a subgroup
\begin{equation}
 {\cal C}las \left ( {\cal M}_{scalar} \right ) \, \subset \,
 {\cal I} \left ( {\cal M}_{scalar} \right )
 \label{classym}
\end{equation}
   }
 \item{{\it The perturbative symmetries}, namely those isometries
$\xi \, \in \, {\cal I} \left ( {\cal M}_{scalar} \right )$
whose image in the symplectic group is lower triangular:
\begin{equation}
\iota_\delta ( \xi ) \, = \, \twomat{A_ \xi}{  0}{  C_ \xi}
{(A^T_ \xi)^{-1}}
\label{persym}
\end{equation}
 These transformations map the electric field strengths
 into linear combinations of the electric field strengths and can be
 be reduced to   linear transformations of the
 gauge potentials. They are almost  invariances of the action.  Indeed
 the only non--invariance comes from the transformation of the period
 matrix
\begin{equation}
{\cal N} \, \longrightarrow  \, (A^T_ \xi)^{-1} \, {\cal N}
(A_ \xi)^{-1}  \, + \, C_ \xi \, (A^T_ \xi)^{-1}
\end{equation}
Denoting collectively all the fields of the theory by  $\Phi$
and utilizing eq.s~\ref{gaiazuma},~\ref{gaiazumadue},~\ref{scripten},
{}~\ref{lagrapm},~\ref{covarianza}, under a perturbative transformation
the action changes as follows:
\begin{eqnarray}
\int \, {\cal L}(\Phi) \, d^4 x  &  \to &
\int \, {\cal L}(\Phi^\prime) \, d^4 x \nonumber \\
&& + \, 4 \Delta\theta_{\Lambda\Sigma} \, \int F^\Lambda \wedge F^\Sigma
\nonumber\\
\Delta\theta _{\Lambda\Sigma} & = &  \left [ C_\xi \, (A^T_\xi)^{-1}
\right ]_{\Lambda\Sigma}
\end{eqnarray}
\par
The added term is a total derivative and does not affect the field
equations. Quantum mechanically, however, it is relevant. It
corresponds to a redefinition of the $theta$--angle. It yields
a symmetry of the path--integral as long as the added term
is an integer multiple of $2 \pi  \hbar $. This consideration
will restrict the possible perturbative transformations to
a discrete subgroup. In any case the group
of perturbative isometries defined by eq.~\ref{persym}  contains
the group of classical isometries as a subgroup:
$ {\cal I} \left ( {\cal M}_{scalar} \right ) \supset
  {\cal P}ert \left ( {\cal M}_{scalar} \right ) \supset
 {\cal C}las \left ( {\cal M}_{scalar} \right )$. }
\item{{\it The non--perturbative symmetries} namely those isometries
$\xi \, \in \, {\cal I} \left ( {\cal M}_{scalar} \right )$
whose image in the symplectic group is of the form:
\begin{equation}
\iota_\delta ( \xi ) \, = \, \twomat{  A_ \xi}{   B_ \xi }{  C_ \xi}
{  D_\xi}
\label{nonpersym}
\end{equation}
with $B_\xi \ne 0$.
These transformations are neither a symmetry of the classical
action nor of the perturbative path integral. Yet they are a
symmetry of the quantum theory. They exchange electric field
strengths with magnetic ones, electric currents with magnetic ones
and hence elementary excitations with soliton states. }
\end{enumerate}
The above discussion of duality symmetries may have left the
audience intrigued about the following point. How can I talk
about non--perturbative symmetries that  exchange electric charges
with magnetic charges if, so far, in the abelian theories described
by eq.s~\ref{ungausugra} and ~\ref{ungaugauga} there are neither
electric nor magnetic couplings? The answer is that the same general
form of abelian theories encoded in these equations can be taken to
represent two quite different things:
\begin{enumerate}
\item {The fundamental theory  prior to the gauging. It is neutral
and abelian since the  non--abelian
interactions and the electric charges are introduced only by the gauging,
but it contains all the fundamental fields.}
\item {The effective theory of the massless modes of the non--abelian
theory. It is abelian and neutral because the only fields which
remain massless are, apart from the graviton, the multiplets in the
Cartan subalgebra ${\cal H} \subset {\cal G}$ of the gauge group and
the neutral hypermultiplets corresponding to flat directions of the
scalar potential.}
\end{enumerate}
What distinguishes the two cases is the type of scalar manifolds
and their isometries.
\par
In case 1) of the above list we have:
\begin{eqnarray}
\mbox{dim}_{\bf C} \, {\cal SM}&= & n \equiv \mbox{dim} \, {\cal G}
\nonumber\\
{\o{1}{4}}\,\mbox{dim}_{\bf R} \, {\cal HM}&= &
{\hat m} \equiv \# \, \mbox{of all hypermul.} \nonumber\\
{\cal I} \left ({\cal SM} \right ) & = & \mbox{cont. group} \,
\supset \, {\cal G} \nonumber\\
{\cal I} \left ({\cal HM} \right ) & = & \mbox{cont. group} \,
\supset \, {\cal G} \nonumber\\
\label{microscop}
\end{eqnarray}
where ${\cal G}$ denotes the gauge group.
\par
In case 2) we have instead:
\begin{eqnarray}
\mbox{dim}_{\bf C} \, {\cal SM}&= & r \equiv \mbox{rank} \, {\cal G}
\nonumber\\
{\o{1}{4}}\,\mbox{dim}_{\bf R} \, {\cal HM}&= &
{  m} \equiv \# \, \mbox{of moduli hypermul.} \nonumber\\
{\cal I} \left ({\cal SM} \right ) & = & \mbox{discr. group} \nonumber\\
{\cal I} \left ({\cal HM} \right ) & = & \mbox{discr. group} \nonumber\\
\label{macroscop}
\end{eqnarray}
In the first instance (corresponding to eq.~\ref{microscop})
the spaces ${\cal SM}$ and ${\cal HM}$ are
chosen with a large continuous group of duality invariances that
correspond to the concept of {\bf global symmetry} of ordinary
field theory. A subgroup of this global symmetry group can be
made into a {\bf local symmetry} of the theory by associating its
generators with the vector fields already present in the theory.
This is the above mentioned gauging procedure. Since ${\cal G}$
is a subgroup of both  isometry groups
${\cal I}   \left ( {\cal SM} \right )$ and
${\cal I} \left ( {\cal HM} \right )$, it follows that ${\cal G}$
acts non--trivially both on the vector multiplets and on the
hypermultiplets. The former action introduces the non--abelian
interactions, the latter action introduces the electric charges
of the matter fields. For consistency of the gauging procedure
the gauge group must be a subgroup of the group of classical
symmetries \footnotemark
\footnotetext{at the price of modifying the Lagrangian by the
introduction of Chern--Simons like terms, de Wit and Van Proeyen
have obtained the gauging also of perturbative transformations
\cite{pertgauging}. The gauge group in this case is non
semisimple and the gauge algebra contains solvable subalgebras.
In these lectures I confine my attention to the block diagonal
gauging}:
\begin{equation}
{\cal G} \, \subset \,
{\cal C}las \left ( {\cal M}_{scalar} \right ) \, \subset \,
Sp(2\bar n, \IR)
\label{gaugediag}
\end{equation}
As a consequence of gauging the lagrangians in eq.s~\ref{ungausugra} and
{}~\ref{ungaugauga} get  modified by the replacement of
ordinary derivatives with covariant derivatives
and by the introduction of new terms that are
of two types:
\begin{enumerate}
\item {fermion--fermion bilinears with scalar field dependent
coefficients}
\item{A scalar potential ${\cal V}$ }
\end{enumerate}
It is particularly nice and rewarding that all the modifications
of the lagrangian and of the supersymmetry transformation rules
can be described in terms of a very general geometric construction
associated with the action of Lie--Groups on manifolds that
admit a symplectic structure: {\it the momentum map}.
In supersymmetry  indeed,
the geometric notion of {\it momentum map} has an exact correspondence
with the notion of {\it gauge multiplet auxiliary fields}
or {\it $D$--fields}. The next section
is devoted to a review of the momentum map and to its
applications in N=2 theories.
\par
Prior to that, however, let me finish with my comparison of cases  1) and
2).
\par
In the microscopic theory (eq.~\ref{microscop}) the manifolds
we start from to do the gauging are typically coset manifolds
${\cal G}/{\cal H}$ for the supergravity case and flat manifolds
for the rigid case. This comes from the need of an
ample duality--symmetry group where the gauge group should be
embedded. Indeed continuous isometries are allowed by non trivial
Special K\"ahler manifolds of the local type and by Quaternionic manifolds.
On the other hand Special K\"ahler manifolds of the rigid type or
HyperK\"ahler manifolds admit continuous isometries only if they
are flat manifolds. Which particular choice of homogeneous
Special K\"ahler manifold or quaternionic manifold is appropriate
is predicted by {\it tree--level string theory} when we deal
with the effective lagrangians of superstrings. In particular
for the class of models dealt with in the recent literature
on string--string duality the choice is:
\begin{eqnarray}
{\cal SM} & = & {\cal ST}[2,n]  \nonumber \\
{\cal HM} & = & {\o{SO(4,\hat m)}{SO(4) \otimes SO(\hat m)}}
\label{bellescelte}
\end{eqnarray}
After the gauging a scalar potential is introduced and most
of the scalar fields  acquire mass by Higgs effect. Yet the
scalar potential admits flat directions. Those scalar fields that
span the flat directions are the {\it moduli} and
they fill the {\it classical moduli space}. For the vector multiplet
sector this classical moduli space is easily deduced.  It is:
\begin{equation}
{\cal SM}_{moduli} \, = \, {\cal ST}[2,r]  \, \subset \, {\cal ST}[2,n]
\label{modclass}
\end{equation}
Indeed it suffices to restrict one's attention to the
$r$ fields in the Cartan's subalgebra of the gauge group.
For the hypermultiplet moduli space we have:
\begin{eqnarray}
& {\cal HM}_{moduli} \, = \,&\nonumber \\
& \mbox{some hypersurface}
\subset \, {\o{SO(4,\hat m)}{SO(4) \otimes SO(\hat m)}} & \nonumber \\
\label{modhyper}
\end{eqnarray}
\par
When we deal with the effective lagrangian of the massless modes,
which is the situation envisaged by eq.s~\ref{macroscop}, the
two manifolds ${\cal SM}$ and ${\cal HM}$ are respectively
identified with
${\hat {\cal SM}}_{moduli}$  and  ${\hat {\cal HM}}_{moduli}$, the
quantum moduli spaces of vector multiplets and hypermultiplets.
There is no need of continuous isometries since there is no
gauging to be done. This allows the manifolds to be non homogeneous,
non flat manifolds depending on the case. Some discrete isometries
however are usually present. They are of the utmost interest. Their
symplectic lifting provides the non--perturbative duality symmetries
discussed above that exchange elementary electric states with
magnetic soliton states.
\section{The Momentum Map}
\label{LL8}
The momentum map is a construction that applies to all manifolds
with a symplectic structure, in particular to K\"ahler, HyperK\"ahler
and Quaternionic manifolds.
\par
Let us begin with the K\"ahler case, namely with the momentum
map of holomorphic isometries. The HyperK\"ahler and quaternionic
case  correspond, instead, to the momentum map of triholomorphic
isometries.
\subsection{Holomorphic momentum map on K\"ahler manifolds}
Let  $g_{i {j^\star}}$ be the K\"ahler metric of a K\"ahler
manifold ${\cal M}$. As many time emphasized, it appears in the kinetic
term of the scalar fields: the Wess--Zumino multiplet scalars in N=1
theories, the vector multiplet scalars in N=2 theories.
If the metric $g_{i {j^\star}}$ has a non trivial group of
continuous isometries ${\cal G}$
generated by Killing vectors $k_\Lambda^i$ ($\Lambda=1, \ldots, {\rm dim}
\,G )$, then the kinetic
lagrangian  admits ${\cal G}$ as a group of global space--time
symmetries. Indeed under an infinitesimal variation
\be
z^i \to z^i + \epsilon^\Lambda k_\Lambda^i (z)
\ee
${\cal L}_{kin}$ remains invariant. Furthermore if all the
couplings of the scalar fields are performed in a diffeomorphic
invariant way, then any isometry of $g_{i {j^\star}}$ extends
from a symmetry
of ${\cal L}_{kin}$ to a symmetry of the whole lagrangian.
Diffeomorphic invariance means that the scalar fields can appear only
through the metric, the Christoffel symbol in the covariant derivative
and through the curvature. Alternatively they can appear through
sections of vector bundles constructed over ${\cal M}$. Typical
case is the dependence on the scalar fields introduced by the
{\it period matrix} ${\cal N}$.
\par
Let $k^i_{\Lambda} (z)$ be a basis of holomorphic Killing vectors for
the metric $g_{i{j^\star}}$.  Holomorphicity means the following
differential constraint:
\be
\partial_{j^*} k^i_{\Lambda} (z)=0
\leftrightarrow \partial_j k^{i^*}_{\Lambda} (\bar z)=0 \label{holly}
\ee
while the generic Killing equation (suppressing the
gauge index $\Lambda$):
\be
\nabla_\mu k_\nu +\nabla_\mu k_\nu=0
\ee
in holomorphic indices reads as follows:
\ba
\nabla_i k_{j} + \nabla_j k_{i}&=&0 \nn\\
\nabla_{i^*} k_{j} + \nabla_j k_{i^*} &=& 0
\label{killo}
\ea
where the covariant components are defined as
$k_{j }=g_{j i^*} k^{i^*}$ (and similarly for
$k_{i^*}$).
\par
The vectors $k_{\Lambda}^i$ are generators of infinitesimal
holomorphic coordinate transformations:
\be
\delta z^i = \epsilon^\Lambda k^i_{\Lambda} (z)
\ee
which leave the metric invariant.
In the same way as the metric is the derivative of a more fundamental
object, the Killing vectors in a K\"ahler manifold are the
derivatives of suitable prepotentials. Indeed the first of
eq.s~\ref{killo}  is automatically satisfied by holomorphic vectors
and the second equation reduces to the following one:
\be
k^i_{\Lambda}=i g^{i j^*} \partial_{j^*} {\cal P}_{\Lambda},
\quad {\cal P}^*_{\Lambda} = {\cal P}_{\Lambda}\label{killo1}
\ee
In other words if we can find a real function ${\cal P}^\Lambda$ such
that the expression $i g^{i j^*} \partial_{j^*}
{\cal P}_{(\Lambda)}$ is holomorphic, then eq.~\ref{killo1} defines a
Killing vector.
\par
The construction of the Killing prepotential can be stated in a more
precise geometrical formulation which involves the notion of
{\it momentum
map}. Let me review this construction which reveals
another  deep connection between supersymmetry and
geometry.
\par
 Consider a K\"ahlerian manifold ${\cal M}$
of real dimension $2n$.
Consider a compact Lie group ${\cal G}$ acting on
 ${\cal M}$  by means of Killing vector
fields ${\bf X}$ which are holomorphic
with respect to the  complex structure
${ J}$ of ${\cal M}$; then these vector
fields preserve also the K\"ahler 2-form
\begin{eqnarray}
&\matrix{
{\cal L}_{\scriptscriptstyle{\bf X}}g = 0 &\leftrightarrow &
\nabla_{(\mu}X_{\nu)}=0 \cr
{\cal L}_{\scriptscriptstyle{\bf X}}{  J}= 0 &\null & \cr }
  \Biggr \} \,\,\Rightarrow & \nonumber \\
& 0={\cal L}_{\scriptscriptstyle{\bf X}}
K = i_{\scriptscriptstyle{\bf X}}
dK+d(i_{\scriptscriptstyle{\bf X}}
K) = d(i_{\scriptscriptstyle{\bf X}}K) & \nonumber\\
\label{holkillingvectors}
\end{eqnarray}
Here ${\cal L}_{\scriptscriptstyle{\bf X}}$ and
$i_{\scriptscriptstyle{\bf X}}$
denote respectively the Lie derivative along
the vector field ${\bf X}$ and the contraction
(of forms) with it.
\par
If ${\cal M}$ is simply connected,
$d(i_{{\bf X}}K)=0$ implies the existence
of a function ${\cal P}_{{\bf X}}$ such
that
\be
-\o{1}{2\pi}d{\cal P}_{{\bf X}}=
i_{\scriptscriptstyle{\bf X}}K
\label{mmap}
\ee
The function ${\cal P}_{{\bf X}}$ is
defined up to a constant,
which can be arranged so as to make it equivariant:
\be
{\bf X} {\cal P}_{\bf Y} =
{\cal P}_{[{\bf X},{\bf Y}]}
\label{equivarianza}
\ee
\par
  ${\cal P}_{{\bf X}}$ constitutes
then a {\it momentum
map}.
This can be regarded as a map
\be
{\cal P}: {\cal M} \, \longrightarrow \,
\IR \otimes
{\IG }^*
\ee
where ${\IG}^*$ denotes the dual of the Lie algebra
${\IG }$ of the group ${\cal G}$.
Indeed let $x\in {\IG }$ be the Lie algebra element
corresponding to the Killing
vector ${\bf X}$; then, for a given
$m\in {\cal M}$
\be
\mu (m)\,  : \, x \, \longrightarrow \,  {\cal P}_{{\bf X}}(m) \,
\in  \, \IR
\ee
is a linear functional on  ${\IG}$.
 If we expand
${\bf X} = a^\Lambda k_\Lambda$ in a basis of Killing vectors
$k_\Lambda$ such that
\be
[k_\Lambda, k_\Gamma]= f_{\Lambda \Gamma}^{\ \ \Delta} k_\Delta
\label{blio}
\ee
we have also
\be
{\cal P}_{\bf X}\, = \, a^\Lambda {\cal P}_\Lambda
\ee
In the following we  use the
shorthand notation ${\cal L}_\Lambda, i_\Lambda$ for
the Lie derivative
and the contraction along the chosen basis
of Killing vectors $ k_\Lambda$.
\par
{}From a geometrical point of view the prepotential,
or momentum map, ${\cal P}_\Lambda$
is the Hamiltonian function providing the Poissonian
realization  of the Lie algebra on the K\"ahler manifold. This
is just another way of stating the already mentioned
{\it  equivariance}.
Indeed  the  very  existence  of the closed 2-form $K$ guarantees that
every K\"ahler space is a symplectic manifold and that we can define  a
Poisson bracket.
\par
Consider Eqs.~\ref{killo1}.
To every generator of the abstract  Lie algebra
${\IG}$ we have associated a function  ${\cal P}_\Lambda$ on
${\cal M}$; the Poisson bracket of
${\cal P}_\Lambda$ with ${\cal P}_\Sigma$ is defined as
follows:
\be
\{{\cal P}_\Lambda , {\cal P}_\Sigma\} \equiv 4\pi K
(\Lambda, \Sigma)
\ee
where $K(\Lambda, \Sigma)
\equiv K (\vec k_\Lambda, \vec k_\Sigma)$ is
the value of $K$ along the pair of Killing vectors.
\par
We now prove the following lemma.
\blem
{\it{The following identity is true}}:
\be
\{{\cal P}_\Lambda, {\cal
P}_\Sigma\}=f_{\Lambda\Sigma}^{\ \ \Gamma}{\cal
P}_\Gamma + C_{\Lambda \Sigma} \label{brack}
\ee
{\it{where $C_{\Lambda \Sigma}$ is a constant fulfilling the
cocycle condition}}
\be
f^{\ \ \Gamma}_{\Lambda\Pi} C_{\Gamma \Sigma} +
f^{\ \ \Gamma}_{\Pi\Sigma} C_{\Gamma \Lambda}+
f_{\Sigma\Lambda}^{\ \  \Gamma} C_{\Gamma \Pi}=0
\label{cocy}
\ee
\elem
{\underbar{Proof}}: Let us set $f_{\Lambda\Sigma}\equiv K(\Lambda,
\Sigma)$. Using eq.~\ref{holkillingvectors}  we get
\ba
4\pi f_{\Lambda\Sigma}&=& 4\pi  f _{\Sigma\Lambda} =2\pi i_\Sigma
i_\Lambda K  \nonumber\\
&=&- i_\Sigma d {\cal P}_\Lambda =i_\Lambda
d {\cal P}_\Sigma=\nn\\
&=&{1\over 2} ({\cal L}_\Lambda {\cal P}_\Sigma - {\cal L}_\Sigma {\cal
P}_\Lambda)\label{2.32}
\ea
Let us now calculate $df_{\Lambda\Sigma}$. Since the exterior
derivative commutes with the Lie derivative we find
\ba
d f_{\Lambda \Sigma}&=& {1\over 8\pi} ({\cal L}_\Lambda d {\cal
P}_\Sigma-{\cal L}_\Sigma d {\cal P}_\Lambda)=\\
&=&{1\over 4} (-{\cal L}_\Lambda i{_\Sigma} K +
{\cal L}_\Sigma i{_\Lambda}
K)
\ea
Using now the identity
\ba
[i_\Lambda , {\cal L}_\Sigma] = i_{[\Lambda, \Sigma]}
\ea
and eq.~\ref{blio}, from eq.~\ref{2.32} we obtain
\ba
d f_{\Lambda \Sigma}&=& {1\over 2} i_{[\Sigma, \Lambda]} K =
-{1\over 2} f^{\ \  \Gamma}_{\Lambda\Sigma} i_{\Gamma} K=\\
& =& {1\over 4\pi} f_{\Lambda\Sigma}^{\  \  \Gamma} d {\cal
P}_\Gamma
\ea
\noindent
It follows that the difference
\be
C_{\Lambda \Sigma} = \{{\cal P}_\Lambda, {\cal P}_\Sigma\}
- f^{\  \  \Gamma}_{\Lambda \Sigma} {\cal P}_\Gamma
\ee
is a constant since we have shown that its exterior derivative
vanishes: $dC_{\Lambda \Sigma}= 0$. The cocycle condition
in eq.~\ref{cocy}
follows from the Jacobi identities fulfilled by the Poisson bracket
of eq.~\ref{brack}.
This concludes the proof of the lemma.
If the Lie algebra ${\IG}$ has a trivial second cohomology group
$H^2({\IG})=0$, then the cocycle $C_{\Lambda \Sigma}$ is a
coboundary; namely we have
\be
C_{\Lambda \Sigma} = f^{\ \ \Gamma}_{\Lambda \Sigma} C_\Gamma
\ee
where $C_\Gamma$ are suitable constants. Hence, assuming
$H^2 (\IG)= 0$
we can reabsorb $C_\Gamma$ in  the definition of ${\cal
P}_\Lambda$:
\be
{\cal P}_\Lambda \rightarrow {\cal P}_\Lambda+ C_\Lambda
\ee
and we obtain the stronger equation
\be
\{{\cal P}_\Lambda, {\cal P}_\Sigma\} =
f_{\Lambda\Sigma}^{\ \  \Gamma} {\cal P}_\Gamma
\label{2.39}
\ee
Note that $H^2({\IG}) = 0$ is true for all semi-simple Lie
algebras.
Using eq.~\ref{brack}, eq.~\ref{2.39}
can be rewritten in components as follows:
\be
{i\over 2} g_{ij^*}(k^i_\Lambda k^{j^*}_\Sigma -
k^i_\Sigma k^{j^*}_\Lambda)=
{1\over 2} f_{\Lambda \Sigma}^{\  \  \Gamma} {\cal
P}_\Gamma
\label{2.40}
\ee
Equation~\ref{2.40} is identical with the equivariance condition
in eq.~\ref{equivarianza}.
\par
Comparing the definition of the K\"ahler
potential in eq.~\ref{popov} with the definition of the momentum
function in eq.~\ref{killo1}, we obtain an expression for
the momentum map function in terms of derivatives of the K\"ahler
potential:
\begin{eqnarray}
{\rm i} \, {\cal P}_\Lambda &=  &{\o{1}{2}}
\left ( k^{i}_\Lambda \, \partial_i {\cal K} -
k^{i^\star}_\Lambda \, \partial_{i^\star} {\cal K}\right ) \nonumber \\
& = &k^{i}_\Lambda \, \partial_i {\cal K} \nonumber \\
& = & - k^{i^\star}_\Lambda \, \partial_{i^\star} {\cal K}
\label{kexpresp}
\end{eqnarray}
Eq.~\ref{kexpresp} is true if the K\"ahler potential is exactly
invariant under the transformations of the isometry group ${\cal G}$
and not only up to a K\"ahler transformation as defined
in eq.~\ref{041}. In other words eq.~\ref{kexpresp} is true if
\begin{equation}
0\, = \, {\cal L}^\Lambda \, {\cal K} \, = \,
k^{i}_\Lambda \, \partial_i {\cal K} +
k^{i^\star}_\Lambda \, \partial_{i^\star} {\cal K}
\label{kpotinvariant}
\end{equation}
Not all the isometries of a general K\"ahler manifold have such
a property, but those that in a suitable coordinate frame  display
a linear action on the coordinates certainly do. However,
in Hodge K\"ahler manifolds, eq.~\ref{kpotinvariant} can be replaced
by the following one which is certainly true:
\begin{eqnarray}
0 & = & {\cal L}^\Lambda \, G \, = \,
k^{i}_\Lambda \, \partial_i {\cal K} +
k^{i^\star}_\Lambda \, \partial_{i^\star} G \nonumber \\
G(z,\bar z) & \equiv &   \mbox{log } \, \parallel  W(z) \parallel^2
\nonumber \\
& = & {\cal K}(z,\bar z ) \, + \, \mbox{Re} \, W(z)
\label{normsect}
\end{eqnarray}
where the {\it superpotential} $W(z)$ is any holomorphic section
of the Hodge line--bundle. Indeed the transformation under
the isometry of the K\"ahler potential is compensated by the
transformation of the superpotential. Consequently, in Hodge--K\"ahler
manifolds eq.~\ref{kexpresp} can be rewritten as
\begin{eqnarray}
{\rm i} \, {\cal P}_\Lambda &=  &{\o{1}{2}}
\left ( k^{i}_\Lambda \, \partial_i G -
k^{i^\star}_\Lambda \, \partial_{i^\star} G\right ) \nonumber \\
& = &k^{i}_\Lambda \, \partial_i G \nonumber \\
& = & - k^{i^\star}_\Lambda \, \partial_{i^\star} G
\label{gexpresp}
\end{eqnarray}
and holds true for any isometry.
\par
In $N=1$ supersymmetry the K\"ahlerian momentum maps
${\cal P}_\Gamma$ appear as auxiliary fields of the
vector multiplets. For $N=1$ supergravity the scalar manifold
is of the Hodge type and eq.~\ref{gexpresp} can always be employed.
\par
On the other hand,
in $N=2$ supersymmetry the auxiliary fields of the vector multiplets,
that form an $SU(2)$ triplet,
are given by the momentum map of triholomorphic isometries on the
hypermultiplet manifold (HyperK\"ahlerian or quaternionic depending
on the local or rigid nature of supersymmetry). The triholomorphic
momentum map is discussed in the  subsection after the next.
Yet, although not
identified with the auxiliary fields, the holomorphic momentum map
plays a role also in $N=2$ theories in the {\it gauging} of the
$U(1)$ connection ~\ref{u1conect}, as I show shortly from now.
\subsection{Holomorphic momentum map on Special K\"ahler manifolds}
Here the K\"ahler manifold is not only Hodge but it is special.
Correspondingly we can write a formula for  ${\cal P}_\Lambda$
in terms of symplectic invariants. In this context, to distinguish
the holomorphic momentum map from the triholomorphic one
${\cal P}_\Lambda^x$ that carries an $SU(2)$ index $x=1,2,3$,
we adopt the notation ${\cal P}_\Lambda^0$.
The request that the isometry group should be embedded into the
symplectic group is formulated by writing:
\begin{eqnarray}
{\cal L}_\Lambda \, V \, & \equiv  & k^i_\Lambda \,\partial_i V \, +   \,
k^{i^\star}_\Lambda \, \partial_{i^\star} V    \nonumber \\
& = & T_\Lambda \, V \, + V \, f_\Lambda(z)
\label{pullsectiso}
\end{eqnarray}
where $V$ is the covariantly holomorphic section of the vector bundle
${\cal H} \, \longrightarrow \, {\cal M}$ defined in eq.~\ref{specpotuno},
\begin{equation}
{\bf Sp}({\bf 2n+2},\IR) \, \ni \, T_\Lambda \, = \, \left (\matrix
{a_\Lambda & b_\Lambda \cr c_\Lambda & d_\Lambda \cr } \right )
\label{alcyon}
\end{equation}
is some element of the real symplectic Lie algebra and $f_\Lambda(z)$
corresponds to an infinitesimal K\"ahler transformation.
\par
The classical or perturbative isometries:
\begin{equation}
b_\Lambda = 0
\end{equation}
that are relevant to the gauging procedure
are normally characterized by
\begin{equation}
 f_\Lambda(z)=0
 \label{parcondicio}
\end{equation}
Under condition~\ref{parcondicio},
recalling eq.s~\ref{specpot} and ~\ref{covholsec},
from eq.~\ref{pullsectiso}
we  obtain:
\begin{equation}
  {\cal L}^\Lambda \, {\cal K} \, = \,
k^{i}_\Lambda \, \partial_i {\cal K} +
k^{i^\star}_\Lambda \, \partial_{i^\star} {\cal K} \, = \, 0
\label{kpotinvariantuno}
\end{equation}
that is identical with eq.~\ref{kpotinvariant}. Hence we can use
eq.~\ref{kexpresp}, that we rewrite as:
\begin{equation}
{\rm i} \, {\cal P}_\Lambda^0 \, = \,
k^{i}_\Lambda \, \partial_i {\cal K} \, =
\, - k^{i^\star}_\Lambda \, \partial_{i^\star} {\cal K}
\label{kexprespprime}
\end{equation}
Utilizing the definition in eq.~\ref{uvector} we easily obtain:
\begin{equation}
k_\Lambda^i \, U^i \, = \, T_\Lambda \, V \, \mbox{exp}[f_\Lambda(z)]
+ {\rm i} \, {\cal P}_\Lambda^0 \, V
\label{passino}
\end{equation}
Taking the symplectic scalar product of eq.~\ref{passino} with ${\bar
V}$ and recalling eq.~\ref{specpotuno} we finally get
\footnotemark
\footnotetext{The following and the next two formulae have been
obtained in private discussions of the author with A. Van Proeyen and
B. de Wit}:
\begin{eqnarray}
{\cal P}_\Lambda^0 &= &
\langle {\bar V} \, \vert \, T_\Lambda \, V \rangle \, = \,
\langle {  V} \, \vert \, T_\Lambda \, {\bar V } \rangle \nonumber \\
& = & \mbox{exp}\left [ {\cal K}\right ] \,
\langle { \bar \Omega} \, \vert \, T_\Lambda \, { \Omega } \rangle
\label{passetto}
\end{eqnarray}
In the gauging procedure we are interested in groups the symplectic
image of whose generators is block--diagonal and coincides with
the adjoint representation in each block. Namely
\begin{equation}
T_\Lambda \, = \, \left ( \matrix { f^{\Sigma}_{\phantom{\Sigma}
\Lambda\Delta} & {\bf 0} \cr {\bf 0} &-
f^{\Sigma}_{\phantom{\Sigma}\Lambda\Delta}
\cr} \right )
\label{aggiungirep}
\end{equation}
Then eq.~\ref{passetto} becomes
\begin{equation}
{\cal P}_\Lambda^0\, = \, e^{\cal K} \, \left ( \, F_\Delta \,
f^\Delta_{\phantom{\Delta}\Lambda\Sigma} \, {\bar X}^\Sigma
\, + \, {\bar F}_\Delta \,
f^\Delta_{\phantom{\Delta}\Lambda\Sigma} \, {  X}^\Sigma \right )
\label{pippopluto}
\end{equation}
\subsection{The triholomorphic momentum map on HyperK\"ahler and
Quaternionic manifolds}
Next I turn to a discussion of
isometries of the manifold ${\cal HM}$ associated with hypermultiplets.
As we know it can be either HyperK\"ahlerian or quaternionic.
For applications to $N=2$ theories we must assume that on ${\cal HM}$
we have an action by triholomorphic isometries of the same
 Lie group ${\cal G}$ that acts on the Special K\"ahler
manifold ${\cal SM}$. This means  that on ${\cal HM}$ we
have Killing vectors
\begin{equation}
\vec k_\Lambda = k^u_\Lambda {\vec \partial\over \partial q^u}
\label{2.71}
\end{equation}
\noindent
satisfying the same Lie algebra  as the corresponding Killing
vectors on ${\cal SM}$. In other words
\begin{equation}
\hat{\vec{k}}_\Lambda ={\hat {\vec k}}_\Lambda =
k^i_\Lambda \vec \partial_i + k^{i^*}_\Lambda
\vec\partial_{i^*} + k_\Lambda^u \vec\partial_u
\label{2.72}
\end{equation}
\noindent
is a Killing vector of the block diagonal metric:
\begin{equation}
\hat g = \left (
\matrix { g_{ij^*} & \quad 0 \quad \cr \quad 0
\quad & h_{uv} \cr } \right )
\label{2.73}
\end{equation}
defined on the product manifold ${\cal SM}\otimes{\cal HM}$.
Triholomorphicity means that the Killing vector fields leave
the HyperK\"ahler structure invariant up to $SU(2)$
rotations in the $SU(2)$--bundle defined by eq.~\ref{su2bundle}.
Namely:
\begin{eqnarray}
{\cal L}_\Lambda K^x& = &\epsilon^{xyz}K^y
W^z_\Lambda \nonumber\\
{\cal L}_\Lambda\omega^x&=& \nabla W^x_\Lambda \nonumber\\
\label{cambicchio}
\end{eqnarray}
where $W^x_\Lambda$ is an $SU(2)$ compensator associated with the
Killing vector $k^u_\Lambda$. The compensator $W^x_\Lambda$ necessarily
fulfils   the cocycle condition:
\begin{equation}
{\cal L}_\Lambda W^{x}_\Sigma - {\cal L}_\Sigma W^x_\Lambda + \epsilon^{xyz}
W^y_\Lambda W^z_\Sigma = f_{\Lambda \Sigma}^{\cdot \cdot \Gamma}
W^x_\Gamma
\label{2.75}
\end{equation}
In the HyperK\"ahler case the $SU(2)$--bundle is flat and the
compensator can be reabsorbed into the definition of the
HyperK\"ahler forms. In other words we can always find a
map
\begin{equation}
{\cal HM} \, \longrightarrow \, L^x_{\phantom{x}y} (q)
\, \in \, SO(3)
\end{equation}
that trivializes the ${\cal SU}$--bundle globally. Redefining:
\begin{equation}
K^{x\prime} \, = \, L^x_{\phantom{x}y} (q) \, K^y
\label{enfantduparadis}
\end{equation}
the new HyperK\"ahler form  obeys the stronger equation:
\begin{equation}
{\cal L}_\Lambda K^{x\prime} \, = \, 0
\label{noncambio}
\end{equation}
On the other hand,
in the quaternionic case,   the non--triviality of the
${\cal SU}$--bundle forbids to eliminate the $W$--compensator
completely. Due to the identification between HyperK\"ahler
forms and $SU(2)$ curvatures eq.~\ref{cambicchio} is rewritten
as:
\begin{eqnarray}
{\cal L}_\Lambda \Omega^x& = &\epsilon^{xyz}\Omega^y
W^z_\Lambda \nonumber\\
{\cal L}_\Lambda\omega^x&=& \nabla W^x_\Lambda \nonumber\\
\label{cambiacchio}
\end{eqnarray}
In both cases, anyhow, and in full analogy with the case of
K\"ahler manifolds, to each Killing vector
we can associate a triplet ${\cal
P}^x_\Lambda (q)$ of 0-form prepotentials.
Indeed we can set:
\begin{equation}
{\bf i}_\Lambda  K^x =
- \nabla {\cal P}^x_\Lambda \equiv -(d {\cal
P}^x_\Lambda + \epsilon^{xyz} \omega^y {\cal P}^z_\Lambda)
\label{2.76}
\end{equation}
where $\nabla$ denotes the $SU(2)$ covariant exterior derivative.
\par
As in the K\"ahler case eq.~\ref{2.76} defines a momentum map:
\be
{\cal P}: {\cal M} \, \longrightarrow \,
\IR^3 \otimes
{\IG }^*
\ee
where ${\IG}^*$ denotes the dual of the Lie algebra
${\IG }$ of the group ${\cal G}$.
Indeed let $x\in {\IG }$ be the Lie algebra element
corresponding to the Killing
vector ${\bf X}$; then, for a given
$m\in {\cal M}$
\be
\mu (m)\,  : \, x \, \longrightarrow \,  {\cal P}_{{\bf X}}(m) \,
\in  \, \IR^3
\ee
is a linear functional on  ${\IG}$.
 If we expand
${\bf X} = a^\Lambda k_\Lambda$ on a basis of Killing vectors
$k_\Lambda$ such that
\be
[k_\Lambda, k_\Gamma]= f_{\Lambda \Gamma}^{\ \ \Delta} k_\Delta
\label{blioprime}
\ee
and we also a choose basis ${\bf i}_x \, (x=1,2,3)$ for $\IR^3$
we get:
\be
{\cal P}_{\bf X}\, = \, a^\Lambda {\cal P}_\Lambda^x \, {\bf i}_x
\ee
Furthermore we need a generalization of the equivariance defined
by eq.~\ref{equivarianza}
\be
{\bf X} \circ {\cal P}_{\bf Y} \,=  \,
{\cal P}_{[{\bf X},{\bf Y}]}
\label{equivarianzina}
\ee
In the HyperK\"ahler case, the left--hand side of eq.~\ref{equivarianzina}
is defined as the usual action of a vector field on a $0$--form:
\begin{equation}
{\bf X} \circ {\cal P}_{\bf Y}\, =  \, {\bf i}_{\bf X} \, d
{\cal P}_{\bf Y}\, = \,
X^u \, {\o{\partial}{\partial_u}} \, {\cal P}_{\bf Y}\,
\end{equation}
The equivariance condition   implies
that we can introduce a triholomorphic Poisson bracket defined
as follows:
\begin{equation}
\{{\cal P}_\Lambda, {\cal P}_\Sigma\}^x \equiv 2 K^x (\Lambda,
\Sigma)
\label{hykapesce}
\end{equation}
leading to the triholomorphic Poissonian realization of the Lie
algebra:
\begin{equation}
\left \{ {\cal P}_\Lambda, {\cal P}_\Sigma \right \}^x \, = \,
f^{\Delta}_{\phantom{\Delta}\Lambda\Sigma} \, {\cal P}_\Delta^{x}
\label{hykapescespada}
\end{equation}
which in components reads:
\begin{equation}
K^x_{uv} \, k^u_\Lambda \, k^v_\Sigma \, = \, {\o{1}{2}} \,
f^{\Delta}_{\phantom{\Delta}\Lambda\Sigma}\, {\cal P}_\Delta^{x}
\label{hykaide}
\end{equation}
In the quaternionic case, instead, the left--hand side of
eq.~\ref{equivarianzina}
is interpreted as follows:
\begin{equation}
{\bf X} \circ {\cal P}_{\bf Y}\, =  \, {\bf i}_{\bf X}\,  \nabla
{\cal P}_{\bf Y}\, = \,
X^u \, {\nabla_u} \, {\cal P}_{\bf Y}\,
\end{equation}
where $\nabla$ is the $SU(2)$--covariant differential.
Correspondingly, the triholomorphic Poisson bracket is defined
as follows:
\begin{equation}
\{{\cal P}_\Lambda, {\cal P}_\Sigma\}^x \equiv 2 K^x (\Lambda,
\Sigma)  - {\o{1}{\lambda}} \, \varepsilon^{xyz} \,
{\cal P}_\Lambda^y  \, {\cal P}_\Sigma^z
\label{quatpesce}
\end{equation}
and leads to the Poissonian realization of the Lie algebra
\begin{equation}
\left \{ {\cal P}_\Lambda, {\cal P}_\Sigma \right \}^x \, = \,
f^{\Delta}_{\phantom{\Delta}\Lambda\Sigma} \, {\cal P}_\Delta^{x}
\label{quatpescespada}
\end{equation}
which in components reads:
\begin{equation}
K^x_{uv} \, k^u_\Lambda \, k^v_\Sigma \, - \,
{\o{1}{2\lambda}} \, \varepsilon^{xyz} \,
{\cal P}_\Lambda^y  \, {\cal P}_\Sigma^z\,= \,  {\o{1}{2}} \,
f^{\Delta}_{\phantom{\Delta}\Lambda\Sigma}\, {\cal P}_\Delta^{x}
\label{quatide}
\end{equation}
Eq.~\ref{quatide} which is the most convenient way of
expressing equivariance in a coordinate basis plays a fundamental
role in the construction of the supersymmetric action, supersymmetry
transformation rules and of the superpotential for $N=2$ supergravity
on a general quaternionic manifold. It is also very convenient
to retrieve the rigid supersymmetry limit. Indeed, when $\lambda \,
\to \, \infty$ eq.~\ref{quatide} reduces to eq.~\ref{hykaide}.
Eq.~\ref{quatide} was introduced in the physical literature in
\cite{specnonspec2} where the general form of $N=2$ supergravity
beyond the limitations of tensor calculus was given.
\subsection{The $N=2$ supergravity lagrangian}
Using the concepts and the geometric structures introduced
so far the form of the bosonic lagrangian for $N=2$ supergravity
anticipated in eq.s~\ref{gausugra} and ~\ref{filastrocca} is
now explained. In particular the scalar potential ${\cal V}$
is expressed, as one sees, in terms of the Killing vectors and
of the momentum map functions. Indeed, differently from the
$N=1$ theory, in $N=2$ supergravity the scalar potential is
entirely due to the gauging of the theory, no superpotential
being available without gauging.
\par
Fermions are not discussed in these lectures, yet it is
a legitimate curiosity to ask what happens to them in the gauging.
Something very simple. Fermions behave as sections of the bundle
${\cal L} \otimes {\cal TSM}$ in the gaugino case, as sections
of the bundle ${\cal THM}\otimes {\cal SU}^{-1}$ in the hyperino
case and as sections of the bundle ${\cal L}\otimes{\cal SU}$ in
the gravitino case. Correspondingly the covariant derivatives
of the fermions appearing in the action and in the transformation
rules involves the composite connections ${\cal Q}$ ,
$\Gamma^{i}_{\phantom{i}j}$, $\omega^x$
and $\Delta^{\alpha\beta}$ defined on these bundles.
Gauging just modifies these composite connections by means of
Killing vectors and momentum map functions. Explicitly we
have:
\begin{eqnarray}
{\cal TSM} & : & \mbox{tangent bundle} ~: \nonumber\\
 \Gamma^{i}_{\phantom{i}j}& \to & \Gamma^{i}_{\phantom{i}j} +
 g\, A^\Lambda\, \partial_j k^i_\Lambda \nonumber\\
{\cal L} & : & \mbox{line bundle} ~: \nonumber\\
{\cal Q} &\to & {\cal Q} + g\, A^\Lambda\, {\cal
P}^0_\Lambda \nonumber \\
{\cal SU} & : & \mbox{$SU(2)$ bundle} ~: \nonumber\\
\omega^x &\to & \omega^x + g\, A^\Lambda\, {\cal
P}^x_\Lambda \nonumber \\
{\cal SU}^{-1}\otimes{\cal THM} & : & \mbox{$Sp(2m)$ bundle} ~: \nonumber\\
\Delta^{\alpha\beta} &\to & \Delta^{\alpha\beta}\nonumber\\
&& + g\, A^\Lambda\,
 \partial_u k_\Lambda^v \, {\cal U}^{u \vert  \alpha A}
 \, {\cal U}^\beta_{v \vert A}\nonumber\\
\label{compogauging}
\end{eqnarray}
\section{An example of non--perturbative rigid special
geometry: $SU(r+1)$ gauge theories}
\label{LL9}
As an example of non--perturbative special geometry
and of the associated electric--magnetic duality
rotations, I choose the example of rigid  $N=2$ Yang--Mills
theories for the gauge group $SU(r+1)$.
\par
Here the essential idea that  allows for the non--perturbative
solution of the model is the introduction of an auxiliary
{\it dynamical Riemann surface}. The moduli space of the gauge theory
is identified with the moduli space of such a surface.
\par
This is the realization of the correspondence already anticipated in
eq.s~\ref{rieperiod}. There we conceived the possibility that
the components of the derivative
$
{\hat U}_i   =   \partial_i {\hat \Omega}   \equiv
\twovec{f^{\alpha}_{i} }{h_{\beta\vert i}}
$ of the holomorphic symplectic section ${\hat \Omega}$ leading to rigid
special geometry might be  regarded as the periods of holomorphic
differentials on a Riemann surface. The complete moduli--space
of genus $g$ Riemann surface is not a rigid special manifold. Yet
a sublocus of this moduli space has such a property. The sublocus
is spanned by the surfaces solving the dynamical problem under
consideration.
\par
In this lecture I summarize
the results obtained for pure N=2
gauge theories without hypermultiplets coupling in
\cite{SW1,kltold,faraggi,kltnew}. My presentation follows~\cite{cynoi}.
For the N=2 {\it microscopic
gauge theory} of a group $G$, with Lie algebra $\IG$
the rigid special geometry is encoded in a
``minimal coupling" quadratic prepotential
of the form:
\begin{eqnarray}
{\cal F}^{(micro)}(Y)&=&g_{IJ}^{(K)} \, Y^I \,
Y^J\nonumber\\
g_{IJ}^{(K)}&=&\mbox{Killing metric on}~{\IG}
\label{mincoup_1}
\end{eqnarray}
This choice is motivated by renormalizability, positivity of
the energy and canonical normalization of the kinetic terms.
It is also motivated by the need to have a continuous group
of isometries where to embed the gauge group, as I have already
explained.
Consider next the effective lagrangian describing the dynamics
of the massless modes. This  is an abelian
 N=2 gauge theory that admits the maximal torus
${H}\subset{G}$
as new gauge group and is based on a new rigid special geometry:
\begin{eqnarray}
{\cal F}^{(eff)}(Y)&=&g_{\alpha\beta}^{(K)} \, Y^\alpha \,
Y^\beta + \Delta{\cal F}^{(eff)}\left ( Y^\alpha \right)\nonumber\\
Y^\alpha&\in& {\IH} \subset {\IG}
\end{eqnarray}
In general the effective prepotential ${\cal F}^{(eff)}(Y)$ has
a transcendental dependence on the scalar fields $Y^\alpha$
of the Cartan subalgebra multiplets, due to the
correction $\Delta{\cal F}^{(eff)}\left ( Y^\alpha \right)$.
The main problem is the determination of this correction.
Perturbatively one can get information on
$\Delta{\cal F}^{(eff)}\left ( Y^\alpha \right)$ and discover its
logarithmic singularity for large values of the scalar fields
$Y^\alpha$.
In particular one has a logarithmic correction to the gauge coupling
{\it period matrix}
\begin{eqnarray}
&\Delta {\bar {\cal N}}_{\alpha \beta}=
{{\partial}^2 \over{ \partial Y^\alpha \partial Y^\beta}}
\Delta {\cal F}\ & \nonumber \\
&\stackrel{Y \to \infty}\sim \ \sum_{\bf\alpha} \
{\bf\alpha}^\alpha {\bf\alpha}^\beta
\log{(Y \cdot {\bf\alpha})^2\over \Lambda^2}&
\end{eqnarray}
where ${\bf\alpha}$ are the root vectors of the gauge Lie algebra and
$\Lambda^2$ is the dynamically generated scale.
The perturbative monodromy following from
\begin{eqnarray}
{\cal N}_{\alpha \beta} &\rightarrow &
[(C+ D{\cal N})(A + B{\cal N})^{-1}]_{\alpha\beta} \nonumber \\
Sp(2r, \IR) &\ni & \twomat{A}{B}{C}{D} \nonumber\\
 & \sim &
\twomat{1}{0}{ \sum_{\bf\alpha} {\bf\alpha}^\alpha {\bf\alpha}^\beta}
{1}
\end{eqnarray}
is assumed to be a part of the monodromy group
of a genus $r$ Riemann surface having a symplectic action on
 the periods
of the surface. Guessing such a dynamical Riemann surface gives
the nonperturbative structure $\Delta {\cal F}^{(eff.)}$.
\par
Denoting by $r$ the rank of the original
gauge group ${G}$, one derives the structure of the
effective gauge theory of the maximal torus ${\IH}$ from
the geometry of an $r$--parameter family ${\cal M}_{1}[r]$
of dynamical genus $r$ Riemann surfaces. The essential steps
of the procedure are as follows: naming $u_i$
(i=1,\dots\,r) the $r$ gauge invariant moduli of the family,
(described as the vanishing locus of an appropriate polynomial)
one makes the identifications:
\begin{eqnarray}
u_i &\to &\langle \, d_{\alpha_1\dots\alpha_{i+1}} Y^{\alpha_1}
\,\dots\, Y^{\alpha_{i+1}} \rangle \nonumber\\
&&(i=1,\dots, r)
\label{rie_1}
\end{eqnarray}
where $Y^{\alpha}$ are the special coordinates of rigid special
geometry and $d_{\alpha_1\dots\alpha_{i+1}}$ is the restriction
to the Cartan subalgebra of the rank $i+1$ symmetric tensor
defining the $(i+1)$--th Casimir operator. The identification
in eq.~\ref{rie_1} is only an asymptotic equality for large values
of $u_i$ and $Y^{\alpha}$; at finite values, the relation between
the moduli $u_i$ and the special coordinates
(namely the elementary fields appearing in the
lagrangian) is much more complicated. One considers the
derivatives :
\begin{equation}
\Omega_{u_i}\,{\stackrel{\rm def}{=}}\,
\partial_{u_i} \,\Omega\, =\,
\partial_{u_i}\left ( \matrix { Y^{\alpha}\cr
{\o{\partial {\cal F}}{\partial Y^{\alpha}}}\cr} \right)
\label{rie_2}
\end{equation}
where  $\Omega (u_i)$ is a section
of the flat $Sp(2r,\IR)$
holomorphic vector bundle  whose existence
is encoded in the definition of rigid special K\"ahler geometry.
This is eq.~\ref{uvectorrig} written in a special coordinate basis.
On one hand the K\"ahler metric is given by the
general formula of eq.~\ref{sympinvrig}.
 On the other hand, one identifies the symplectic vectors
$\Omega_{u_{i}}$
with the period vectors:
\begin{equation}
 \Omega_{u_{i}}~=~\left ( \matrix {\int_{A^\alpha} \,
\omega^{i}\cr \int_{B^\alpha}\omega^{i}\cr} \right )
\quad (i=1,\dots\, r=\mbox{ genus})
\label{rie_3bis}
\end{equation}
of the $r$ holomorphic 1-forms $\omega^{i}$
along a canonical homology basis, defined as in eq.~\ref{intsecrie},
 of a genus $r$ dynamical Riemann surface ${\cal M}_{1}[r]$.
The generic moduli space $M_r$ of genus $r$ surfaces is $3r-3$
 dimensional.
The dynamical Riemann surfaces ${\cal M}_1[r]$ fill an
$r$-dimensional sublocus $L_R[r]$.
The problem is that of characterizing intrinsically this locus.
Let
\begin{equation}
i \ : \ L_R[r] \rightarrow M_r
\end{equation}
be the inclusion map of the wanted locus and let
\begin{equation}
\label{hb}
{H}\stackrel{\pi}{\rightarrow} M_r
\end{equation}
be the Hodge bundle on $M_r$, that is the rank $r$ vector bundle whose
sections are the holomorphic forms on the Riemann surface
$\Sigma_r\in M_r$. As fibre metric on
this bundle one can take the imaginary part of the period matrix:
\begin{equation}
\label{imn}
{\rm Im}\, {\cal N}_{\alpha\beta} =\int_{\Sigma_r} \omega^{\alpha}\wedge
{\bar \omega}^{\beta^*}
\end{equation}
where $\omega^\alpha$ is a basis holomorphic one-forms. The locus $L_R[r]$
is defined by the following equation:
\begin{equation}
\label{boh}
i^*\partial \bar\partial ||\omega||^2=i^*{  K}
\end{equation}
where $||\omega||^2$ $=\int_{\Sigma_r}\omega\wedge{\bar\omega}$ is the norm
of any section of the Hodge bundle and ${\cal K}$ is the K\"ahler class
of $M_r$.
\par
Using very general techniques of algebraic geometry, the dependence
of the periods (see eq.s\ref{rie_2}) on the moduli parameters can be
determined through the solutions of the Picard--Fuchs differential
system, once ${\cal M}_{1}[r]$ is  explicitly described
as the vanishing locus of a holomorphic superpotential
${\cal W}(Z,X,Y;u_i)$. In particular one can study the monodromy
group $\Gamma_M$ of the differential system
and the symmetry group of the potential
$\Gamma_{\cal W}$, that are related to the full group of
{\it electric--magnetic  duality rotations} $\Gamma_D$ as follows:
\begin{equation}
\Gamma_{\cal W}~=~\Gamma_D/\Gamma_M
\label{rie_5}
\end{equation}
The elements of ${\Gamma_D}\supset\Gamma_M$ are given by integer
valued
symplectic matrices $\gamma \in Sp(2r,\ZZ)$
that act on the symplectic
section $\Omega$. Given the geometrical interpretation
of these sections provided by eq.s~\ref{rie_3bis},
the elements $\gamma \in
\Gamma_D \subset Sp(2r,\ZZ)$ correspond to changes of the
canonical homology basis respecting the intersection matrix
in eq.~\ref{intsecrie}.
\par
To be specific we mention the results obtained for the gauge
groups ${G}=SU(r+1)$. The rank $r=1$ case, corresponding
to ${G}=SU(2)$, was studied by Seiberg and Witten in their
original paper \cite{SW1}. The extension to the general case,
with
particular attention devoted to the $SU(3)$ case, was
obtained in \cite{kltold,faraggi}. In all these cases the
dynamical Riemann
surface ${\cal M}_{1}[r]$ belongs to the
hyperelliptic locus of genus $r$ moduli space,
the general form of a hyperelliptic surface
being described (in inhomogeneous coordinates)
by the following algebraic equation:
\begin{equation}
w^2~=~P_{(2+2r)} (z)~=~\prod_{i=1}^{2 +2r} \, (z - \lambda_i)
\label{ipere}
\end{equation}
where $\lambda_i$ are the $2+2r$ roots of a degree $2+2r$
polynomial. The hyperelliptic locus
\begin{equation}
L_H [r]\subset M_r\quad ,\quad \mbox{dim}\,  L_H[r]=2r-1
\label{hlocus}
\end{equation}
is a closed submanifold of codimension
$r-2$ in the $3r-3$ dimensional moduli space of genus
$r$ Riemann surface\footnote{For genus 1, the moduli space is also
1--dimensional and the hyperelliptic locus is the full moduli
space.}. The $2r-1$ hyperelliptic moduli
are the $2r+2$ roots of the polynomial appearing in
eq.~\ref{ipere}, minus three of them that
can be fixed at arbitrary points by means of fractional
linear transformations on the variable $z$.
Because of their definition, however, the dynamical Riemann
surfaces ${\cal M}_1[r]$, must have $r$ rather than
$2r-1$ moduli. We conclude that the $r$--parameter
family  ${\cal M}_{1}[r]$ fills a locus $L_R[r]$ of
codimension $r-1$ in the hyperelliptic locus:
\begin{eqnarray}
&L_R[r]\subset L_H [r],\hskip 0.4cm
\mbox{codim}\, L_R[r]=r-1 & \nonumber \\
& \hskip 0.4cm
\mbox{dim} L_R[r]=r &
\label{dynlocus}
\end{eqnarray}
This fact is expressed by additional conditions imposed on
the form of the degree $2+2r$ polynomial of
eq.\ref{ipere}. In references \cite{kltold,faraggi}
$P_{(2+2r)} (z)$ was determined to be of the following form:
\begin{equation}
\label{squadra}
 P_{(2+2r)} (z) =  P_{(r+1)}^2 (z)  -
P_{(1)}^2(z)
\end{equation}
where $P_{(r+1)}(z)$ and $P_{(1)}(z)$ are two polynomials
respectively of degree $r+1$ and $1$. Altogether we have
$r+ 3$ parameters that we can identify with the $r+1$
roots of $P_{(r+1)}(z)$ and with the two coefficients
of $P_{(1)}(z)$
\begin{eqnarray}
&P_{(r+1)}(z)~= \prod_{i=1}^{r+1} \, (z - \lambda_i)
& \nonumber\\
& P_{(1)}(z)~=~\mu_1 \, z + \mu_0  &
\end{eqnarray}
Indeed, since the polynomial in eq.~\ref{squadra}  must be effectively
of order $ 2+2r$, the highest order coefficient of
$P_{(r+1)}(z)$ can be fixed to $1$ and the only independent
parameters contained in $P_{(r+1)}(z)$ are the roots. On the
other hand, since $P_{(1)}(z)$ contributes only subleading powers, both
of its coefficients $\mu_1$ and $\mu_0$ are effective
parameters. Then, if we take into
account fractional linear transformations,
three {\it gauge fixing} conditions can be imposed
on the $r+3$ parameters $\{ \lambda_i\},
\{\mu_i\}$. In ref. (\cite{kltold,faraggi}) this freedom was
used to set:
\begin{eqnarray}
\sum_{i=1}^{r+1} \, \lambda_i&=&0 \nonumber\\
\mu_1&=&0\nonumber\\
\mu_0&=&\Lambda^{r+1}
\label{lasceltadiwolf}
\end{eqnarray}
where $\Lambda$ is the dynamically generated  scale. With
this choice the $r$--parameter family of dynamical
Riemann surfaces is described by the equation:
\begin{equation}
w^2=\left ( z^{r+1} - \sum_{i=1}^{r} \, u_i \left (\lambda
\right )
\, z^{r-i} \right )^2 \, - \, \Lambda^{2r+2}
\label{wolf}
\end{equation}
where the coefficients
\begin{equation}
 u_i \left (\lambda_1, \dots ,\lambda_{r+1} \right )
\qquad\qquad (i=1,\dots \, r)
\end{equation}
are symmetric functions of the $r+1$ roots constrained by the first
of eq.s~\ref{lasceltadiwolf} and can be identified with the
moduli parameters introduced in eq.~\ref{rie_1}. In the particular
case $r=1$, the gauge--fixing of eq.~\ref{lasceltadiwolf} leads to
the following quartic form for the elliptic curve studied in \cite{SW1}
\begin{equation}
w^2=\left ( z^2 - u \right )^2 - \Lambda^{4}=
z^4 - 2\, u \, z^2 + u^2 - \Lambda^4
\label{inhomseiwit}
\end{equation}
Of course other gauge fixings give equivalent descriptions of
${\cal M}_1[r]$; however, for our next purposes, it is particularly
important to choose a
gauge fixing of the $SL(2,\IC)$ symmetry such that the equation
${\cal M}_1[r]$ can be recast in the form of a Fermat polynomial
in a weighted projective space deformed by the marginal operators of
its chiral ring. In this way it is quite easy to study the symmetry
group
of the potential $\Gamma_{\cal W}$
identifying the $R$-symmetry group and to
derive the explicit form of the Picard-Fuchs equations satisfied by
the
periods. This is relevant for the embedding of the monodromy
and $R$-symmetry groups in $Sp(2r,\ZZ)$. The alternative gauge-fixing
that we choose is the following:
\begin{eqnarray}
\sum_{i=1}^{r+1} \, \lambda_i&=&0 \nonumber\\
\mu_1 \, \mu_0 + \,
\left ( \sum_{i=1}^{r+1} \, {\o{1}{\lambda_i}} \right )
\prod_{i=1}^{r+1} \, \lambda_i^2 \,
&=&0\nonumber\\
-\mu_0^2 + \prod_{i=1}^{r+1} \, \lambda_i^2&=&1
\label{lamiascelta}
\end{eqnarray}
To appreciate the convenience of this choice
 let us consider the general inhomogeneous form of
the equation of the hyperelliptic surface eq.~\ref{squadra}
and let us (quasi-)homogenize it by setting:
\begin{equation}
\begin{tabular}{ll}
$w={\o{ Z}{Y^{r+1}}}$&$z={\o{ X}{Y}}$ \ .
\end{tabular}
\label{omogeneizzazione}
\end{equation}
With this procedure eq.\ref{squadra} becomes a quasi--homogeneous
polynomial constraint:
\begin{eqnarray}
0&=&{\cal W}\left ( Z,X,Y;\{\lambda\},\{ \mu \} \right )\nonumber\\
{}~&=& - \, Z^2 + \left ( \prod_{i=1}^{r+1} \,
\left ( X - \lambda_i  \, Y\right ) \right )^2 \nonumber\\
&& -
\left ( \mu_1 \, X \, Y^{r} + \mu_0 \, Y^{r+1} \right )^2
\label{homog_1}
\end{eqnarray}
of degree:
\begin{equation}
 \mbox{deg} \, {\cal W}~=~2r+2
\label{gradino}
\end{equation}
in a weighted projective space $W\IC\IP^{2;r+1,1,1}$, where
the quasi--homogeneous coordinates $Z$, $X$, $Y$ have degrees
$r+1$,$1$ and $1$, respectively. Adopting the notations of
\cite{kimetal}, namely denoting by\footnote{Note the difference of
notation: $W\IC\IP^{n;q_1,q_2,\dots,q_{n+1}}$ is the full weighted
projective space, in which eq.~\ref{pesataequazione} is a hypersurface.}
\begin{equation}
WCP^n(d;q_1,q_2,\dots,q_{n+1})_{\chi}
\label{pesataequazione}
\end{equation}
the zero locus (with Euler number $\chi$)
of a quasi--homogeneous polynomial of degree $d$ in an
$n$--dimensional weighted projective space, whose  $n+1$
quasi--homogeneous coordinates have weights $q_1$,$\dots$,$q_{n+1}$:
\begin{eqnarray}
&{\cal W}\left ( \lambda^{q_1} \, X_1,\dots\,
\lambda^{q_{n+1}} \, X_{n+1}\right ) \,= & \nonumber \\
& \lambda^d \,
{\cal W}\left ( X_1,\dots , X_{n+1} \right )&
\label{rie_8}
\end{eqnarray}
we obtain the identification:
\begin{equation}
{\cal M}_{1}[r] ~=~WCP^2(2r+2;r+1,1,1)_{2(1-r)}
\label{agnizione}
\end{equation}
that yields, in particular:
\begin{eqnarray}
{\cal M}_{1}[1] &= &WCP^2(4;2,1,1)_{0}  \nonumber \\
{\cal M}_{1}[2] &= &WCP^2(6;3,1,1)_{2}
\label{cognizione}
\end{eqnarray}
for the $SU(2)$ case studied in
\cite{SW1} and for
the $SU(3)$ case studied in \cite{kltold,faraggi}.
Using the alternative gauge fixing in eq.~\ref{lamiascelta}, the
quasi--homogeneous Landau--Ginzburg superpotential of eq.~\ref{homog_1},
whose vanishing locus defines the dynamical Riemann surface, takes
the standard form of {\it  a Fermat superpotential} deformed
by the marginal operators of its {\it chiral ring}:
\begin{eqnarray}
&{\cal W}\left ( Z,X,Y;\{\lambda\},\{ \mu \} \right ) = & \nonumber \\
& - Z^2 + X^{2r+2} + Y^{2r+2} & \nonumber \\
&+\sum_{i=1}^{2r-1} \, v_i \left ( \lambda \right ) \,
X^{2r+1-i} \, Y^{i+1}&
\label{fermatform}
\end{eqnarray}
The coefficients $v_i \, \quad (i=1,\dots\, 2r-1)$ are the
$2r-1$ moduli of a hyperelliptic curve. In our case, however,
they are expressed as functions of the $r$ independent
roots $\lambda_i$ that remain free after the gauge--fixing
of eq.~\ref{lamiascelta} is imposed.  The coefficients $v_i$
have a simple expression as symmetric functions of
the $r+1$ roots $\lambda_i$ subject to the constraint
that their sum should vanish. For the general form
of these expressions we refer the reader to \cite{cynoi}.
 In particular for the first case $r=1$ we obtain
\begin{eqnarray}
&{\cal M}_1[1] ~\hookrightarrow ~& \nonumber\\
& 0 = {\cal W}\left ( Z,X,Y;v=2u\right )~=~& \nonumber \\
&-Z^2 + X^4 + Y^4 +  v (\lambda ) \, X^2 Y^2 & \label{mamma}
\end{eqnarray}
where
\begin{eqnarray}
& \lambda_1 + \lambda_2 =0 & \nonumber\\
& \mu_1=0 & \nonumber\\
& \mu_0=\sqrt{\lambda_1^2 \lambda_2^2 - 1} & \nonumber\\
& v = \lambda_1^2 + \lambda_2^2 +4\lambda_1 \, \lambda_2
{}~=~& \nonumber \\
&-2\, \lambda_1^2\,
 {\stackrel{\rm def}{=}}\, 2 \, u&
\label{seiwitequa}
\end{eqnarray}
 Alternatively, using as independent parameters the coefficients
$u_i(\lambda)$ appearing in eq.~\ref{wolf}, we can characterize
the locus $L_R[r]$ of dynamical Riemann surfaces
by means of the following
equations on the hyperelliptic moduli $v_i$:
\begin{eqnarray}
\label{n2}
v_k &=& -2 u_k + \sum_{i+j = k -1} u_i u_j, \quad ( k = 1,\dots,r)
\nonumber\\
v_{r+k} &=& \sum_{i+j = r+k-1}u_i u_j - \delta_{r-1,k} \mu_1^2
\end{eqnarray}
Considering now the Hodge filtration of the middle cohomology group
$H^{(1)}_{DR}({\cal M}_1[r])$:
\begin{eqnarray}
&{\cal F}^{0} \,  \subset  \, {\cal F}^{1}\, & \nonumber\\
&{\cal F}^{0}~=~H^{(1,0)}& \nonumber \\
& {\cal F}^{1}\equiv H^{(1)}_{DR} = H^{(1,0)} +
H^{(0,1)} \nonumber\\
\label{filtraggio}
\end{eqnarray}
the Griffiths residue map (\cite{griffith,petropaolo})
provides an association between elements of ${\cal F}^{k}$
and  polynomials $P^{I}_{k|(2r+2)}(X)$
of the chiral ring ${\cal R} ({\cal W} )~{\stackrel{\rm def}{=}}~
 \IC [X]/\partial {\cal W}$ of degree $k|(2r+2) \equiv
(k+1)(2r+2)-r-3$
according to the following pattern:
\begin{equation}
\begin{tabular}{llll}
$\mbox{cohom.}$&$\mbox{deg}$&$\mbox{polynom.}$&$~$\\
$~$&$~$&$~$&$~$\\
${\cal F}^0$&$r-1$&$P^i_{0|(2r+2)}~$&$i=1,\ldots,r$\\
${\cal F}^1$&$3r+1$&$P^{i^*}_{1|(2r+2)}$&$i^*=1,\ldots,r$
\end{tabular}
\end{equation}
Explicitly, the periods of eq.~\ref{rie_3bis}
are represented by:
\begin{eqnarray}
\label{n3}
\int_C \omega^i &=& \int_C {X^{r-i} Y^{i -1}\over {\cal W}}\, \omega
\nonumber\\
\int_C \omega^{i^*} &=& \int_C {X^{r+i} Y^{2r- i + 1}\over {\cal W}^2}
\,\omega
\end{eqnarray}
where $C$ denotes any of the homology cycles and
$\omega= Z\, \dop X\wedge
\dop Y + {\rm cycl}$. Using standard reduction techniques
\cite{petropaolo}
one can obtain the first-order Picard-Fuchs differential system
\begin{equation}
\label{n4}
\left({\partial\over\partial v^I} \bfone - A_I(v)\right) V=0
\hskip 0.5cm I = 1,\ldots 2r-1
\end{equation}
satisfied by the $2r$-component vector:
\begin{equation}
\label{period}
V =
\left ( \matrix { \int_C \omega^i\cr
\int_C \omega^{i^*} \cr} \right ) \label{V2r}
\end{equation}
in the $2r-1$-dimensional moduli space of elliptic surfaces.
Using the explicit embedding of the locus $L_R[r]\subset L_H[r]$
described by eq.s~\ref{n2}, we obtain the Picard-Fuchs
differential system of rigid special geometry by a trivial
pull-back of eq.~\ref{n4}:
\begin{equation}
\label{n5}
\left({\partial\over\partial u^i} \bfone - A_I(v){\partial v^I\over
\partial u^i} \right) V=0.
\label{picf}
\end{equation}
The explicit solution of the Picard--Fuchs
equations  for $r=1,2$ has been given respectively in \cite{CDF,kltnew}.
The solution of the Picard--Fuchs equations for generic $r$ determines in
principle the period of the surface and the monodromy group.
I do not attempt to solve eq.s~\ref{picf} for generic $r$, I will
mostly concentrate on the $r=1$ case as a pedagogical example.
\par
Before plunging into the details of the $r=1$ case I want
to discuss the symmetry
group of $\cM_1(r)$, which, together with the monodromy group
$\G_M$ defines
the duality group $\G_D$ according to eq.~\ref{rie_5}.
This symmetry group can be defined by considering those linear
transformations ${\bf X} \, \to \, M_A{\bf X}$ of
the quasi--homogeneous
coordinate vector ${\bf X}=(X,Y,Z)$ such that
\begin{equation}
{\cal W}(M_A{\bf X};v)=f_A(v) {\cal W}({\bf X} ;\phi_A(v))
\label{diedro_2}
\end{equation}
where $\phi_A(v)$ is a (generally non--linear) transformation
of the moduli and $f_A(v)$ is a compensating  overall
rescaling of the superpotential that depends both on the moduli $v$
and on the chosen transformation $A$.
Let us restrict our attention to the sublocus  of
the dynamical Riemann surface ${\cal W}({\bf X};u)=0$
defined by eq.~\ref{dynlocus}, so that the moduli
space geometry is a special K\"ahler geometry with K\"ahler potential
\begin{equation}
K = {\rm i} (Y^\alpha \bar\cF_\alpha-\bar Y^\alpha\cF_\alpha)\ .
\end{equation}
In this case, only the subgroup
$\Gamma_{\cal W}^0 \subset \Gamma_{\cal W}$
given by the transformations
that have a compensating rescaling factor $f(u)=e^{i\ \theta}$
acts as an isometry  group for the moduli space,
in contrast with curved
special geometry, where the whole $\Gamma_{\cal W}$ generates
isometries.
The hyperelliptic superpotential in eq.~\ref{fermatform} admits a
$\Gamma_{\cal W}^0$ symmetry
group which is isomorphic to the dihedral group $D_{2r+2}$, defined
by the following relations on two generators $A,B$:
\begin{equation}
\label{n6}
A^{2r+2} = \bfone \hskip 0.5cm ; \hskip 0.5cm B^2 = \bfone
\hskip 0.5cm ; \hskip 0.5cm (AB)^2 = \bfone.
\end{equation}
The action of the generators on the moduli is the following.
Let $\alpha^{2r+2}=1$ be a $(2r+2)^{\rm th}$ root of the unit and
let the moduli $v_i$ be arranged into a column vector ${\bf v}$.
Then we have:
\begin{eqnarray}
\label{ab}
{\bf v}^\prime = A {\bf v}, \hskip 1cm
A = \left(\matrix{\alpha^2 & 0 & \ldots & 0 \cr 0 & \alpha^3 & \ldots & 0
\cr \vdots & \vdots & \ddots & \vdots \cr 0 & 0 & \ldots & \alpha^{2r}}
\right)\nonumber\\
{\bf v}^{\prime\prime} = B {\bf v}, \hskip 1cm
B = \left(\matrix{0 & 0 & \ldots & 1
\cr \vdots & \vdots & \ddots & \vdots
\cr 0 & 1 & \ldots & 0 \cr 1 & 0 & \ldots & 0}\right)
\end{eqnarray}
For the transformations $A$ and $B$ the compensating transformations
on the homogeneous coordinates $M_A$ and $M_B$ are
\begin{eqnarray}
M_A &=& \left ( \matrix{\alpha & 0 & 0 \cr 0 &1 &0 \cr 0& 0&1 }\right )
\nonumber \\
M_B &=& \left ( \matrix{0 &1 &0 \cr  1 & 0 &0 \cr 0 &0 &1 }\right )
\end{eqnarray}
Consequently the differential Picard--Fuchs system for the period
(see eq.~\ref{period}) of the
generic hyperelliptic surface has a $\Gamma_{\cal W}^0=D_{2r+2}$
symmetry as defined above
and the generators $A$ and $B$ act by means of
suitable $Sp(2 r, \ZZ)$ matrices on the period vector given by
eq.\ref{rie_3bis}. However  eq.s~\ref{picf}
are invariant only under the cyclic subgroup $\ZZ_{2r +2}
\subset D_{2r+2}$ generated by $A$.
Hence the potential $\tilde {\cal W}(u)={\cal W}\left(v(u)\right)$
of the $r$-dimensional locus
$L_R[r]$ of dynamical Riemann surfaces and the Picard-Fuchs first
order system admits only the duality symmetry  $\Gamma_{\tilde{\cal W}}^0=
\ZZ_{2r + 2}$.
\par
The physical interpretation of this group is R-symmetry.
Indeed, recalling eq.~\ref{rie_1} we see that when each
of the elementary fields $Y^\alpha$ appearing in the lagrangian
is rescaled as $Y^\beta \to \alpha Y^\beta$, then the first $u_i$ moduli
are rescaled with the powers of $\alpha$ predicted by eq.~\ref{rie_1}.
According to the analysis of reference \cite{noi} this is precisely the
requested R-symmetry for the
topological twist. All the scalar components of the
vector multiplets have the same R-charge ($q_R=2$) under a
$U_R(1)$ symmetry of the classical action,
which is broken to a discrete subgroup
in the quantum theory. Henceforth the integer symplectic matrix
that realizes $A$ yields the R-symmetry matrix of rigid special
geometry for $SU(r+1)$ gauge theories.
An important problem is the derivation of the corresponding
R-symmetry matrix in $Sp(2r +4, \ZZ)$, when the gauge theory
is made locally supersymmetric by coupling
it to supergravity including also
the dilaton-axion vector multiplet suggested by string theory.
\par
Another very important allied topic to electro--magnetic dualities
is indeed that of topological field theories: a topic that possibly
captures their most profound meaning. Yet within the scope of
these lectures I cannot open this Pandora's vase and therefore
I rather turn to discuss the details of Seiberg Witten
example of rigid special geometry.
\par
For convenience of later normalizations
let us rewrite eq.~\ref{mamma} in the equivalent form:
\begin{eqnarray}
&0~=~{\cal W}(X,Y,Z;u)~=~& \nonumber \\
& -Z^2  +
{\o {1}{4}}
\left ( X^4 + Y^4 \right )
   +   {\o{u}{2}}  X^2 Y^2&
\label{diedro_1}
\end{eqnarray}
One realizes that this potential has a
$\Gamma_{\cal W}=D_3$ symmetry group \cite{giveon,lopez}
defined by the following generators and relations
\begin{equation}
{\hat A}^2= \bfone\quad , \quad C^3=\bfone \quad, \quad
(C\hat  A)^2 =\bfone
\end{equation}
with the following action on the homogeneous coordinates
and the modulus $u$\footnotemark \footnotetext{We forget about
the action on the $Z$
coordinate, which is immaterial, since it contributes only
with a quadratic term to the polynomial  ${\cal W}$}:
\begin{equation}
  \begin{array}{cc}
{\hat A}\, :&\cases{ M_{\hat A} =
\left (\matrix {i & 0 \cr 0 & 1 \cr}
\right ) \cr  \phi_{\hat A} (u)\, = \, - u \cr
f_{\hat A}(u)\,
= \, 1 \cr } \\
{ C}\, :&\cases { M_C = {\o{1}{\sqrt{2}}} \,
\left (\matrix {i & 1 \cr -i & 1 \cr}
\right ) \cr
\phi_C (u)\, = \,{\o{u-3}{u+1}} \cr
  f_C(u)\, = \,
{\o{1+u}{2}}\cr } \\
\end{array}   \\
\label{diedro_5}
\end{equation}
Eq.~\ref{diedro_5} is given in reference \cite{lopez}, where the
authors posed themselves the question why only the
$\ZZ_2$ cyclic group generated by $\hat A$ is actually realized as an
isometry group of the rigid special K\"ahlerian metric.
The answer is contained in the previous general discussion:
\begin{equation}
\begin{array}{ccccccc}
\ZZ_2 &=& \Gamma_{\cal W}^{rig} & \subset &
\Gamma_{\cal W} & = & D_3 \\
\end{array}
\label{diedro_7}
\end{equation}
Namely it is only $\ZZ_2$ that preserves the potential
with a unit rescaling factor.
The natural question at this point is
what is the relation of this
$\ZZ_2 \subset D_3$ with the dihedral $D_4$ symmetry
expected for $r=1$. The answer is simple:
the $\ZZ_4$ action in $D_4$ becomes a $\ZZ_2$ action on the $u$
variable, $u\to \alpha^2 u$, ($\alpha^4=1$).
\par
As it has been shown in \cite{CDF} the Picard--Fuchs equation
associated, in the $SU(2)$ case, with the symplectic section:
\begin{equation}
\begin{array}{cccc}
\Omega_u~ =~ &\partial_u \Omega = &
\partial_u \, \left ( \matrix{Y\cr{\o{\partial {\cal F}}{\partial Y}}
\cr} \right )~=~ & \left ( \matrix{\int_A \, \omega\cr\int_B \,\omega
\cr} \right )
\end{array}
\label{diedro_8}
\end{equation}
is
\begin{equation}
\label{diedro_10}
\left ( \partial_u \bfone - A_u \right ) V = 0\ , \label{mamma2}
\end{equation}
where $V$ is defined in eq.~\ref{V2r}, and the $2\times 2$ matrix connection
$A_u$ is given by:
\begin{equation}
\label{diedro_11}
A_u = \left(\matrix{0 & -{1\over 2}
\cr {- 1/2\over 1 - u^2}  &
{2 u\over 1 - u^2} }\right)
\end{equation}
with solutions
\begin{eqnarray}
\label{k3t2t2}
& {\del_u Y} \equiv f^{(1)}(u) =& \nonumber \\
& F({1\over 2},{1\over 2},1;{1 + u\over 2}) +
{\rm i} F({1\over 2},{1\over 2},1;{1 -u\over 2})\nonumber \\
& \partial_u \,\frac{\del {\cal F}}{\del Y} \equiv
f^{(2)}(u) = & \nonumber\\
& {\rm i} F({1\over 2},{1\over 2},1;{1 -u\over 2})&
\end{eqnarray}
As it is obvious, $f^{(1)}(u)$ and $f^{(2)}(u)$ just provide a
basis of two independent solutions of the linear second
order differential equation derived from the linear system
in eq.~\ref{diedro_10}. Any other pair of linear combinations of
the above functions would solve the same linear system.
The reason why precisely $f^{(1)}(u)$ and $f^{(2)}(u)$ are
respectively
identified with ${\del_u Y}$ and
$\partial_u \,\frac{\del {\cal F}}{\del Y}$  is given
by the boundary conditions imposed at infinity. When
$ u \, \rightarrow \, \infty $, the
special coordinate $Y(u)$ must approach the value it
has in the original microscopic $SU(2)$ gauge theory.
There the parameter $u$ was defined as the restriction to
the Cartan  subalgebra of the gauge invariant quadratic
polynomial
$ {\rm tr} \left ( Y^x {\o{\rm i}{2}} \sigma_x \right )^2$
so that $u={\rm const} (Y)^2$, the special coordinate
$Y(u)$ of the effective theory being $Y^z$ of the microscopic
one. Correspondingly the boundary condition at infinity
for $Y(u)$ is
\begin{equation}
Y(u) \approx 2\, \sqrt{2u}+\ldots \qquad\mbox{for }u\to\infty
\label{asintuno}
\end{equation}
At the same time when $ u \, \rightarrow \, \infty $ the
non perturbative rigid special geometry
must approach its perturbative limit  defined by the following
prepotential:
\begin{equation}
{\cal F}_{pert} (Y) \, \equiv \, {{\rm i}\over {2\pi}} \,
Y^2 \, \mbox{log} \, Y^2
\label{asintpert}
\end{equation}
Combining eq.~\ref{asintuno} and eq.~\ref{asintpert} we obtain
\begin{equation}
\frac{\del {\cal F}}{\del Y} \, \approx \, {{\rm i}\over {\pi}}
2\, \sqrt{2u} \, {\rm log } \, u +\ldots \qquad\mbox{for }u\to\infty
\label{asintdue}
\end{equation}
so that for $u\to\infty$ we have:
\begin{eqnarray}
& {\del_u Y}   \approx   \sqrt{{2 \over u}}
 +\ldots \qquad   \mbox{for } & \nonumber\\
& \partial_u \,\frac{\del {\cal F}}{\del Y}   \approx
 {{\rm i}\over {\pi}} \, 2\left ( {{1}\over {\sqrt{2 u}}}
 \, {\rm log} u \, + \, \sqrt{{2 \over u}} \right )
 +\ldots  &
\label{bordini}
\end{eqnarray}
The boundary conditions in eq.s~\ref{bordini} are just realized by
the choice of eq.s~\ref{k3t2t2}. Indeed using the relation
between hypergeometric functions and elliptic integrals:\footnote{We
use the notation $K(x)$ for what is usually denoted as $K(k)$ where
$x=k^2$, and similar for other elliptic integrals.}
\begin{eqnarray}
{\pi \over 2} \, F({1\over 2},{1\over 2},1;{x})&=& K(x) \nonumber \\
\, \equiv \, \int_{0}^{\pi \over 2} \,
\left (1 - x \, \mbox{Sin}^2 \theta  \right )^{-{1\over 2}}\,  d \theta & &
\nonumber\\
{\pi \over 2} \, F({1\over 2},-{1\over 2},1;{x})&=& E(x) \nonumber \\
\, \equiv \, \int_{0}^{\pi \over 2} \,
\left (1 - x \, \mbox{Sin}^2 \theta  \right )^{{1\over 2}} \, d \theta
& & \nonumber\\
{\pi \over 4} \, F({1\over 2},{1\over 2},2;{x})&=& B(x)= \nonumber \\
\left ( {E(x) \over x}\,  + \, {{x-1}\over {x}} \, K(x) \right )  & &
\end{eqnarray}
and the relations
\begin{eqnarray}
\int_{0}^{x} \, K(t) \, dt &=& 2\,x\, B(x)\nonumber\\
K(\ft{1}{x}) &=&\sqrt{x}\left(K(x)+iK(1-x)\right)\nonumber\\
 \int_0^x \ft1{\sqrt{t}} K(\ft1t) dt &=& 2\,\sqrt{x}\, E(\ft1x)+2i
\ ,
\end{eqnarray}
the solutions of eq.s~\ref{k3t2t2} can also be written as
\begin{eqnarray}
\label{periodandoprime}
 & {\del_u Y} \equiv f^{(1)}(u) & \nonumber \\
& =\frac{2}{\pi}\left[
K \left ({{1+u} \over 2} \right ) +
 {\rm i} \, K \left ({{1-u} \over 2} \right )\right] & \nonumber \\
& = \frac{2}{\pi}
\sqrt{{2 \over {1+u}}}
\, K \left (
 {2 \over {1+u}} \right )
 & \nonumber \\
& \partial_u \,\frac{\del {\cal F}}{\del Y} \equiv
f^{(2)}(u) =  \frac{2}{\pi}  & \nonumber \\
& = {\rm i} K \left ({{1-u} \over 2} \right )&
\end{eqnarray}
By means of an integration one then obtains:
\begin{eqnarray}
\label{integrando}
 &
Y(u) = \frac{2}{\pi}\int_{u_0}^u \,\sqrt{{2 \over {1+t}}} \, K \left (
 {2 \over {1+t}} \right ) \, dt \, = \,& \nonumber \\
& {8\over {\pi}} \, \sqrt{\frac{ 1+u}{2}} \,
 E \left ( {2 \over {1+u}} \right ) \, + \, {\rm const}
 & \nonumber \\
 & \frac{\del {\cal F}}{\del Y}  =\frac{2}{\pi}
{\rm i} \, \int_{0}^{u} \, K \left ({{1-t} \over 2} \right )
\, dt \, = & \nonumber\\
& -\frac{4i}{\pi} \,(1-u)\, B\left( \frac{1-u}{2}\right)
\, + \, {\rm const.}&
\end{eqnarray}
Choosing zero for the integration constants, the result
in eq.~\ref{integrando} coincides with the integral representations
originally given by Seiberg and Witten \cite{SW1}:
\begin{equation}
\label{paragonando}
\left\{\begin{array}{l}
Y(u) = 2a(u)=\frac{2\sqrt{2}}{\pi}
\int_{-1}^{1} \,\sqrt{{u-x} \over {1-x^2}} \, dx
 \\ \null\\
 \frac{\del {\cal F}}{\del Y}  =  2\,a_D(u)=2i\frac{\sqrt{2}}{\pi}
 \int_{1}^{u} \,
\sqrt{{u-x} \over {1-x^2}} \, dx
\ .\end{array}\right.
\end{equation}
Equipped with the above explicit solutions we can discuss
duality, monodromy R symmetry and the explicit special
metric on the rigid special manifold.
The duality group of
electric--magnetic rotations is, in this case \cite{SW1}:
\begin{eqnarray}
\Gamma_D~& \equiv &{\bar G}_\theta \, \subset \, PSL(2,\ZZ)
\nonumber\\
{\bar G}_\theta & \equiv & \left(\matrix{1 & -1 \cr 0 &
1 \cr }\right)^{-1}  \, G_\theta \left(\matrix{1 & -1 \cr 0 &
1 \cr }\right)
\label{diedro_24}
\end{eqnarray}
namely the conjugate, via the translation matrix
$\left(\matrix{1 & -1 \cr 0 & 1}\right)$
of that subgroup $G_\theta$ of the elliptic modular group which is
generated by the two matrices $S=\left(\matrix{0 & 1 \cr -1 &
0 \cr }\right)$ and $T^{(-2)}=\left(\matrix{1 & -2 \cr 0 &
1 \cr }\right)$ (\cite{gunning}). Indeed the group ${\bar G}_\theta$
is defined by its action on the symplectic section
$\Omega_u $ which is generated by the two matrices:
\begin{eqnarray}
\label{diedro_25}
R &=& \left(\matrix{- 1 & 2 \cr -1 & 1 \cr}\right) \, \nonumber\\
&=&
  \left(\matrix{1 & -1 \cr 0 &
1 \cr }\right)^{-1}  \, S \left(\matrix{1 & -1 \cr 0 &
1 \cr }\right) \nonumber \\
T_1 &=& \left(\matrix{1& -2\cr 0& 1 \cr }\right)
\,  \nonumber \\
&=&
\, \left(\matrix{1 & -1 \cr 0 &
1 \cr }\right)^{-1}  \, T^{(-2)} \left(\matrix{1 & -1 \cr 0 &
1 \cr }\right)
\end{eqnarray}
where $R$ is the $R$-symmetry generator and
$T_1$ is the monodromy matrix associated with the singular point
$u=1$ of the Picard--Fuchs system in eq.~\ref{mamma2}. This
is explained as follows.
The isometry group should be symplectically embedded. Hence,
naming $u_i \, \to \phi_i(u)$  the transformations belonging
to the {\it discrete isometry group}  there must exist
$M_\phi \, \in \, Sp(2r,\IR)$ such that
\begin{equation}
\begin{array}{cccc}
\null\\
{}~&\Omega  \left (\phi (u)\right )\,
 &=&
e^{i\theta_\phi} \, M_\phi \, \Omega
\left (u\right )\,\\
\null\\
{}~&\Omega_{u_i} \left (\phi (u)\right )\,
{\o {\partial \phi^{i}}{\partial u_j}} &=&
e^{i\theta_\phi} \, M_\phi \, \Omega_{u_j}
\left (u\right )\,\\
\end{array}
\label{diedro_26}
\end{equation}
The isometry $u\to -u$, corresponding to $R$-symmetry  (
$Y \to \, {\rm  i} \, Y$ in the perturbative limit )
induces, in this theory,
the transformation
\begin{eqnarray}
-\Omega_{u} ( - u)\, &= &
\mbox{i} \, \pmatrix { -1 &  2 \cr -1 &  1 \cr} \,
\pmatrix{ f^{(1)}(u) \cr f^{(2)}(u) \cr } \nonumber \\
&=& \mbox{i} \, R\,  \Omega_{u}(u)
\end{eqnarray}
while the monodromy transformation around $u=1$ gives ($r$ small)
\begin{eqnarray}
\Omega_{u} \left (1+r\, e^{2\pi i} \right )\, & = & \,
 \pmatrix { 1 & -2 \cr 0 & 1 \cr } \,
\pmatrix{ f^{(1)}(1+r) \cr f^{(2)}(1+r) \cr }\nonumber \\
&=& T_1\,  \Omega_{u}(1+r)
\label{diedro_27quat}
\end{eqnarray}
Having recalled the explicit form of the isometry--duality group
let us now study the structure of the rigid special metric.
To this effect let us introduce the ratio
of the two solutions to eq.~\ref{mamma2},
\begin{equation}
\label{k3t2t5}
{ {\bar {\cal N}}}(u) = {f^{(2)}(u)\over f^{(1)}(u)}.
\end{equation}
Recalling eq:~\ref{rieperiod}, we know that
such a ratio is identified with the matrix ${ {\bar {\cal N}}}$ appearing
in the vector field kinetic terms:
\begin{equation}
{\cal L}^{vector}_{kin}\,=\, { \o{{\rm i}}{2} } \,
[ {\bar {\cal N}}(u) \, F^{-}_{\mu\nu} \,F^{-}_{\mu\nu} \, - \,
{\cal N}({\bar u}) \, F^{+}_{\mu\nu} \,F^{+}_{\mu\nu} ]
\label{diedro_27}
\end{equation}
Recalling  eq.\ref{sympinvrig}  we can now write the explicit
form of the rigid special K\"ahler metric in the variable $u$:
\begin{eqnarray}
ds^2 &=& g_{u{\bar u}} \, |du|^2 \nonumber \\
  g_{u{\bar u}} &=&
2 \, {\rm Im}\, {\bar {\cal N}}(u) \, |f^{(1)}(u)|^2
\label{metricozza}
\end{eqnarray}
\par
Calculating the Levi--Civita connection and Riemann tensor of
this metric we obtain:
\begin{eqnarray}
\Gamma^{u}_{u u}&=&-\, g^{u{\bar u}} \,
\partial_u \, g_{u{\bar u}} \nonumber\\
&=&-{\o{1}{2 {\rm i}}} \,
{\o { \partial {\bar {\cal N}}/\partial u}{ {\rm Im}\,
 {\bar {\cal N}}(u) } }\, - \,
\partial_u \mbox{log} f^{(1)}(u) \nonumber \\
R^{u}_{u{\bar u }u}&=& \,
\partial_{\bar u} \,\Gamma^{u}_{u u}\nonumber\\
&=&{\o{1}{4 }} \,
{\o {1 }{ ({\rm Im}\, {\bar {\cal N}}(u)})^2 } \,
|\partial {\bar {\cal N}}/\partial u |^2 \nonumber \\
R_{{\bar u}u{\bar u} u}&=& \,
g_{u{\bar u}} R^{u}_{u{\bar u}u }\nonumber \\
&=&{\o{1}{2 }} \,
{\o {1 }{ {\rm Im} \,{\bar {\cal N}}(u)} } \,
|\partial {\bar {\cal N}}/\partial u |^2 \, |f^{(1)}(u)|^2
\label{curvelievissime}
\end{eqnarray}
so that we can verify that the above metric is
indeed {\it rigid special
K\"ahlerian}, namely that it satisfies the constraint:
\begin{equation}
R_{{\bar u}u{\bar u} u}\, - \, C_{uuu} \,
{\bar C}_{{\bar u}{\bar u}{\bar u}} \, g^{u{\bar u}}~=~0
\label{costretto_1}
\end{equation}
by calculating the Yukawa coupling or anomalous magnetic moment
tensor:
\begin{equation}
C_{uuu}~=~ \partial_u {\bar {\cal N}} \,
\left ( f^{(1)}(u) \right )^2
\label{ctensore}
\end{equation}
As one can notice from its explicit form (see eq.~\ref{metricozza}),
the K\"ahler metric of the rigid N=2 gauge theory of rank
$r=1$ is not the Poincar\'e metric in the variable
${\bar {\cal N}}$, as one might naively expect from the fact that
${\bar {\cal N}}=\tau$ is the standard modulus of a torus and that
$G_{\theta}\, \subset \, PSL(2,\ZZ)$ linear fractional
transformations are isometries. Indeed eq.~\ref{metricozza}
is to be contrasted with the expression for the
Poincar\'e metric:
\begin{equation}
ds^2~=~g^{Poin}_{{\cal N}{\bar{\cal N}}}\,
 |d {\bar {\cal N}} \, |^2
{}~=~ \, {\o{1}{4}} \,
{\o{1}{({\rm Im}\, {\bar {\cal N}})^2}} \, |d {\bar {\cal N}} \, |^2
\label{metricapoincare}
\end{equation}
{}From eq.~\ref{curvelievissime} however it is amusing to note
that the Ricci
form of the rigid metric is precisely the Poincar\'e metric.
\begin{equation}
R^{Ricci}_{~~~~{{\cal N}}{\bar { {\cal N}}}}~=~
g^{Poin}_{~~~~{ {\cal N}}{ {\bar {\cal N}}}}
\label{masterequation}
\end{equation}
This is a consequence of the general equation eq.~\ref{boh} in the case of
 one modulus where the period matrix $\cal N$ can be used as a parameter.
\subsection{The rigid special coordinates}
In the special coordinate basis the anomalous magnetic moment tensor
is given by:
\begin{eqnarray}
C_{YYY}&=& C_{uuu} \, \left ( {\o{\partial u}{\partial Y}}
\right )^3 \nonumber\\
&=&-{{2{\rm i}} \over \pi}
{\o{1}{1-u^2}} \, \left ( {\o{\partial u}{\partial Y}} \right )^3
\label{cyyy}
\end{eqnarray}
The second of eq.s~\ref{cyyy} follows from the comparison
between eq.~\ref{ctensore} and the Picard--Fuchs
eq.~\ref{diedro_10} satisfied by the periods that yields:
\begin{equation}
C_{uuu}~=~-{{2{\rm i}} \over \pi} \, {\o{1}{1-u^2}}
\end{equation}
In the large $u$ limit the asymptotic behaviour of the special
coordinate is given by eq.~\ref{asintuno}:
\begin{equation}
C_{YYY}(u)~=~{\o{\partial^3 {\cal F}}{\partial Y^3}}(u)~
 \approx~{\o{{\rm i}
 }{\sqrt{2}\pi}} \,  u^{-1/2}+\ldots
 \end{equation}
and by triple integration one verifies consistency with
the asymptotic behaviour
of the prepotential ${\cal F}(Y)$ (see eq.~\ref{asintpert}):
\begin{equation}
{\cal F}(Y)~\approx~
\mbox{const} \, Y^2  \log {Y^2}  \, +
\ldots \qquad\mbox{for }Y\to\infty
\label{asintoto_2}
\end{equation}
The formula in eq.~\ref{asintoto_2} contains the leading classical
form of ${\cal F}(Y)$ plus the first perturbative correction
calculated with standard techniques of quantum field--theory.
Eq.~\ref{asintoto_2} was the starting point of the analysis
of Seiberg and Witten who from the perturbative
singularity structure inferred the monodromy group and then
conjectured the dynamical Riemann surface. The same procedure
has been followed to conjecture the dynamical Riemann surfaces
of the higher rank gauge theories.
The nonperturbative solution is given by
\begin{equation}
{\cal F}(Y)~=~  \frac{i}{2\pi} Y^2  \log \frac{Y^2}{\Lambda^2}
+Y^2\sum_{n=1}^{\infty}C_n  \left(
\frac{\Lambda^2}{Y^2}\right) ^{2n}
\label{generalformofeffe}
\end{equation}
The infinite series in eq.~\ref{generalformofeffe} corresponds
to the sum over instanton corrections of all instanton--number.
\par
 The important thing to note is that the special coordinates $Y^\alpha(u)$
of rigid special geometry approach for large values of $u$ the
Calabi--Visentini coordinates of the manifold $ST[2,n] $
discussed in  previous lectures.
As stressed there, the $Y^\alpha$ are
not special coordinates for local special geometry.


\end{document}